\documentclass[useAMS,usenatbib]{mn2e}
\usepackage{graphicx}
\usepackage{mathptmx}
\def \msol  {\hbox{M$_{\odot}$}}
\def \lsol  {\hbox{L$_{\odot}$}}
\def \arcdeg {\hbox{$^{\circ}$}}                        
\def \arcmin {\hbox{$^\prime$}}                         
\def \arcsec {\hbox{$^{\prime\prime}$}}                 
\def \kms    {\hbox{${\rm km\,s}^{-1}$}}                

\title[GBS HARP Survey: Orion B]{The JCMT Legacy Survey of the Gould Belt: a
first look at Orion B with HARP}
\author[The GBS Orion Team]{J.V. Buckle$^{1,2}$\thanks{E-mail:
j.buckle@mrao.cam.ac.uk},
E.I. Curtis$^{1,2}$,
J.F. Roberts$^{3}$,
G.J. White$^{4,5}$,
J. Hatchell$^{6}$,
C. Brunt$^{6}$, \newauthor
H.M. Butner$^{7,8}$,
B. Cavanagh$^{8}$,
A. Chrysostomou$^{8,9}$,
C.J. Davis$^{8}$,
A. Duarte-Cabral$^{10}$,\newauthor
M. Etxaluze$^{4,5}$,
J. Di Francesco$^{11,12}$,
P. Friberg$^{8}$,
R. Friesen$^{11,12}$,
G.A. Fuller$^{10}$,\newauthor
S. Graves$^{1,2}$, 
J.S. Greaves$^{13}$,
M.R. Hogerheijde$^{14}$,
D. Johnstone$^{11,12}$,
B. Matthews$^{11}$,\newauthor
H. Matthews$^{15}$,
D. Nutter$^{16}$,
J.M.C. Rawlings$^{17}$,
J.S. Richer$^{1,2}$,
S. Sadavoy$^{11,12}$,\newauthor
R.J. Simpson$^{16}$,
N.F.H. Tothill$^{6}$,
Y.G. Tsamis$^{18}$,
S. Viti$^{17}$,
D. Ward-Thompson$^{16}$,\newauthor
J.G.A. Wouterloot$^{8}$,
J. Yates$^{17}$  \\
$^{1}$Astrophysics Group, Cavendish Laboratory, J J Thomson Avenue, Cambridge, CB3 0HE\\
$^{2}$Kavli Institute for Cosmology, c/o Institute of Astronomy, University of Cambridge, Madingley Road, Cambridge, CB3 0HA \\ 
$^{3}$Centro de Astrobiolog\'{i}a (CSIC/INTA), Ctra de Torrej\'{o}n a Ajalvir km 4, E-28850 Torrej\'{o}n de Ardoz, Madrid, Spain\\
$^{4}$Department of Physics and Astronomy, Open University, Walton Hall, Milton
Keynes, UK\\
$^{5}$Science and Technology Facilities Council, Rutherford Appleton
Laboratory, Chilton, Didcot, UK\\
$^{6}$School of Physics, University of Exeter, Stocker Road, Exeter, UK\\
$^{7}$Department of Physics and Astronomy, James Madison University, 901
Carrier Drive, Harrisonburg, VA 22807, USA\\
$^{8}$Joint Astronomy Centre, 660 N. A'Ohoku Place, University Park, Hilo,
Hawaii 96720, USA\\
$^{9}$School of Physics, Astronomy and Mathematics, University of
Hertfordshire, College Lane, Hatfield, UK\\
$^{10}$Jodrell Bank Centre for Astrophysics, School of Physics and Astronomy, The University of Manchester, Oxford Road, Manchester M13 9PL, UK\\
$^{11}$National Research Council Canada, Herzberg Institute of Astrophysics, 5071 West Saanich Rd, Victoria, BC, V9E 2E7\\
$^{12}$Department of Physics \& Astronomy, University of Victoria,  3800 Finnerty Rd., Victoria, BC, Canada\\
$^{13}$Scottish Universities Physics Alliance, Physics \& Astronomy, University
of St Andrews, North Haugh, St Andrews, Fife, UK\\
$^{14}$Leiden Observatory, Leiden University, PO Box 9513, 2300 RA, Leiden, The
Netherlands\\
$^{15}$National Research Council of Canada, Dominion Radio Astrophysical
Observatory, 717 White Lake Rd., Penticton, BC, Canada\\
$^{16}$School of Physics \& Astronomy, Cardiff University, 5 The Parade,
Cardiff, UK\\
$^{17}$Dept of Physics \& Astronomy, University College London, Gower Street,
London, UK\\
$^{18}$Instituto de Astrof\'isica de Andaluc\'ia (CSIC), Camino Bajo de Hu\'tor 50, 18008 Granada, Spain}

\begin{document}

\date{Accepted 2009 August 27.  Received 2009 August 3; in original form 2009 July 2}

\pagerange{\pageref{firstpage}--\pageref{lastpage}} \pubyear{2009}

\maketitle

\label{firstpage}

\begin{abstract}
The Gould Belt Legacy Survey will survey nearby star-forming regions (within~500~pc), using HARP (Heterodyne Array Receiver Programme),  SCUBA-2
(Submillimetre Common-User Bolometer Array 2) and POL-2 (Polarimeter 2) on the
James Clerk Maxwell Telescope (JCMT). This paper describes the initial data
obtained using HARP to observe $^{12}$CO, $^{13}$CO and C$^{18}$O
$J~=~3\rightarrow2$ towards two regions in Orion B, NGC 2024 and NGC 2071. We
describe the physical characteristics of the two clouds, calculating
temperatures and opacities utilizing all three isotopologues. We find good
agreement between temperatures calculated from CO and from dust emission in the
dense, energetic regions. We determine the mass and energetics of the clouds,
and of the high-velocity material seen in $^{12}$CO emission, and compare the
relative energetics of the high- and low-velocity material in the two clouds.
We present a {\sc CLUMPFIND} analysis of the $^{13}$CO condensations. The slope of the condensation
mass functions, at the high-mass ends, is similar to the slope of the initial
mass function.
\end{abstract}


\section{Introduction}

\subsection{The Gould Belt Legacy Survey}
The Gould Belt Legacy Survey \citep[GBS,][]{wardthompson2007} has been awarded
612 hours of time on the James Clerk Maxwell Telescope (JCMT) to survey nearby
star-forming regions (within~500~pc), using HARP \citep[Heterodyne Array
Receiver Programme,][]{buckle},  SCUBA-2 \citep[Submillimetre Common-User
Bolometer Array 2,][]{hollandb} and
POL-2 \citep*[Polarimeter 2,][]{bastien}. The HARP component will observe a
large typical sample of prestellar and protostellar sources in three CO
isotopologues. CO $J=3\rightarrow$2 and the same transition in $^{13}$CO and C$^{18}$O are excited in the physical conditions of
star-forming molecular clouds, where typical temperatures and densities are
10--50 K and 10$^4$--10$^5$ cm$^{-3}$. $^{12}$CO observations are used to
search for and map any high-velocity outflows present, while the isotopologues
measure the line widths and velocity profiles in the cores and filaments,
resolving detailed kinematic and density properties of the cores. The GBS aims
to provide a large and unbiased sample of star-forming material in the solar
vicinity at relatively high spatial (8--14 arcsec) and spectral (0.05--1.0
\kms) resolution. The GBS will trace the very earliest stages of star
formation, through submillimetre continuum imaging using SCUBA-2 observations
of low-temperature, high-column density regions, and provide an inventory of
all the protostellar objects contained in the nearby molecular clouds of the
Gould Belt, covering $\sim$~700 deg$^2$. The HARP  observations trace the
kinematics of molecular gas in the cores and clusters at the same spatial
resolution as the 850~$\umu$m SCUBA-2 observations of the dust emission. The
key science goals of the HARP component of the GBS are
\citep{wardthompson2007}:

\begin{itemize}
\item To search for and map high-velocity outflows in the cores in order to
differentiate between starless and protostellar cores.
\item To derive simple constraints on the column density and CO depletion in
these cores.
\item To investigate support mechanisms and cloud/cluster/core evolution.
\item To characterize the cloud kinematics in a wide variety of environments
and investigate the role and evolution of turbulence in star formation.
\end{itemize}

The HARP targets for the GBS are a sample of the cloud regions that will be
observed by SCUBA-2, with sizes $\sim$0.8 deg$^2$, and containing filaments and
clustered star formation. This paper describes the initial HARP data obtained
for two targets in Orion B, NGC 2024 and NGC 2071.

\subsection{Orion B}
\begin{figure}
\includegraphics[width=8cm,angle=0]{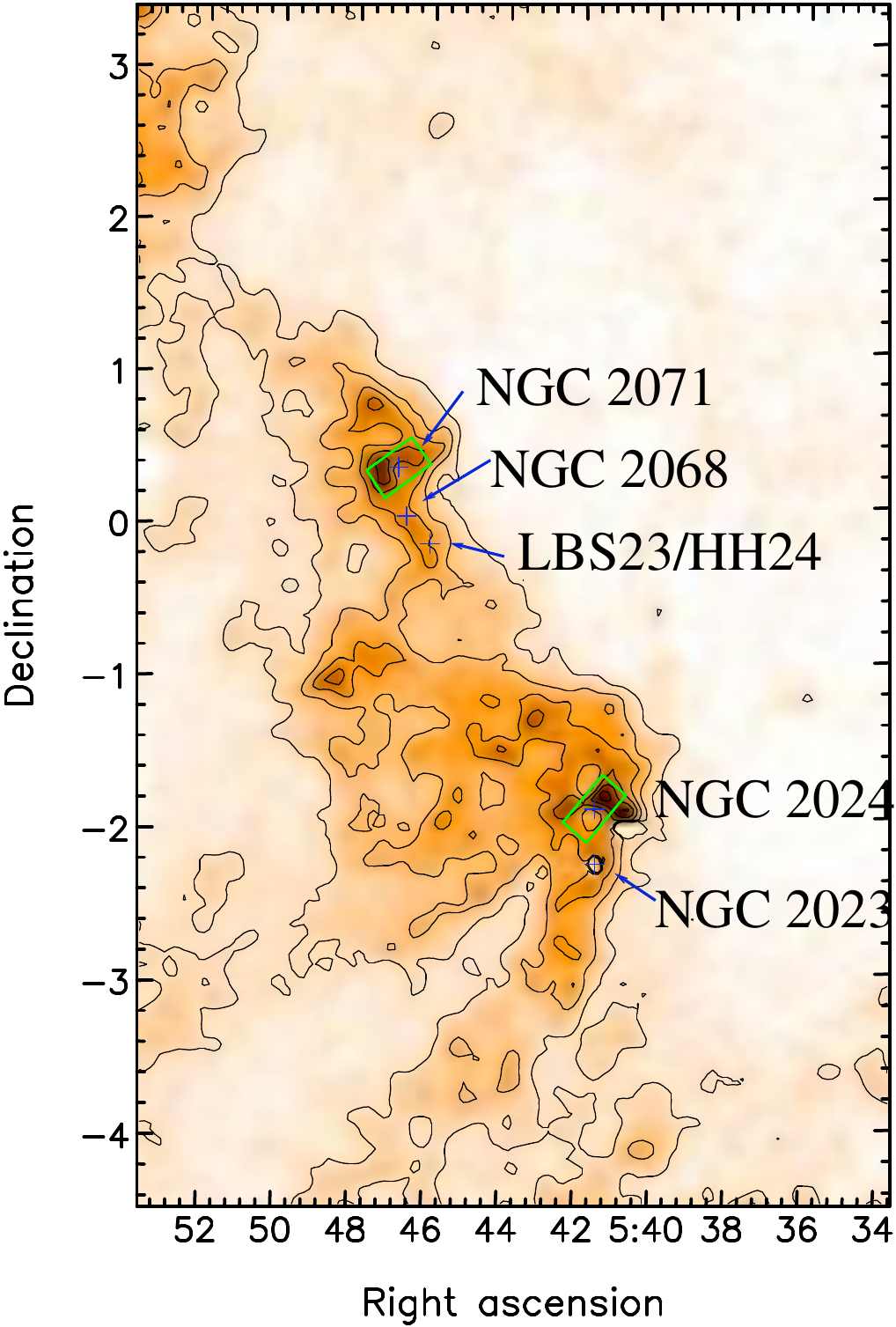}
\caption{\label{fig-orion} NGC 2024 and NGC 2071 shown within the Orion B cloud
complex seen in visual extinction, A$_\nu$ from the atlas of
\citet{dobashi2005}. The contours are drawn at A$_\nu$=1,2,3,4,5,6 mag. }
\end{figure}

The Orion B cloud complex is the closest region of high-mass star formation,
lying at a distance of 415~pc  \citep{anthony1982,menten2007}. Five main star
formation regions were identified from early wide field CO surveys: NCG
2024, NGC 2071, LBS 23 (HH 24), NGC 2068 and NGC 2023
\citep*{tucker1973,kutner1977, white1981a,wilson2005}. The cloud
complex subtends an area of $\sim$ 26 deg$^2$, which corresponds to $\sim$
1.5~kpc$^2$, and has a mass of $\sim$ 8 $\times$ 10$^4$~$\msol$. The main
star forming regions are clearly visible in the unbiased surveys for young
stellar objects \citep*[][hereafter
J06]{lada1991a,lis1999,motte2001,mitchell2001,johnstone2006} and in the
 dense gas tracer CS \citep{lada1991b}. Observations of individual sources
in CO lines revealed a complex structure of outflows and gas motions
\citep*{white1981b,phillips1982}, which are now known to
be associated with extensive star formation activity \citep{flaherty2008}. A
detailed summary of the Orion B regions can be found in the Handbook of Star
Forming Regions \citep[see chapters by][]{gibb2009,bally2009,meyer2009}. The
two GBS Orion B targets, NGC 2024 and NGC 2071, for which we have obtained  the
first data are shown as boxes in Fig.~\ref{fig-orion} overlaid on an extinction
map of the Orion B cloud complex.

\subsubsection{NGC 2024}

NGC 2024 \citep[The Flame Nebula,][]{anthony1982} is a bright emission
nebula that is crossed by a prominent dust lane. Intense radio continuum
and recombination line emission
\citep*{krugel1982,barnes1989,bik2003,rodriguez2003}, plus the detection of an
infrared cluster, suggested the presence of ionizing star(s) of spectral
type in the range O8 V-B2 V
\citep*{lada1991a,comeron1996,giannini2000,haisch2000,bik2003,kandori2007}.

Millimetre and submillimetre wavelength observations revealed a number of
compact condensations that were believed to mark star forming sites. NGC 2024 has also been shown to contain a number of protostars
positioned along a ridge of enhanced star formation activity, and its
associated dense molecular material distributed as a filamentary structure
\citep{mezger1988,mezger1992,visser1998}. Far-infrared and submillimetre polarimetric
observations tracing the
magnetic field structure of the star-forming ridge suggest it contains
helical magnetic
fields surrounding a curved filamentary cloud, possibly driven by the
ionization front of the expanding H{\sc ii} region
\citep*{hildebrand1995,dotson2000,greaves2001,matthews2002}. Previous molecular
line observations of NGC 2024
have included studies of the outflow and thermal properties of
the gas
\citep*{watt1979,white1981a,phillips1988,barnes1989,richer1989,barnes1990,richer1992,chernin1996,mangum1999,snell2000},
revealing a complex and boisterous medium
permeated with outflows, turbulence and complex gas dynamics.

\subsubsection{NGC 2071}

NGC 2071 is associated with a reflection nebula, molecular cloud and
infrared cluster NGC 2071IR \citep[][hereafter
J01]{white1981a,white1981b,persson1981,motte2001,johnstone2001}. The
main infrared cluster has been resolved into eight distinct near-infrared
sources having a total luminosity of 520~$\lsol$
\citep{butner1990,walther1993}. The bipolar molecular outflow from NGC 2071 has
been
extensively studied in CO emission
\citep*{white1981a,snell1984,scoville1986,moriarty1989,kitamura1990,houde2001,eisloffel2000},
and is found to be amongst
the most energetic bipolar outflows currently known, and extends for $\sim$
2~pc \citep*{margulis1989,chernin1992,chernin1995,stojimirovic2008}. This
outflow is also seen
in the shock excited 2.12 $\umu$m H$_2$ line
\citep*{persson1981,lane1986,burton1989,garden1990,aspin1992}.

\citet{snell1986} have reported the presence of several extended radio
continuum sources. Subsequent observations of H$_2$O
masers and the associated radio continuum emission
\citep*{smith1994,torrelles1998,seth2002} suggest that the
infrared sources IRS 1 and IRS 3 have strong winds, and are surrounded by
rotating circumstellar disks. Further outflow activity around IRS 3 was
suggested by its H$_2$ emission \citep{aspin1992} to indicate the presence
of a circumstellar disk oriented perpendicular to the main outflow axis
\citep{walther1993,torrelles1998}, that presumably contains a
massive protostar.

\subsubsection{Outline}

In this paper, we present the initial heterodyne data on Orion B taken for the GBS. We
determine characteristic physical properties for two clouds, NGC 2024 and
NGC 2071, comparing and contrasting regions with varying physical
characteristics.  We present measurements of temperature and opacity across the regions,
using these to provide temperature- and opacity-corrected
masses and energetics of the clouds and the high-velocity
material.  We use our $^{13}$CO data to investigate clumpy structure
within the clouds, calculate the mass function of the
condensations, and compare the values found with core mass functions
in these regions, and to the initial mass function.
Sec.~\ref{sec:obs} provides an overview of the observations and data
reduction.  Sec.~\ref{sec:results} describes the reduced data sets,
highlighting regions of interest within the clouds with reference to
previous observations.  Sec.~\ref{sec:cloud} presents the physical
characteristics of opacity, temperature, energetics and kinematics for
the clouds, while Sec.~\ref{sec:clumps} presents an analysis of the
condensations as traced by $^{13}$CO emission. We summarize the
results presented in the paper in Sec.~\ref{sec:summ}.

\section{Observations}
\label{sec:obs}

\subsection{Overview of HARP}
HARP is a recently-commissioned spectral-imaging receiver for the JCMT operating at
submillimetre wavelengths \citep{buckle,smith2003,smith2008}. It works in conjunction with the new back-end
correlator, ACSIS \citep[Auto-Correlation Spectral Imaging
System,][]{buckle,dent} offering high-spectral and spatial resolution, the
latter matched to that of SCUBA-2 \citep{hollandb}. HARP maps spectral lines in
the 325-375~GHz atmospheric window, the JCMT B-band frequencies, where
transitions of some of the most abundant molecules in the interstellar medium
reside. The CO, $^{13}$CO and C$^{18}$O $J=3 \to 2$ lines in this window are
used by the GBS to trace dense and/or warm gas around star-forming cores. The
HARP imaging array consists of 16 SIS detectors arranged in a $4\times 4$ grid,
separated by 30~arcsec. The beam size is 14~arcsec at 345~GHz, which means this
array arrangement under-samples the focal plane with respect to Nyquist and
further data points must be taken in between the nominal on-sky positions to
produce a fully-sampled map. HARP works in single sideband mode, with typical
receiver noise temperatures under 150~K and main-beam efficiencies of
$\eta_\mathrm{mb}=0.61$ \citep{buckle}. The K-mirror rotates the array with
respect to the plane of the sky to maximize the observing efficiency, for
instance allowing large raster scan maps to be made with the array oriented at
an angle to the scan direction. ACSIS provides either wide-band (up to 1.9~GHz
wide) or high-resolution spectra (up to 31~kHz channels). Additionally the IF
can be separated into two sub-bands to allow simultaneous imaging of two
close-together transitions e.g. the $^{13}$CO and C$^{18}$O $J=3 \to 2$ lines
shown here.

\subsection{Description of Observations}
The observations presented here comprise 31.3 hrs of data taken on multiple
nights from November 2007 to October 2008. The system temperatures, which are
measured for each detector in every observation, varied (across different weather bands) from 344 to 684~K for
the $^{12}$CO and 343 to 575~K for the $^{13}$CO/C$^{18}$O data. Two fields in
Orion B are analyzed: (i) NGC 2024, $(10.8\times 22.5)$~arcmin$^2$ centred on
05$^{\rm h}$41$^{\rm m}$39$^{\rm s}$.8, -01$^{\arcdeg}$54$^{\arcmin}$27$^{\arcsec}$ (J2000) at PA$=-40$~deg (east of north) (ii) NGC 2071,
$(13.5\times21.6)$~arcmin$^2$, centred on 05$^{\rm h}$47$^{\rm m}$00$^{\rm s}$.0, +00$^{\arcdeg}$19$^{\arcmin}$53$^{\arcsec}$ (J2000) angled at
PA$=-55$~deg. The maps were taken in the raster position-switched observing
mode, where spectra are taken ``on-the-fly'' with the telescope constantly
scanning in a direction parallel to the sides of the map, taking spectra
separated at 7.3~arcsec along this direction. The array is inclined at an angle
of $\sim$14~deg to the scan direction which produces rows separated by
7.3~arcsec perpendicular to the scan direction automatically \citep{buckle}. At
the end of a scan row, the map is displaced half the array spacing
perpendicular to the scan direction before a new row is started. This overlaps
half of the new scan row's data points with the first to double the integration
time and even out any noise variations due to missing receptors or intrinsic
differences in receptor performance. The noise is further evened-out by
observing a second map scanning in a perpendicular direction to the first to
``basket-weave'' the field.

Separate absolute off positions were used for each field: (1)
05$^{\rm h}$39$^{\rm m}$00$^{\rm s}$,-01$^{\arcdeg}$00$^{\arcmin}$00$^{\arcsec}$ for NGC 2024 and (2) 05$^{\rm h}$43$^{\rm m}$44$^{\rm s}$, +00$^{\arcdeg}$21$^{\arcmin}$42$^{\arcsec}$.2 for NGC 2071. Each
was verified to be absent of emission by examining short 60~s position-switched
`stare' observations towards each position in $^{12}$CO. A low-resolution
$^{12}$CO mode (1~GHz bandwidth with 977~MHz channel spacing) was used for this
inspection as it provided lower noise and ensured the other weaker
isotopologues would also be free of emission.

\subsubsection{The GBS HARP/ACSIS configuration}

The $^{12}$CO data were taken in an ACSIS dual sub-band
mode: (1) with modest resolution to map the broad line flows, at a rest
frequency of 345.796~GHz, with the 1~GHz bandwidth divided into 1024
channels, separated by 977~kHz
($\sim$0.85~km~s$^{-1}$) and (2), simultaneously, with higher resolution to map the
detail of outflows close to the ambient cloud velocity, again at a rest
frequency of 345.796~GHz, with 250~MHz bandwidth divided into 4096
channels, separated by 61~kHz
($\sim$0.05~km~s$^{-1}$). Similarly, the $^{13}$CO/C$^{18}$O data were
taken in a high-resolution dual sub-band mode, each sub-band has a
rest frequency of 330.588 or 329.331~GHz respectively with the 250~MHz
band separated into 4096 61~kHz channels (at a spacing of
0.055~km~s$^{-1}$).

\subsection{Data Reduction}
The data were reduced using the Starlink project
software\footnote{http://starlink.jach.hawaii.edu/starlink}, with KAPPA
\citep{currie} routines used to mask out poorly-performing detectors and
calculate a self-flat \citep*[see][]{curtis} for each observation. SMURF \citep{jenness} routines were
used to make the cube using a Gaussian gridding technique, with 6~arcsec pixels
and a 9~arcsec beam, giving an effective angular resolution of 16.6~arcsec in
the final datacubes. KAPPA routines were then used to remove a linear baseline
and crop the edges of the cubes spatially and spectrally. These data are
presented as a `first look' at the data being gathered for the GBS towards Orion. Data in the isotopically-substituted species towards
these regions are still being collected, and the final data sets will have a
much increased signal-to-noise.  Table \ref{tab:noise} describes the mean noise
achieved in the datacubes collected to date, and the noise requirements of the
final GBS data products. 
Data are presented in units of antenna temperature \citep[$T_A^*$,][]{kutner1981}, which can be converted to main beam temperature ($T_{\rm mb}$) using $T_{\rm mb}$ = $T_A^* / \eta_{\rm mb}$.

\begin{table}
\caption{\label{tab:noise}Current and required noise levels of GBS Orion B
data}
\begin{tabular}{lrrrrrrrr}
\hline
&\multicolumn{3}{c}{current levels$^a$}&\multicolumn{3}{c}{required
levels$^b$}\\
Source&$^{12}$CO&$^{13}$CO&C$^{18}$CO&$^{12}$CO&$^{13}$CO&C$^{18}$CO\\
&\multicolumn{3}{c}{mean 1$\sigma$ rms /K}&\multicolumn{3}{c}{mean 1$\sigma$
rms /K}\\
\hline
NGC 2024&0.10&0.26&0.28&0.3&0.25&0.3\\
NGC 2071&0.11&0.32&0.43&0.3&0.25&0.3\\
\hline
\end{tabular}\\
{\footnotesize
$^a$6~arcsec pixels, 16.6~arcsec beam, 1.0
($^{12}$CO)/0.1($^{13}$CO/C$^{18}$O)~\kms\ channels.\\
$^b$7.5~arcsec pixels, 14~arcsec beam, 1.0
($^{12}$CO)/0.1($^{13}$CO/C$^{18}$O)~\kms\ channels.}
\end{table}

\section{Results}
\label{sec:results}

\subsection{HARP spectral imaging}

\subsubsection{NGC 2024}

\begin{figure}
\begin{minipage}{85mm}
\vbox{
\includegraphics[width=8cm,angle=-90]{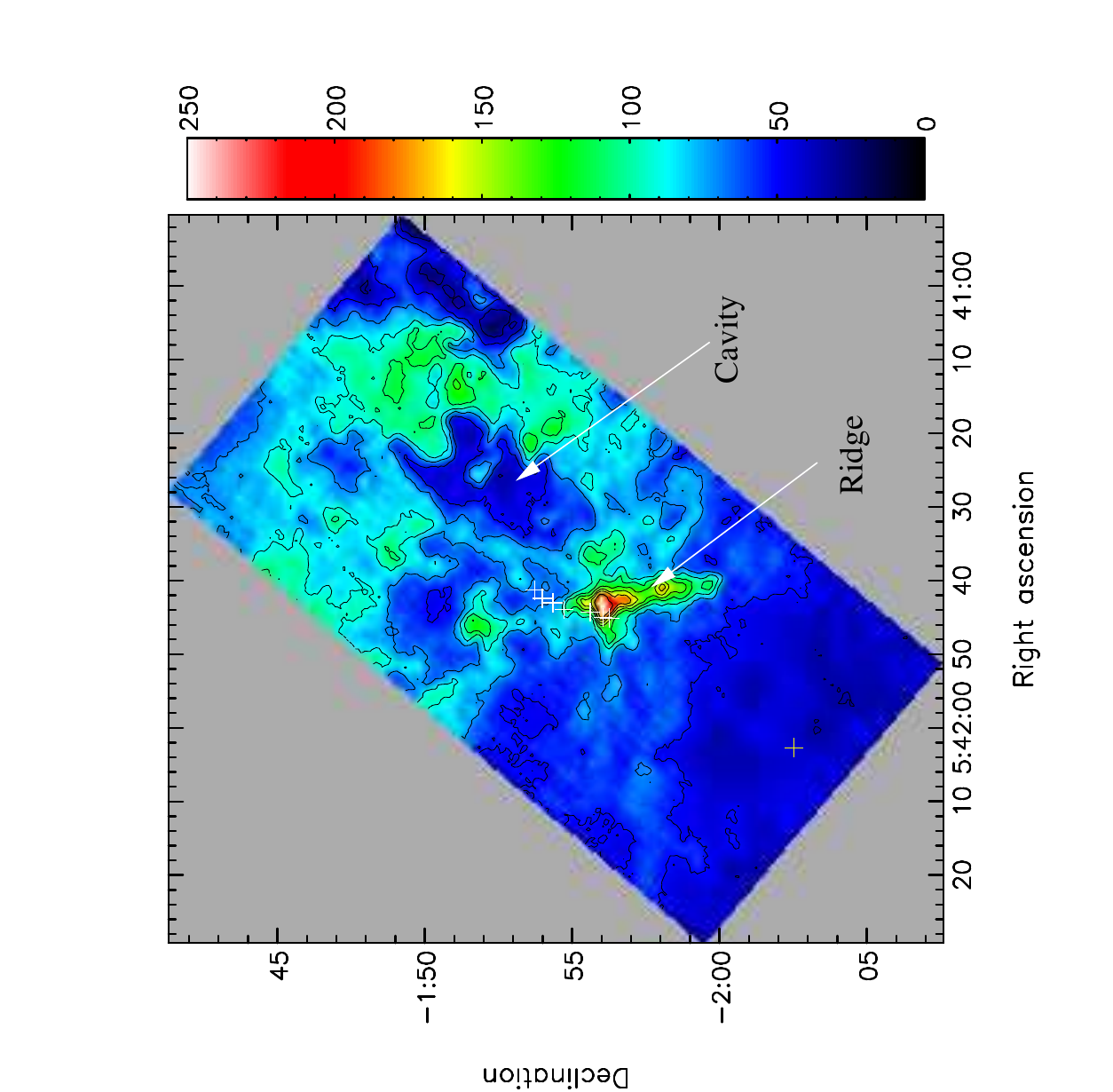}
\includegraphics[width=7.3cm,angle=0]{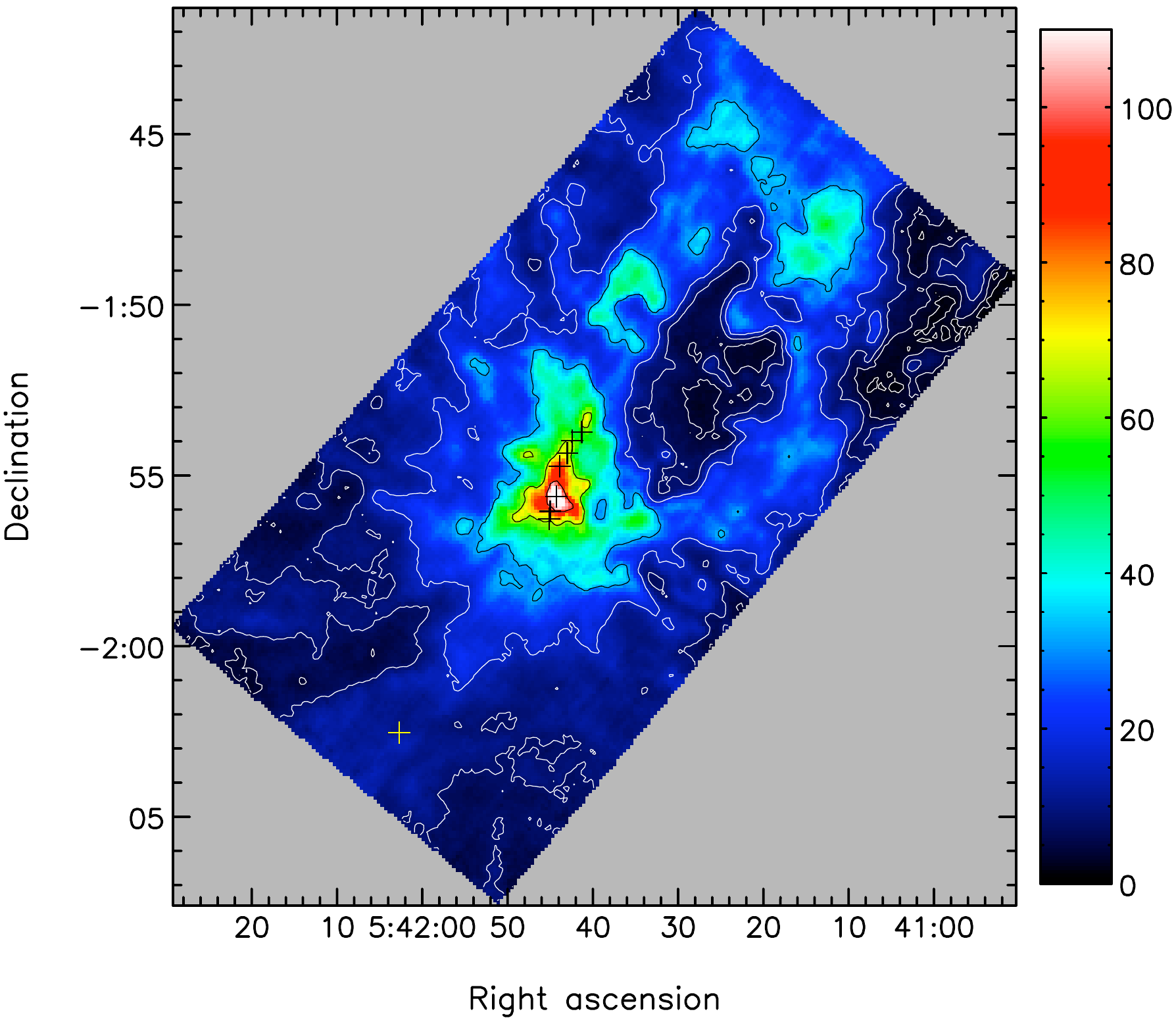}
\includegraphics[width=7.3cm,angle=0]{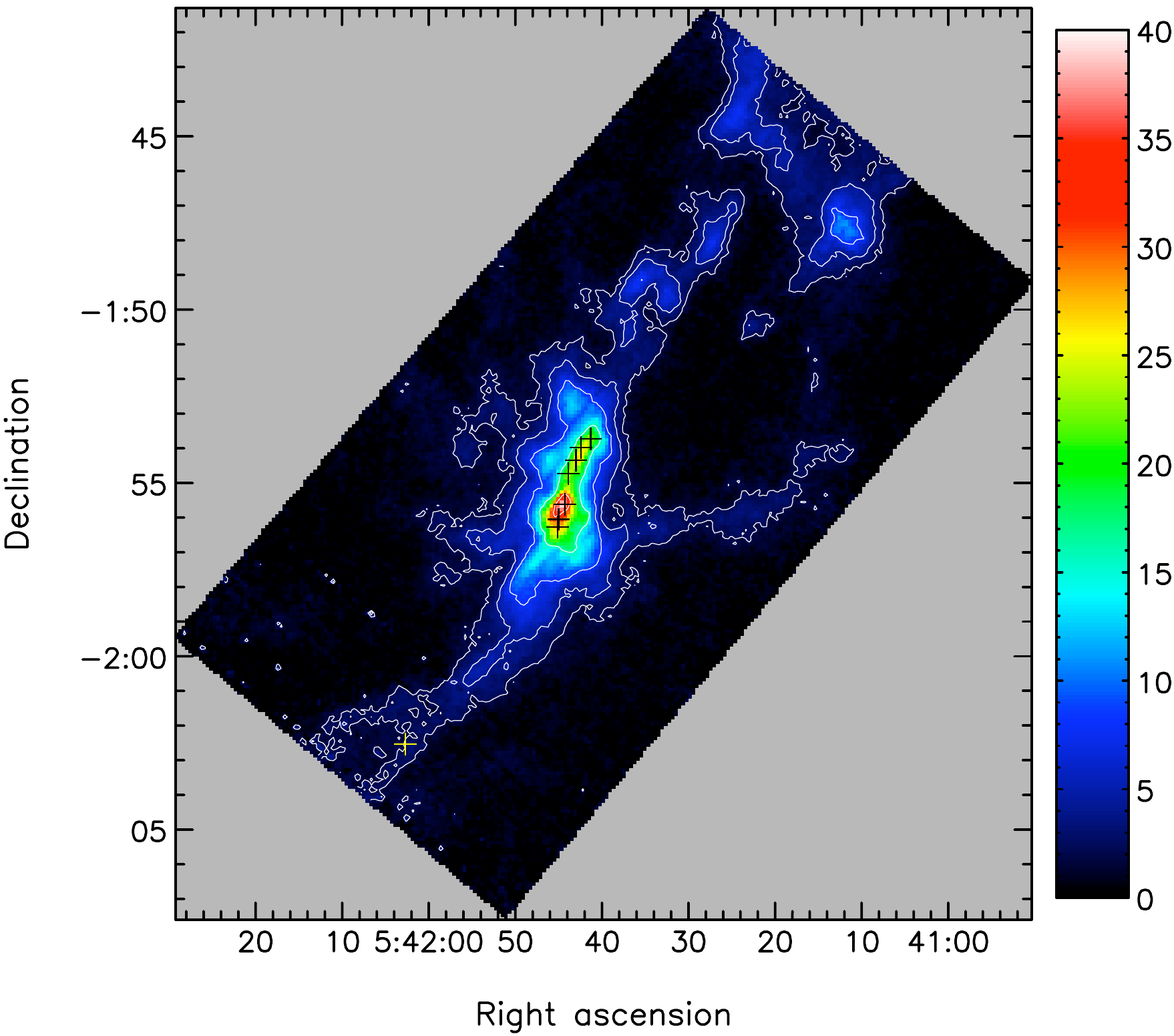}}
\caption{\label{fig-ngc2024-12co-int}Integrated intensity images of $^{12}$CO
(top), $^{13}$CO (middle) and C$^{18}$O (bottom) towards NGC 2024. Crosses mark
the positions of sources FIR1--7, and an isolated continuum source to the far
south. Contours are $^{12}$CO: from 30 K \kms\ with 20 K \kms\ steps; $^{13}$CO
and C$^{18}$O: from 2 K \kms, with contour levels doubling at every step. }
\end{minipage}
\end{figure}

\begin{figure}
\includegraphics[width=8.5cm,angle=0]{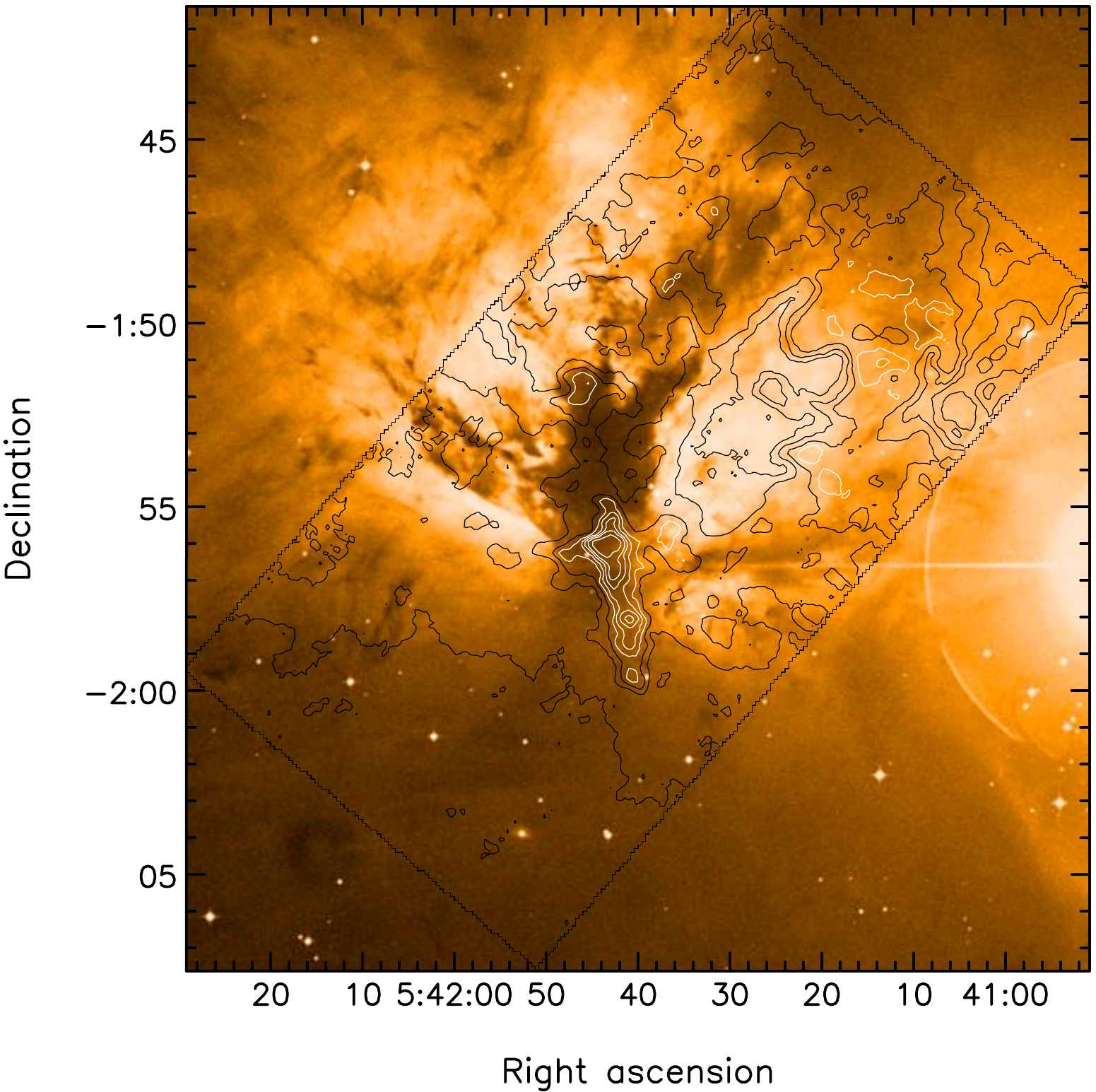}
\caption{\label{ngc2024-optical}DSS infrared image of NGC 2024 overlaid with
$^{12}$CO contours as in Fig.~\ref{fig-ngc2024-12co-int}.}
\end{figure}

Fig.~\ref{fig-ngc2024-12co-int} displays the total integrated intensity images
across the full width of the emission lines towards NGC 2024 in
$^{12}$CO~3$\to$2 (-20.0--50.0~\kms), $^{13}$CO~3$\to$2 (4.0--16.0~\kms) and
C$^{18}$O~3$\to$2 (6.0--14.0~\kms). In this cloud, the integrated intensity
image in $^{12}$CO is dominated by emission to the north of the map,
particularly along a ridge extending south-west of the sources FIR1--7 (marked
in Fig.~\ref{fig-ngc2024-12co-int} with white crosses). A cavity is seen to the
north-west of this position, with the rest of the $^{12}$CO emission
surrounding the cavity in a clumpy ring. The cavity marks a region of bright
optical emission. Fig.~\ref{ngc2024-optical} shows the digitized sky survey
(DSS) infrared image\footnote{from Gaia Skycat,http://archive.eso.org/cms/tools-documentation/skycat} overlaid with $^{12}$CO
contours. The dense ridge where we see the brightest molecular emission appears
as a dark dust lane in the optical image. The bright optical filaments appear
in regions which lack $^{12}$CO emission. In particular, the large cavity
contoured in $^{12}$CO integrated intensity emission clearly follows the shape
of the bright optical emission. The peak of the $^{13}$CO emission is seen in
the ridge containing the FIR sources. Further emission follows the $^{12}$CO
emission, delineating a cavity surrounded by a ring of clumpy, filamentary
structure. The C$^{18}$O emission peaks on the positions of the FIR sources in
the ridge. Weaker, fragmented and clumpy emission surrounds the cavity seen in
$^{12}$CO and $^{13}$CO.

The $^{12}$CO spectra show evidence of several velocity components along the
line of sight. Fig.~\ref{fig-ngc2024-sp} shows the spectra from a single pixel
at the peak of C$^{18}$O emission, from the three  isotopologues. Also shown is
a composite spectrum averaged over the observed region.  The high-resolution
$^{12}$CO spectrum shows four distinct peaks, at 4.6~\kms, 8.3~\kms, 11.2~\kms
and 13.0~\kms. The $^{13}$CO spectrum has components at 4.2~\kms, 8.7~\kms\ and
11.3~\kms, while the C$^{18}$O line has a main component at 11.0~\kms, with a
shoulder extension at 9.2~\kms. These line profiles suggest self absorption in
the $^{12}$CO line at 9.6~\kms and 11.9~\kms. Previous observations have also
found several velocity components towards NGC 2024
\citep[e.g.][]{emprechtinger2009}. At 9~\kms, the dense ridge, thought to be in
the foreground, is seen as a separate component in C$^{18}$O and $^{13}$CO,
while there is a dip in the $^{12}$CO spectrum. This supports the suggestion
that emission from the blue-side of the $^{12}$CO line is being absorbed by
foreground material in the dense ridge \citep{emprechtinger2009,graf1993}. The
main $^{13}$CO and C$^{18}$O components peak near 11~\kms, the velocity of the
extended molecular cloud in the background {\citep[e.g.][]{barnes1989}, and we
also see a $^{12}$CO component at this velocity. The dip in the $^{12}$CO
spectrum at 11.9~\kms, which is part of the main component of the $^{13}$CO and
C$^{18}$O emission, suggests that there may be a further foreground component
at this velocity which is absorbing the $^{12}$CO emission.  The
self-absorption of the $^{12}$CO emission undoubtedly causes its different peak
position to the isotopologues tracing higher column density material. The
multiple line-of-sight components are not clearly seen in spectra averaged over
the whole region, and demonstrate the importance of high-\emph{spectral} and
spatial resolution observations in understanding the kinematics of star
formation/molecular clouds, especially in clustered regions.

\begin{figure}
\vbox{
\includegraphics[width=6cm,angle=-90]{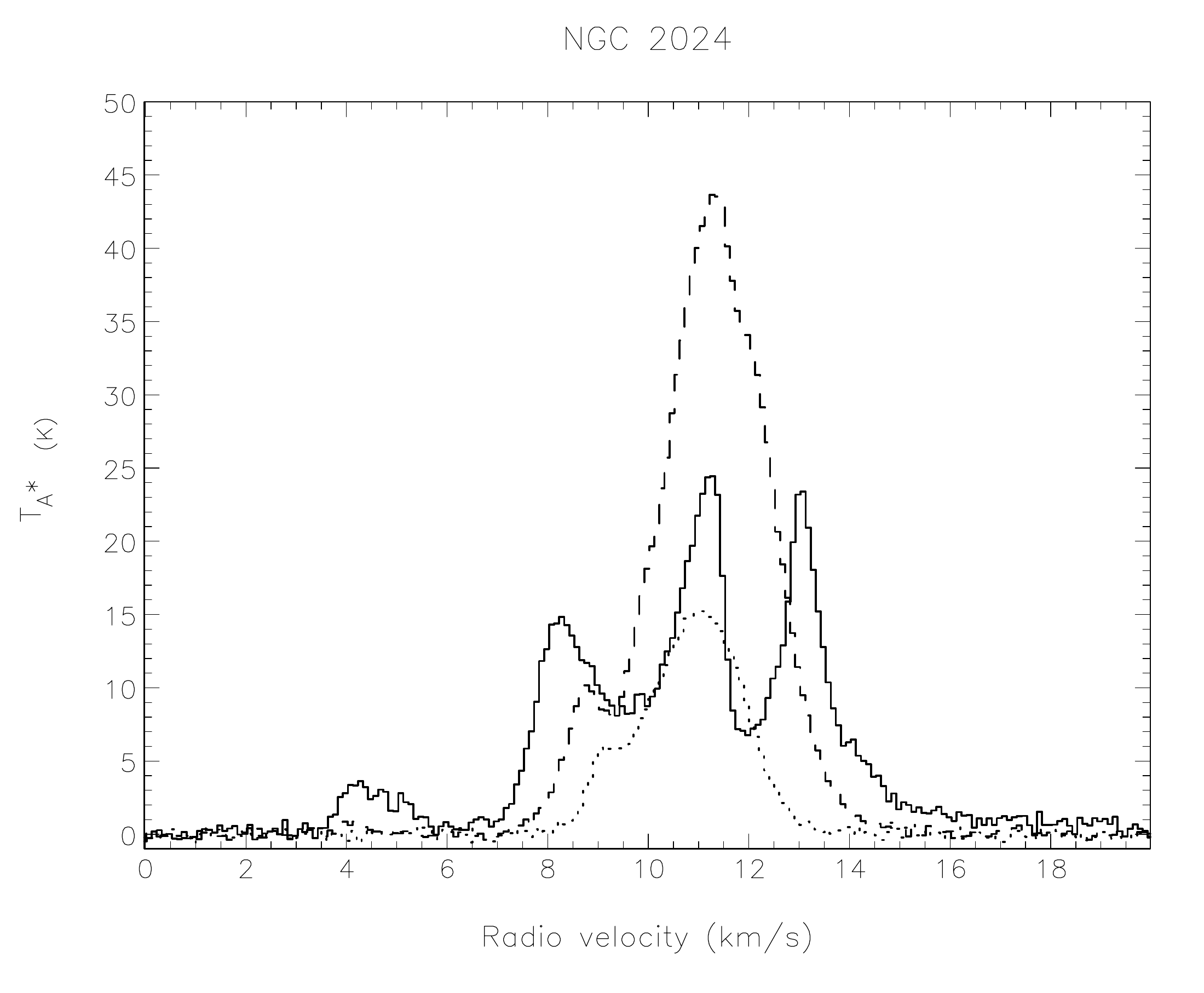}
\includegraphics[width=6cm,angle=-90]{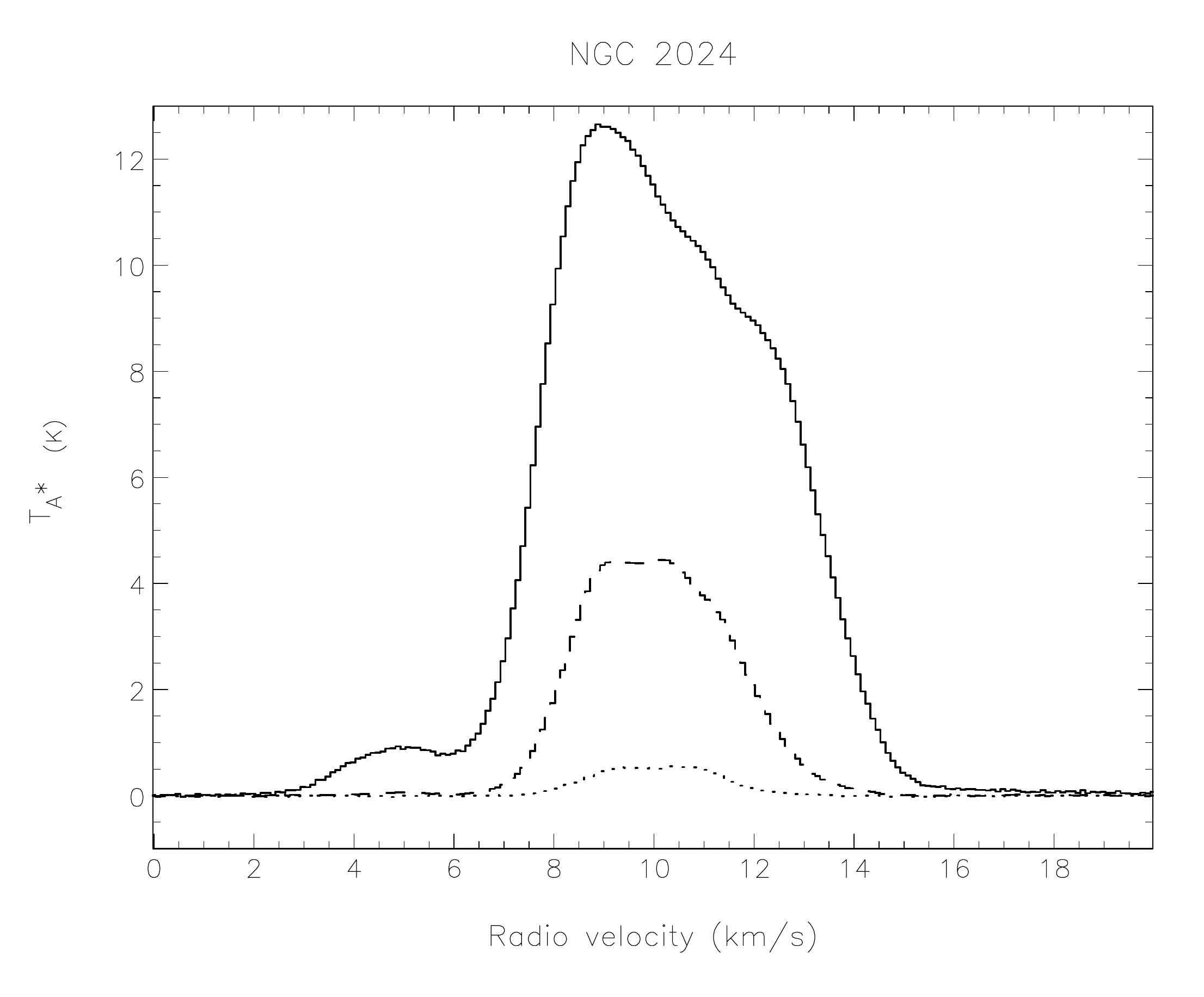}}
\caption{\label{fig-ngc2024-sp}Spectra from a single pixel position (5$^{\rm h}$41$^{\rm m}$44$^{\rm s}$.6
-1$^{\arcdeg}$55$^{\arcmin}$33$^{\arcsec}$, top) and averaged over the whole cloud (bottom) of $^{12}$CO (solid
line) $^{13}$CO (dashed line) and C$^{18}$O (dotted line) towards the C$^{18}$O
peak in NGC 2024. All of the spectra have been observed at high-resolution, and
binned to 0.1~\kms. }
\end{figure}

Fig.~\ref{fig-ngc2024-pv} shows position-velocity (PV) diagrams of the three
isotopologues across a large section of NGC 2024. The integrated intensity image
in the left is marked with two lines showing the right ascension and
declination cuts used. In the $^{12}$CO map, we can see a velocity gradient
running north-south, from $\sim$13.0~\kms\ in the south to $\sim$10.0~\kms\ in
the north. High-velocity material is seen as extended filaments across much of
the PV diagrams, with the brightest high-velocity material seen associated with
the ridge in the declination cut. The cavity previously described is labelled
in the right ascension cut, at two velocity intervals straddling the main bulk
of the molecular cloud at 11.0~\kms. From 12.0--14.0~\kms\ the cavity shows a
distinct velocity gradient along its edge, and is narrower than the outline
seen  at velocities 8.0--10.0~\kms. The lower-velocity component of the cavity
does not have the same velocity gradient along the edge.  The very
high-velocity outflow material is only clearly seen in the $^{12}$CO PV
diagrams, but emission from $^{13}$CO shows extensions in velocity that suggest
it is associated with outflow material. Both the $^{12}$CO and $^{13}$CO show
compact clumps of emission at blue-shifted velocities near 4~\kms, one of the
components revealed in the spectra (Fig.~\ref{fig-ngc2024-sp}). The $^{12}$CO
PV diagrams show filaments extending between the clumps and the bulk of the
emission, suggesting these clumps could be physically associated with the
cloud, possibly through wind-driven activity. The emission at red-shifted
velocities, from 12--14~\kms, shows a much smoother distribution than the
blue-shifted emission.  The C$^{18}$O emission is very compact in velocity,
tracing the bulk of the dense material in the ridge, and no longer contains the
component at  4~\kms.

\begin{figure*}
\begin{minipage}{180mm}
\vbox{
\includegraphics[width=8.1cm,angle=-90]{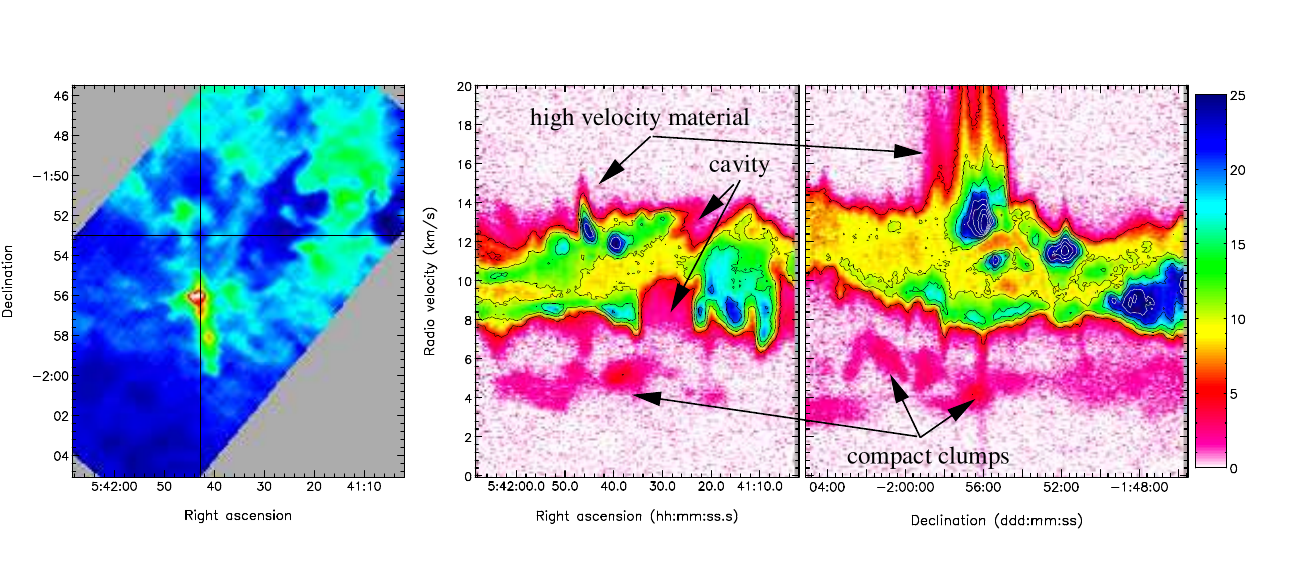}
\includegraphics[width=18cm,angle=0]{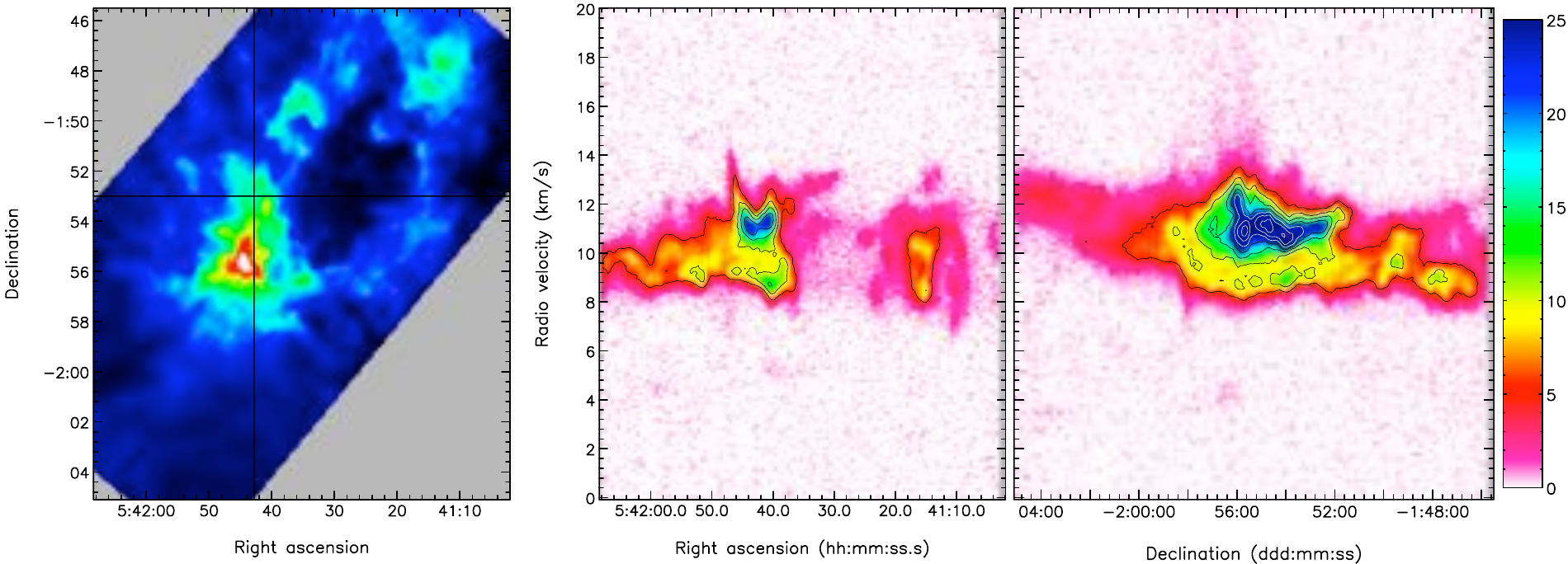}
\includegraphics[width=18cm,angle=0]{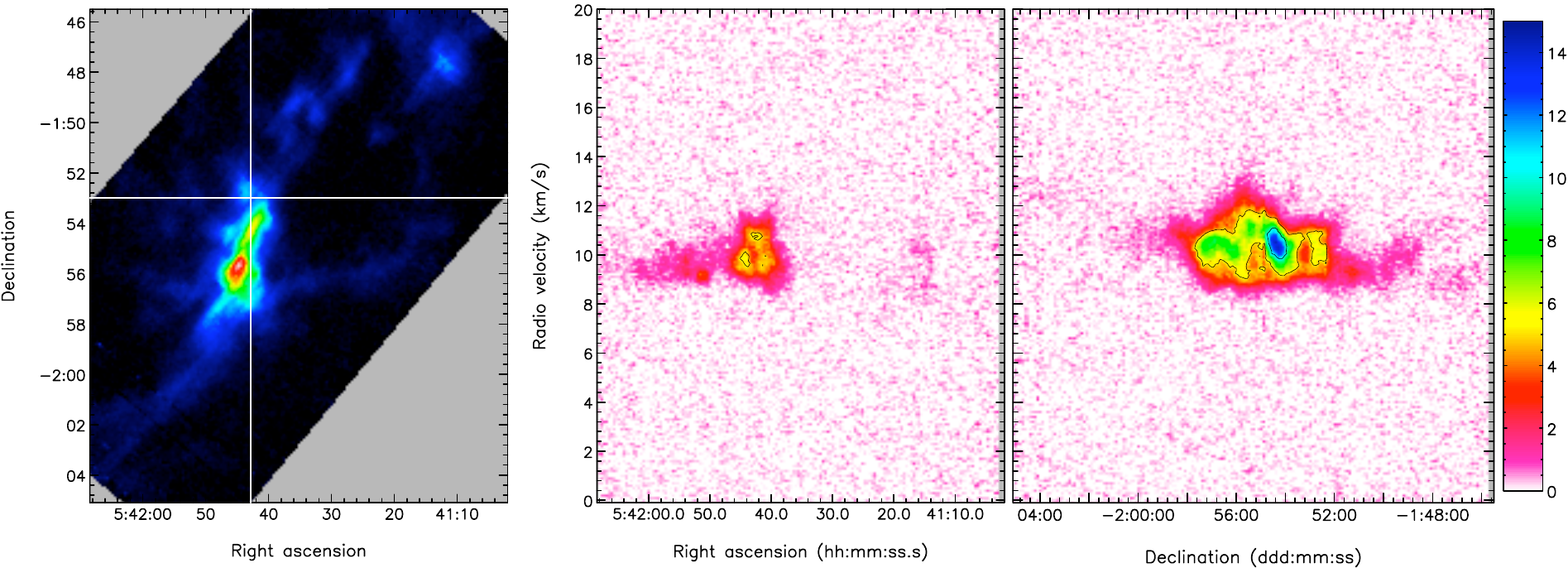}}
\caption{\label{fig-ngc2024-pv}Position velocity maps of NGC 2024 in $^{12}$CO,
$^{13}$CO and C$^{18}$O, through the positions shown in horizontal and vertical
lines in the integrated intensity image on the left. }
\end{minipage}
\end{figure*}

\subsubsection{NGC 2071}

Fig.~\ref{fig-int2} shows total integrated intensity images across the full
width of the emission lines for $^{12}$CO~3$\to$2 (-20.0--35.0~\kms),
$^{13}$CO~3$\to$2 (-3.0--18.0~\kms) and C$^{18}$O~3$\to$2 (5.0--12.0~\kms)
towards NGC 2071. The $^{12}$CO emission is dominated by a curved filament of
emission, surrounding a compact, bright, optical emission region.
Fig.~\ref{ngc2071-optical} shows the DSS infrared image overlaid with $^{12}$CO
contours towards NGC 2071, showing the $^{12}$CO emission nearly surrounding the
bright nebula. The $^{13}$CO and C$^{18}$O emission peaks along the compact
ridge, but both show a different spatial structure to the $^{12}$CO emission at
the southern end of the ridge, at the position of the optical nebula. In
$^{13}$CO and C$^{18}$O, the emission to the south, where the $^{12}$CO
integrated intensity falls off, displays separate arcs.

\begin{figure}
\begin{minipage}{85mm}
\vbox{
\includegraphics[width=8cm,angle=-90]{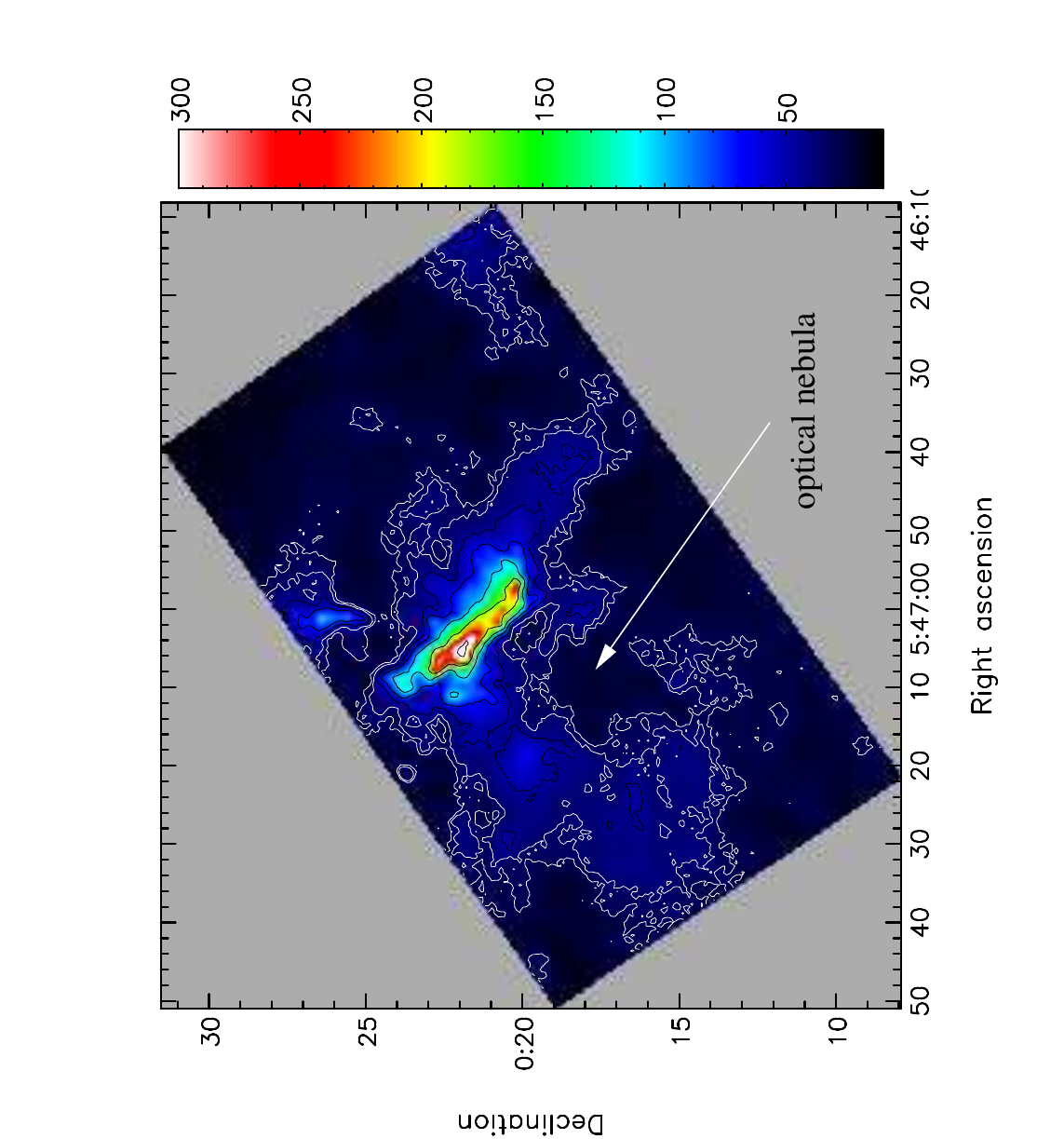}
\includegraphics[width=8cm,angle=0]{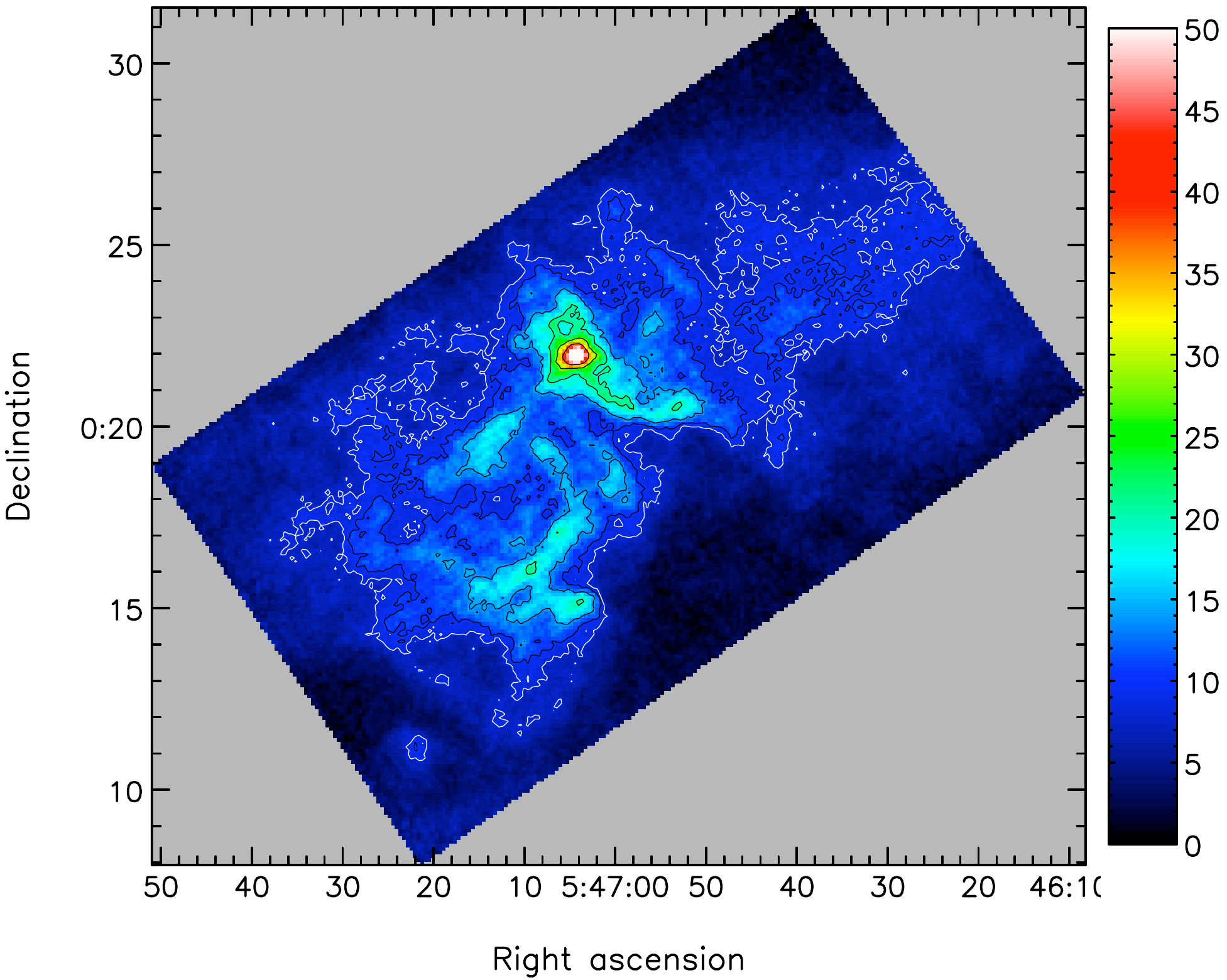}
\includegraphics[width=8cm,angle=0]{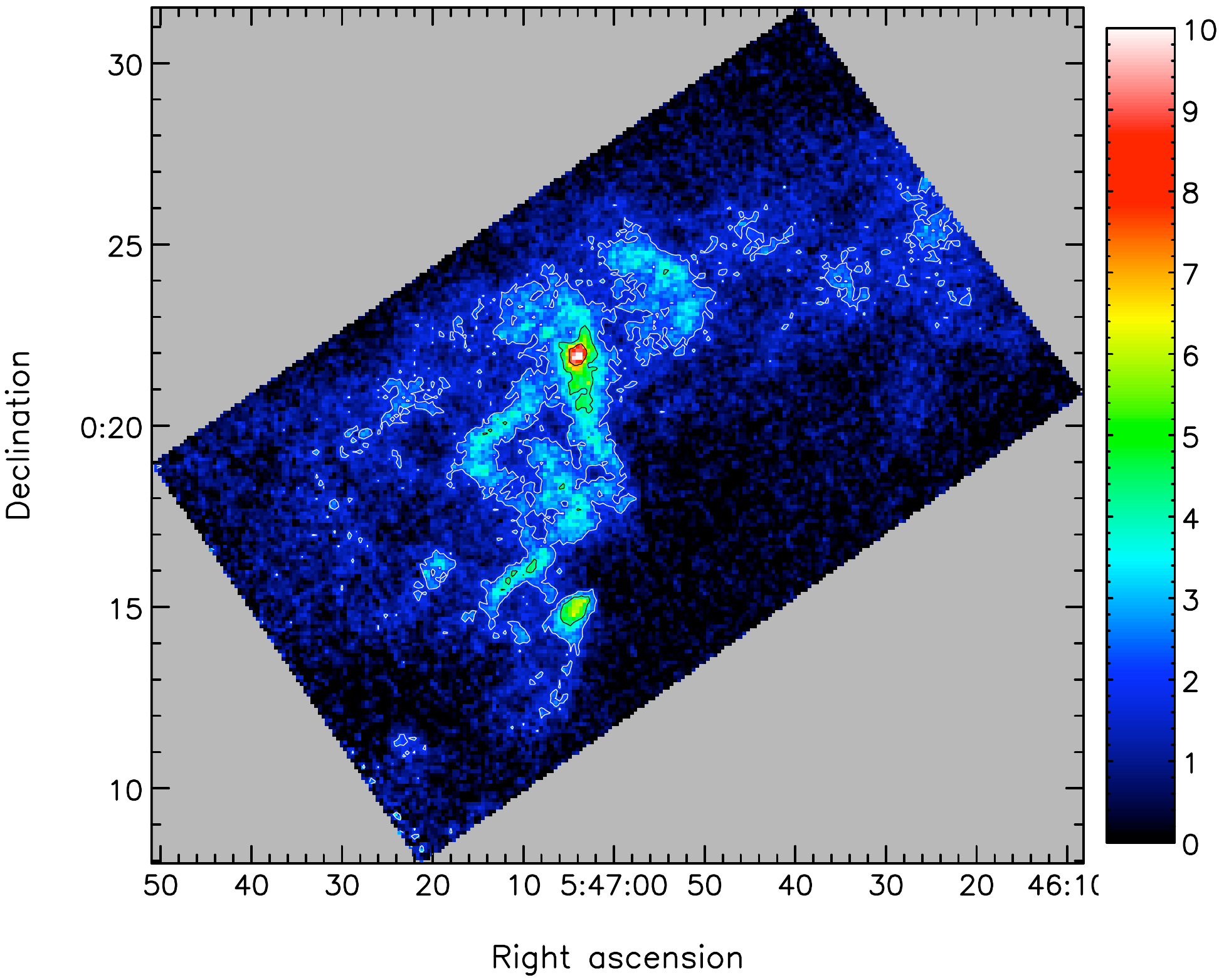}
}
\caption{\label{fig-int2}Integrated intensity images of $^{12}$CO (top),
$^{13}$CO (middle) and C$^{18}$O (bottom) towards NGC 2071. Contours are
$^{12}$CO: 30,35 (white) and 45,65,105,185,345 (black) K \kms; $^{13}$CO 7,9
(white) and 11,15,21,29,39 (black) K \kms; C$^{18}$O: 2 (white) and 4,7 (black)
K \kms. The first C$^{18}$O contour is at 5$\sigma$.  }
\end{minipage}
\end{figure}

\begin{figure}[!h]
\includegraphics[width=8.5cm,angle=0]{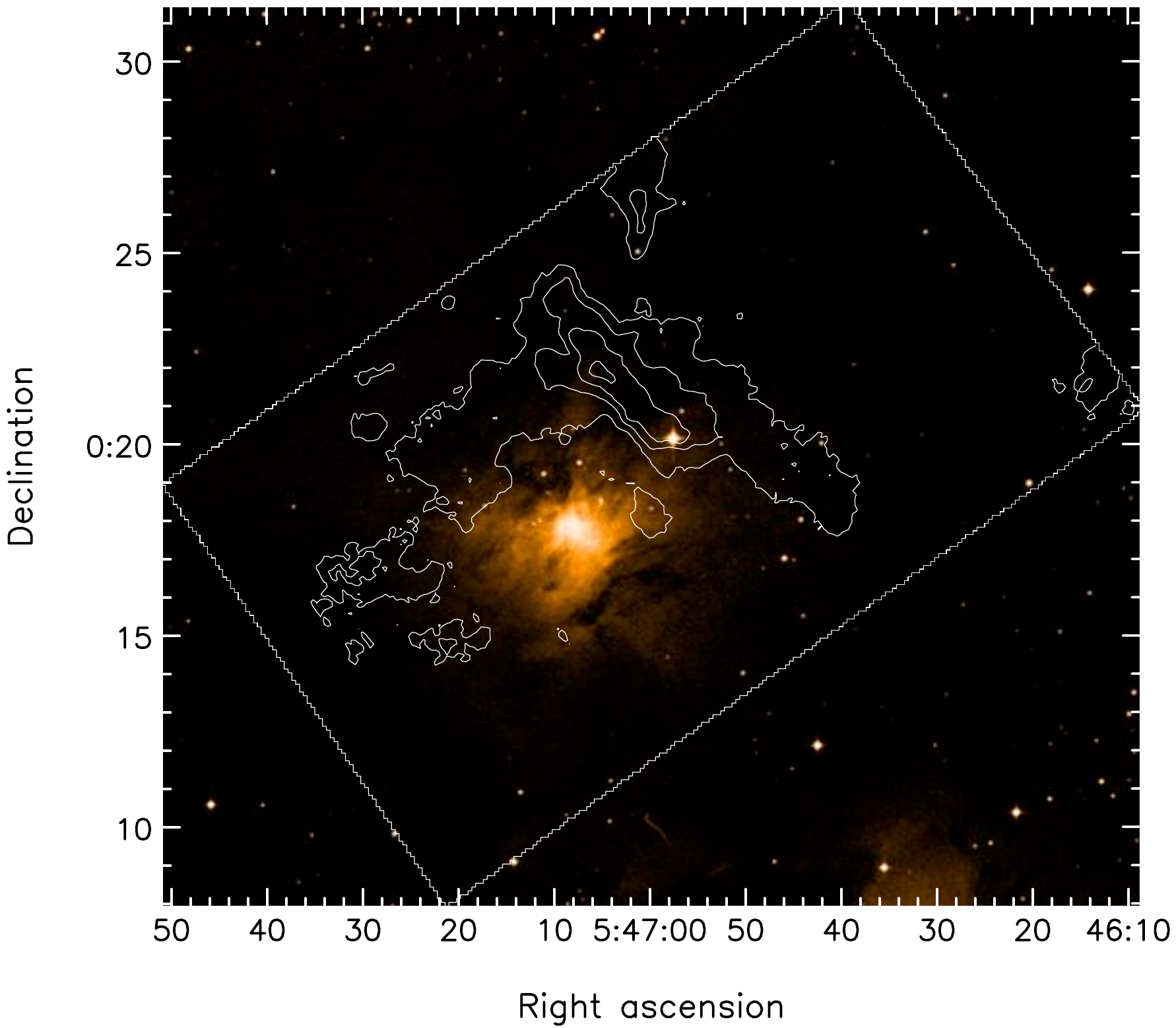}
\caption{\label{ngc2071-optical}DSS infrared image of NGC 2071 overlaid with
$^{12}$CO contours as in Fig.~\ref{fig-int2}.}
\end{figure}

The spectra towards individual positions in NGC 2071 are shown in
Fig.~\ref{fig-sp1}. At the systemic velocity of the cloud, around 10~\kms, the
$^{12}$CO is heavily self-absorbed, even in the low-density regions near the
optical nebula (Fig.~\ref{fig-sp1}, top). Towards the peak of emission from
C$^{18}$O (Fig.~\ref{fig-sp1}, middle), emission from $^{13}$CO is also
self-absorbed. The absorption dip in $^{12}$CO emission is also clear in the
spectra averaged across the whole cloud (Fig.~\ref{fig-sp1}, bottom).

\begin{figure}
\vbox{
\includegraphics[width=6cm,angle=-90]{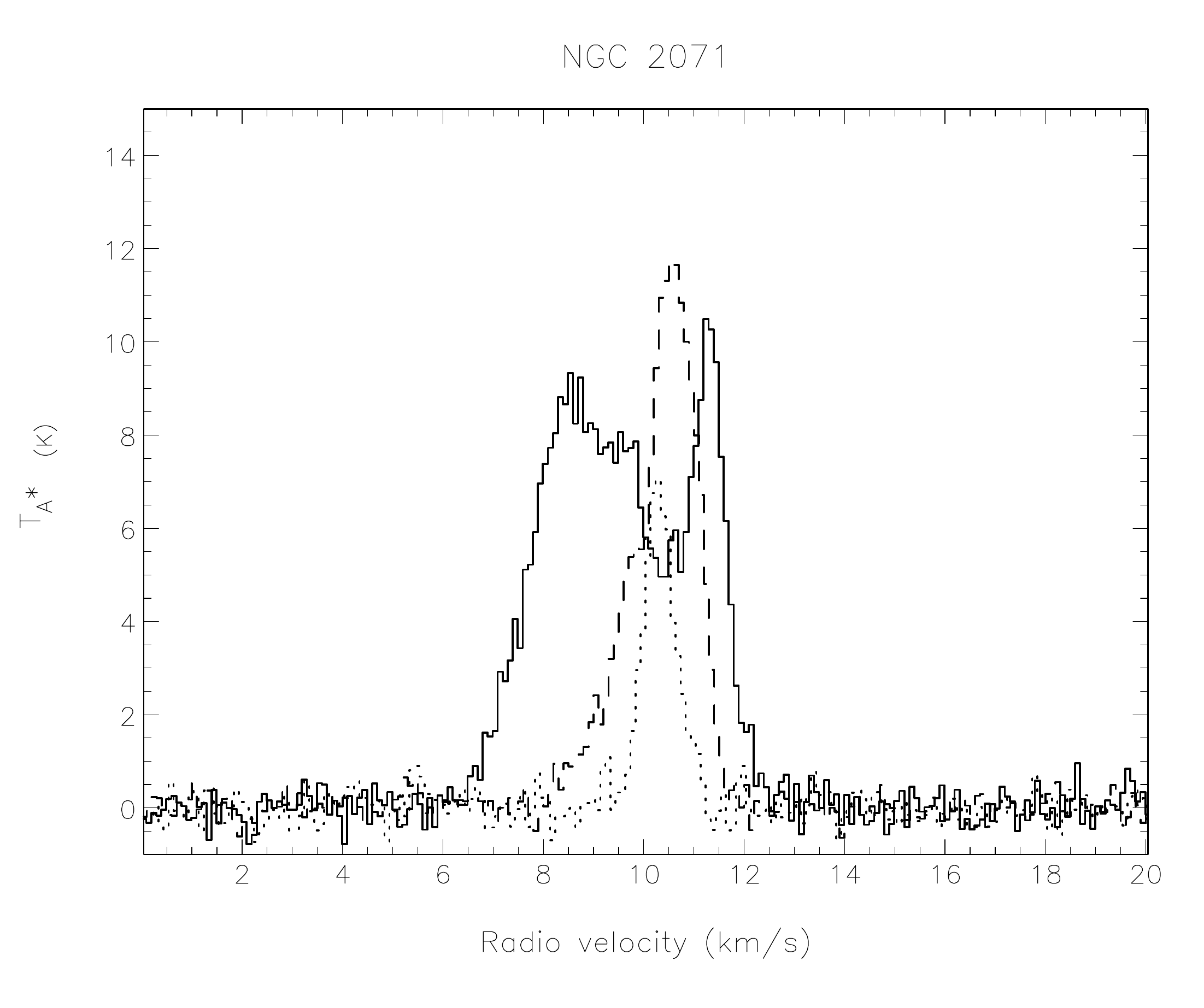}
\includegraphics[width=6cm,angle=-90]{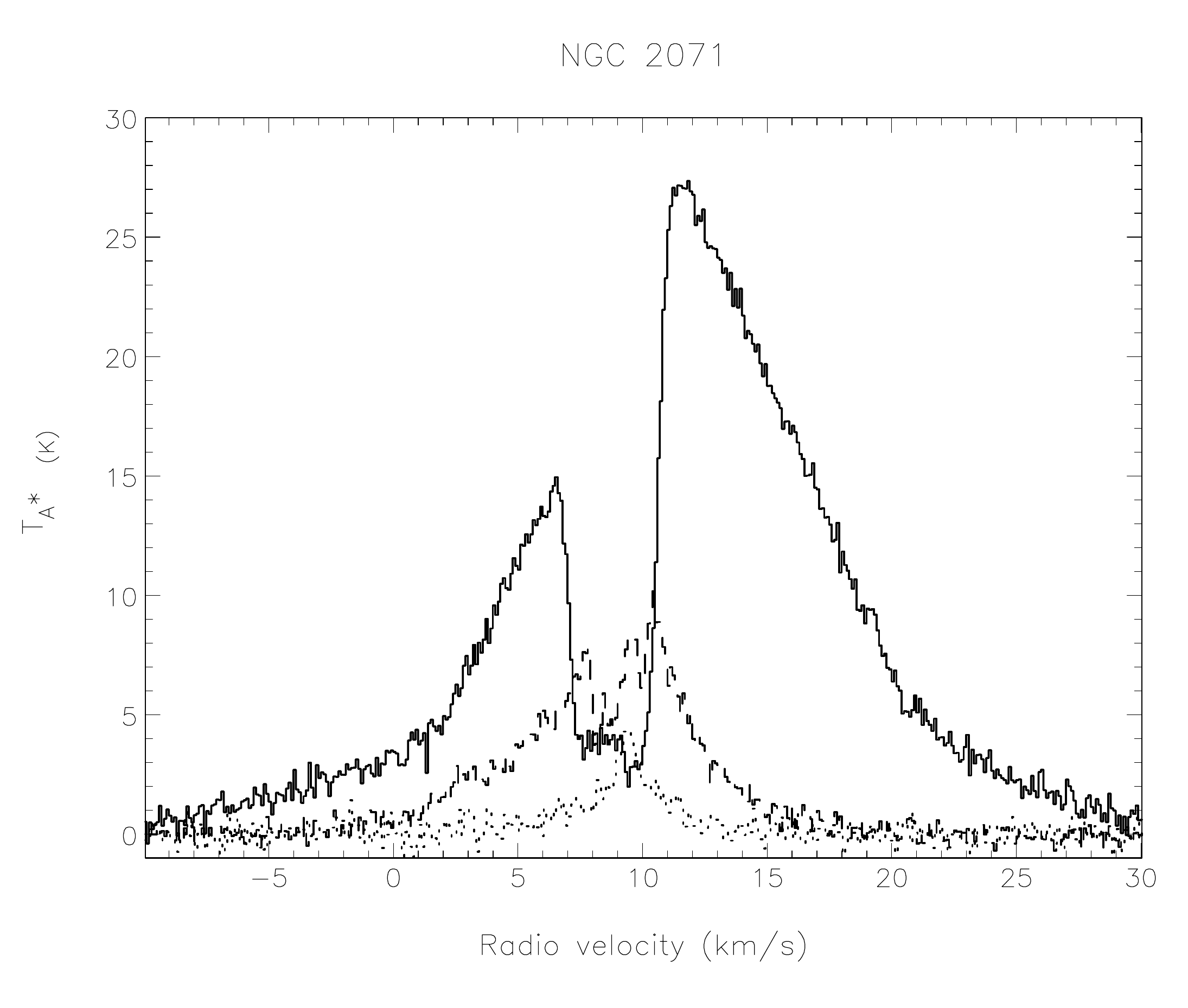}
\includegraphics[width=6cm,angle=-90]{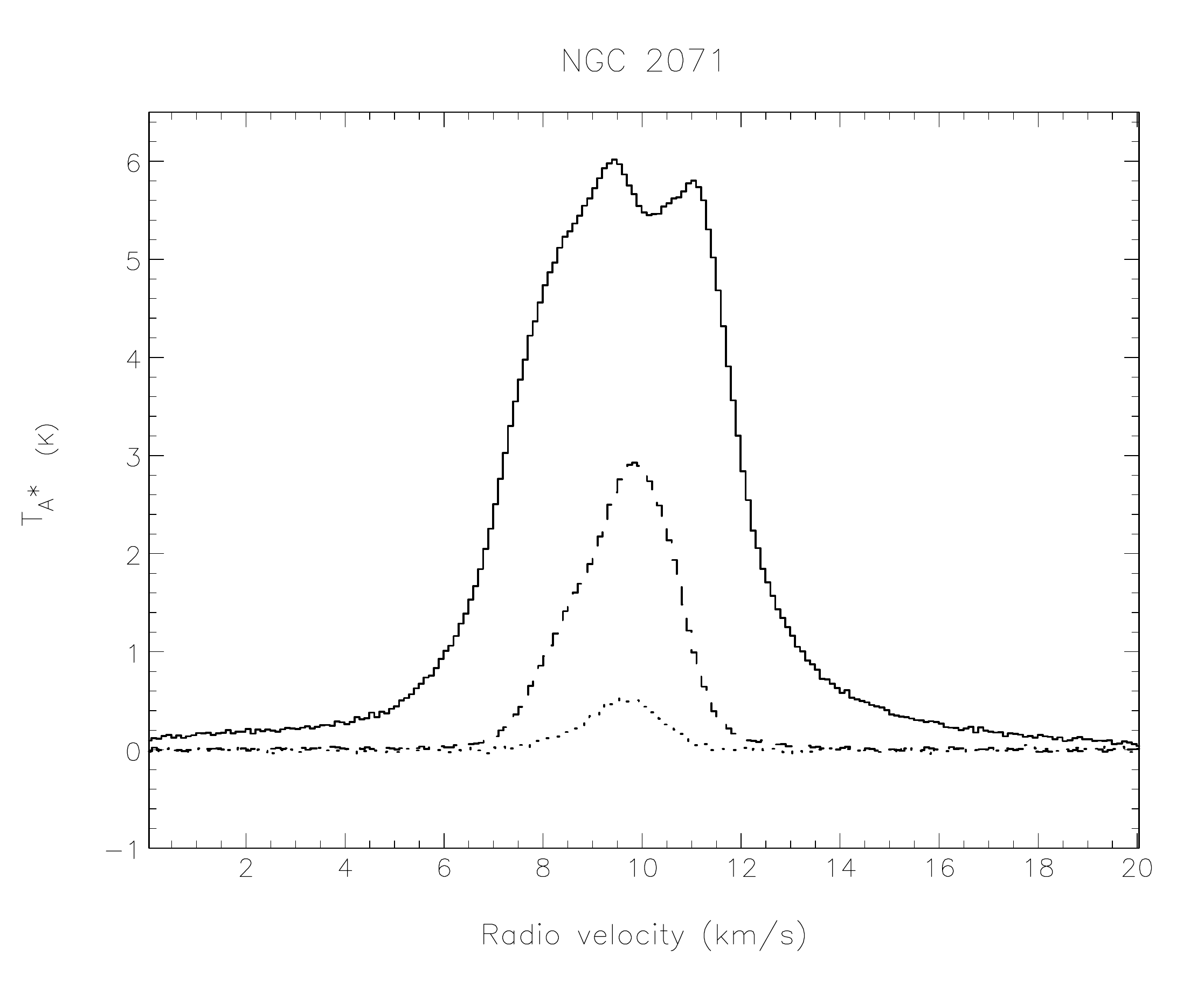}}
\caption{\label{fig-sp1}Spectra from single pixel positions; (top) 5$^{\rm h}$47$^{\rm m}$04$^{\rm s}$.4
+0$^{\arcdeg}$14$^{\arcmin}$53$^{\arcsec}$ towards the optical nebula; (middle) 5$^{\rm h}$47$^{\rm m}$04$^{\rm s}$.0 +0$^{\arcdeg}$21$^{\arcmin}$53$^{\arcsec}$, towards the
C$^{18}$O peak; and averaged over the whole cloud (bottom) of $^{12}$CO (solid
line) $^{13}$CO (dashed line) and C$^{18}$O (dotted line) in NGC 2071.}
\end{figure}

Fig.~\ref{fig-ngc2071-pv} shows the PV diagrams for NGC 2071, as for
Fig.~\ref{fig-ngc2024-pv}. We see high-velocity material from $^{12}$CO and
$^{13}$CO, and evidence for the C$^{18}$O emission tracing lower-velocity
outflow material. There is no clear velocity gradient through the cloud. The
$^{12}$CO and $^{13}$CO PV diagrams along a cut in right ascension suggest that
there may be a second component along the line of sight to the west, or that
the cloud is more extended in velocity in this region.  In contrast to NGC 2024,
towards NGC 2071, we do not see any clear features in the $^{12}$CO PV diagram
that may be associated with cavities or bubbles carving out to the cloud edges
through activity within the cloud. The C$^{18}$O emission shows a compact,
clumpy structure in the PV diagrams.

\begin{figure*}
\begin{minipage}{180mm}
\vbox{
\includegraphics[width=7.3cm,angle=-90]{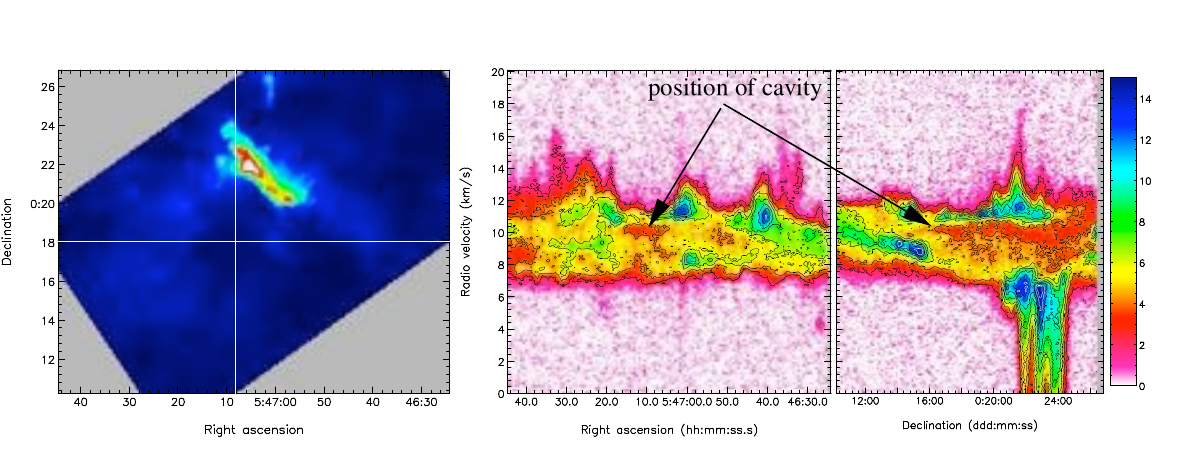}
\includegraphics[width=18cm,angle=0]{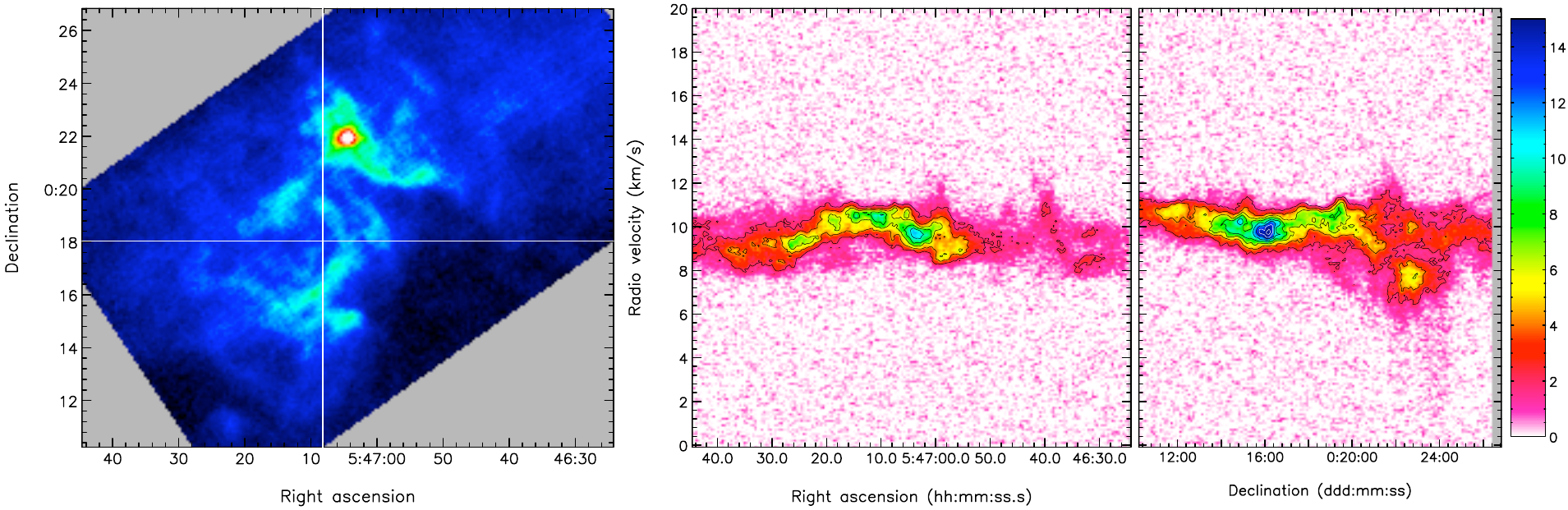}
\includegraphics[width=18cm,angle=0]{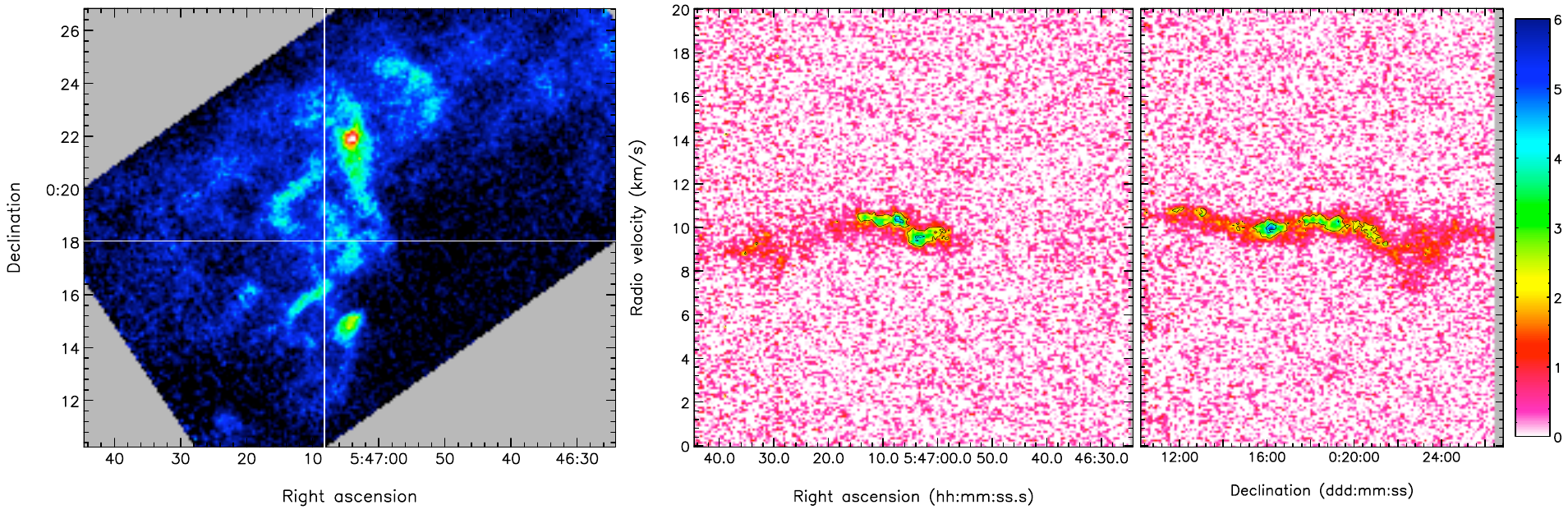}}
\caption{\label{fig-ngc2071-pv}Position velocity maps of NGC 2071 in $^{12}$CO,
$^{13}$CO and C$^{18}$O, as for Fig.~\ref{fig-ngc2024-pv}.}
\end{minipage}
\end{figure*}

\subsection{High-velocity material}
\label{sec-hv}

Figs.~\ref{fig-rgb1} and \ref{fig-rgb2} show false colour red, green and blue
images of the average intensity across the $^{12}$CO line profile towards
NGC 2024 and NGC 2071, where the emission near the systemic velocity of the
clouds is shown in green, and the red- and blue-shifted gas is shown in the
corresponding colour.

\begin{figure}
\includegraphics[width=8.5cm,angle=0]{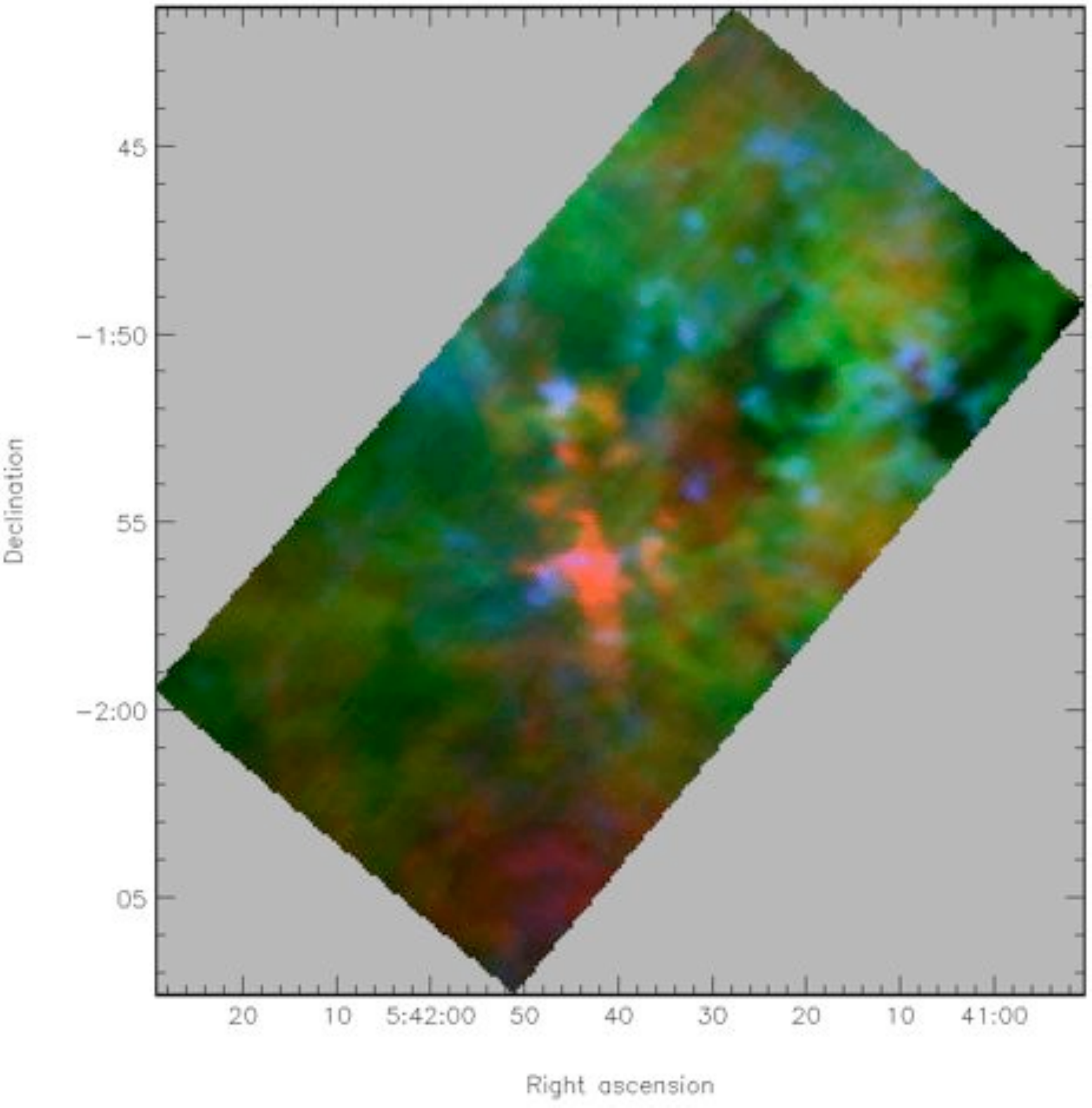}
\caption{\label{fig-rgb1}False colour image of the average $^{12}$CO emission
towards NGC 2024 in red (13.2--19.2~\kms), green (7.2--13.2~\kms) and blue (1.2--7.2~\kms) velocity channels.}
\end{figure}

\begin{figure}
\includegraphics[width=8.5cm,angle=0]{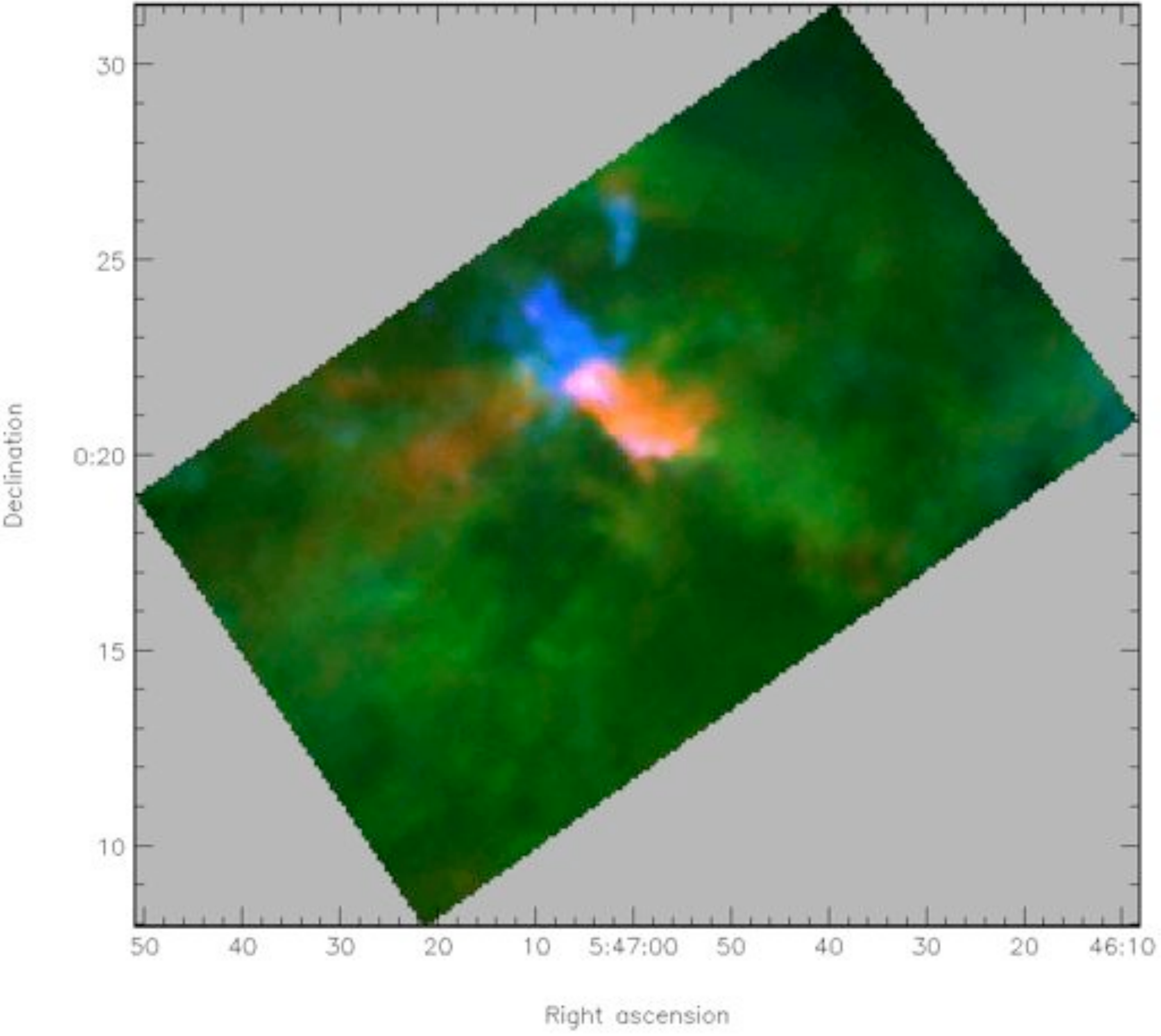}
\caption{\label{fig-rgb2}False colour image of the average $^{12}$CO emission
towards NGC 2071 in red (13.1--19.1~\kms), green (7.1--13.1~\kms) and blue (1.1--7.1~\kms) velocity channels. }
\end{figure}

Towards NGC 2024, a single region dominates the high-velocity red-shifted
material, associated with an extended outflow from FIR5 and a compact outflow
from FIR6 \citep{richer1989}. Additionally, more diffuse components are seen
throughout the map, which probably indicate different components along the
line-of-sight, rather than protostellar outflows. However, there are two
regions to the north, and several regions in the centre of the map where the
relative position of red and blue emission, and the shape of the emission,
suggest there are embedded outflow sources. The PV diagrams
(Fig.~\ref{fig-ngc2024-pv}) show several filamentary structures in velocity
that may be associated with outflows that are less extensive in velocity than
the obvious extremely high-velocity flows.

Fig.~\ref{fig-out1} shows contours of blue- and red-shifted emission overlaid
on the integrated C$^{18}$O intensity map for the main NGC 2024 outflow source.
In the $^{12}$CO $J=3\rightarrow$2 emission, we see the high-velocity
blue-shifted lobe is spatially compact, although extended in velocity. This is
the only region in NGC 2024 where such high-velocity emission is detected.

\begin{figure}[h]
\includegraphics[width=8cm,angle=-90]{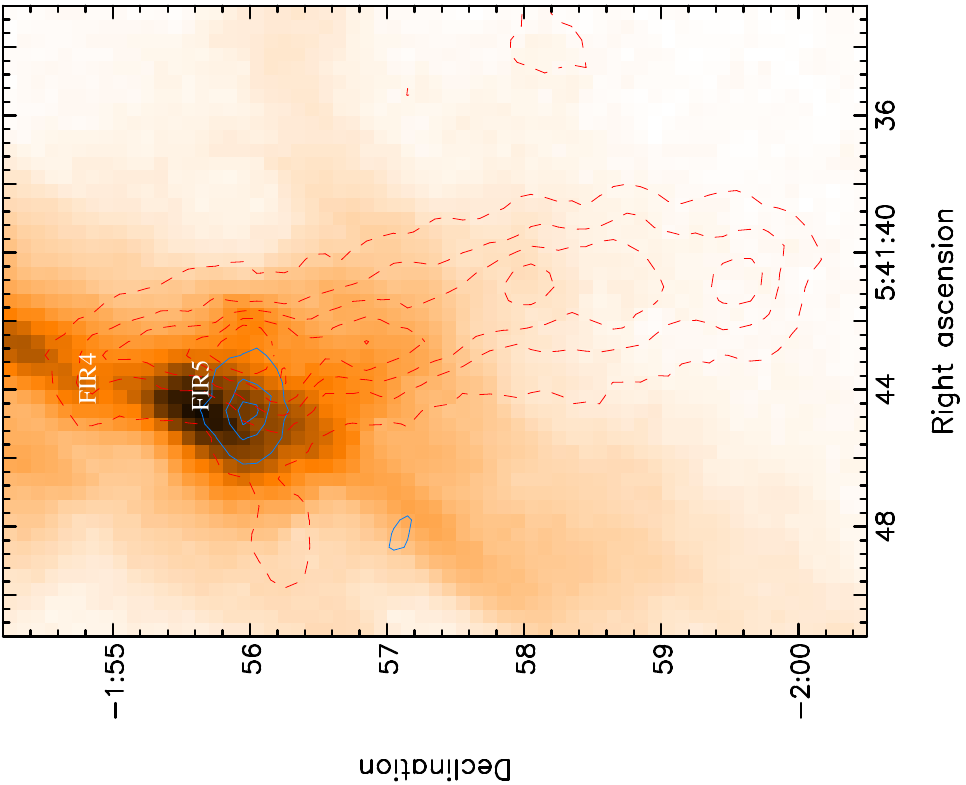}
\caption{\label{fig-out1}NGC 2024 high-velocity red- and blue-shifted $^{12}$CO
emission. Red contours show emission from 14.5 to 40.0~\kms, and  blue contours
show emission from -20.0 to 4.0~\kms, both starting at 5~K\,\kms, in steps of
20~K\,\kms. The image shows the integrated C$^{18}$O intensity. The suspected driving
sources FIR5 and FIR6 have been labelled.}
\end{figure}

NGC 2071 is dominated by a high-velocity outflow, previously identified as one
of the most energetic and collimated outflows ever discovered
\citep[e.g.][]{stojimirovic2008}. The false colour image shows there
are at least two more outflows, as suggested by \citet{stojimirovic2008}. The
PV diagrams (Fig.~\ref{fig-ngc2071-pv}) suggest how powerful this outflow is,
with small extensions in velocity showing in the C$^{18}$O emission. Again, the
velocity-extended filamentary structures in this large-scale PV diagram suggest
that there may be several protostellar outflows embedded within the cloud. In
Sec.~\ref{sec:energy}, we investigate the energy injection from outflows in
these regions.  Fig~\ref{fig-out2} shows the blue- and red-shifted contours
overlaid on the C$^{18}$O emission for the main NGC 2071 outflow region,
overlaid with SCUBA sources (J06). Red- and blue-shifted outflow lobes can be
identified,  for at least three, and possibly five sources.

\begin{figure}[h]
\includegraphics[width=7cm,angle=0]{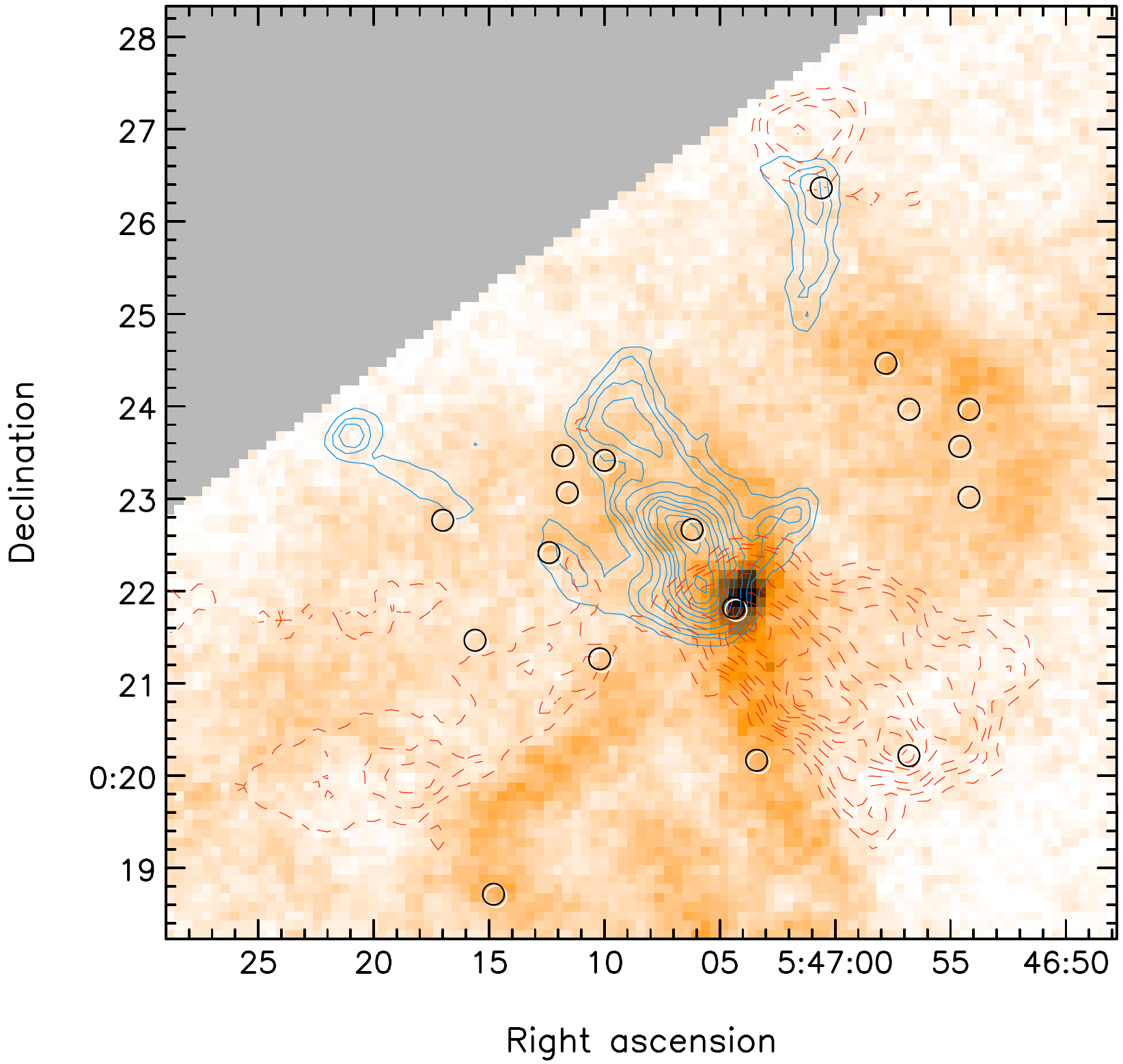}
\caption{\label{fig-out2}NGC 2071 high-velocity red- and blue-shifted $^{12}$CO
emission. Red contours, show emission from 14.5 to 35.0~\kms, blue contours
show emission from -25.0 to 0.5~\kms, contour levels at
5,10,15,25,45,65,85,105,125,145~K\,\kms. The image shows the integrated
C$^{18}$O intensity. Black circles denote positions of SCUBA cores
(J06).}
\end{figure}

\subsection{Low density material}
\label{sec-ld}

Towards both regions, a cavity is seen in emission from $^{12}$CO, associated
with an optically-bright H{\sc ii} region. However the behaviour of the less
abundant isotopologues, and the velocity structure of these regions is
different towards the two clouds.

Towards NGC 2024, emission from all of the CO isotopologues falls off in this
region, as can be seen in the integrated intensity maps
(Fig.~\ref{fig-ngc2024-12co-int}). The cavity is coincident with the bright
optical emission, and it may be that we are seeing the CO disassociated by an
H{\sc ii} region. Fig.~\ref{fig-scuba1} shows the SCUBA 850~$\umu$m continuum
image \citep{difrancesco2008} overlaid with C$^{18}$O integrated intensity
contours.  The C$^{18}$O emission closely follows the structure seen in the
dust, and the clumpiness of both the dust and the C$^{18}$O emission suggest
that cores are present in a region surrounding the cavity. We present an
analysis of all the $^{13}$CO condensations in Sec.~\ref{sec:clumps}.

\begin{figure}[h]
\includegraphics[width=8.5cm,angle=0]{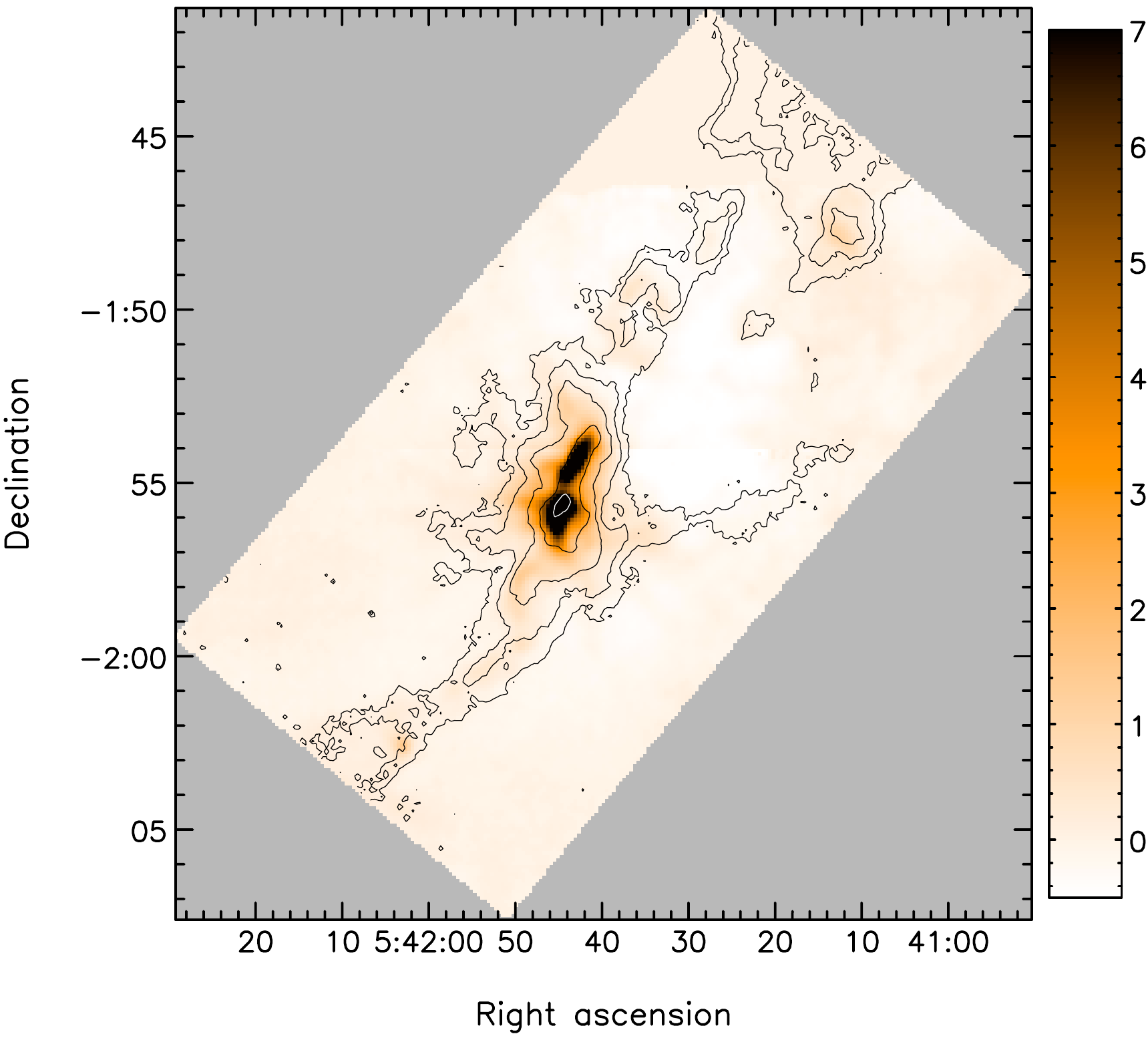}
\caption{\label{fig-scuba1}NGC 2024 SCUBA 850 $\umu$m maps
\citep{difrancesco2008}, overlaid with C$^{18}$O integrated intensity contours.
Contour levels are as Fig.~\ref{fig-ngc2024-12co-int}.}
\end{figure}

In the NGC 2024 $^{12}$CO PV diagram (Fig.~\ref{fig-ngc2024-pv}), the cavity can
clearly be seen, and shows a different spatial structure at high and low
velocities. There is an abrupt cut-off in emission across the lower velocities,
from 8.0 to 10.0 \kms. At the higher velocities, from 12.0 to 14.0 \kms, there
is a velocity gradient, and the eastern edge has a higher gradient in velocity
than the western edge. If we are seeing an H{\sc ii} region, then the different
morphologies between the gas at low }velocities and at high velocities may be
giving information on the three-dimensional spatial structure. The low
velocities, tracing material at the front of the cloud, have an abrupt cut-off
to the emission because at this point, the material has broken out of the
molecular cloud, and there is no more material entrained in the boundary
layers. At higher velocities, if tracing material deeper into the cloud, the
eastern edge is pushing into the dense molecular ridge seen in the C$^{18}$O
emission, while the western edge is pushing into more clumpy and fragmented
material. This would account for the different velocity gradients seen in the
PV diagrams, and supports the three-dimensional models suggested for this
region \citep{emprechtinger2009,barnes1989}.

\begin{figure}[h]
\includegraphics[width=8.5cm,angle=0]{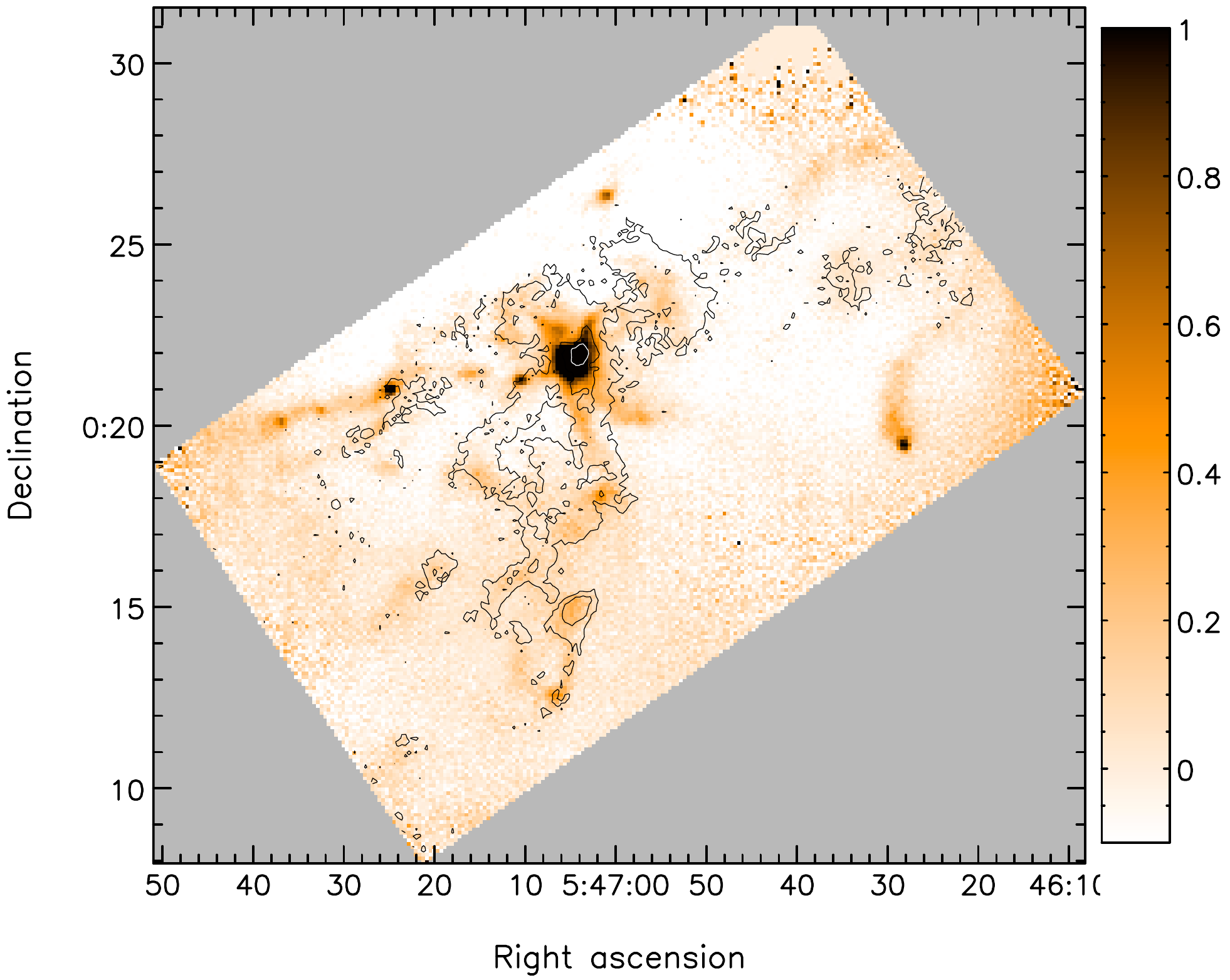}
\caption{\label{fig-scuba3}NGC 2071 SCUBA 850 $\umu$m map
\citep{difrancesco2008}, overlaid with C$^{18}$O integrated intensity contours.
Contour levels are as in Fig.~\ref{fig-int2}.}
\end{figure}

Fig.~\ref{fig-scuba3} shows the NGC 2071 SCUBA 850 $\umu$m map
\citep{difrancesco2008}, overlaid with C$^{18}$O integrated intensity contours.
The C$^{18}$O emission does not follow the structure seen in the dust as
closely as in NGC 2024, and it is in the region co-incident with the optical
nebula that we see the most discrepancy. In the PV diagrams
(Fig.~\ref{fig-ngc2071-pv}), we see that the $^{13}$CO and C$^{18}$O emission
is clumpy and fragmented, both spatially, and in velocity, while the $^{12}$CO
emission has a lower intensity region near the systemic cloud velocity,
embedded within material at much higher velocities.

\section{Analysis of Cloud Parameters}
\label{sec:cloud}

\subsection{Opacity}
\label{sec-opacity}
We calculate the gas opacity based on the line peak ratios of
$^{12}$CO/$^{13}$CO and $^{13}$CO/C$^{18}$O. The intensity ratio is related to
the opacity through:

\begin{eqnarray}
R_{\rm 1318} &=&\frac{T_{\rm A}^*(^{13}{\rm CO})}{T_{\rm A}^*({\rm C^{18}O})} =
\frac{1-\exp(-\tau_{13})}{1-\exp(-\tau_{18})}\\
R_{\rm 1213} &=&\frac{T_{\rm A}^*(^{12}{\rm CO})}{T_{\rm A}^*({\rm C^{13}O})} =
\frac{1-\exp(-\tau_{12})}{1-\exp(-\tau_{13})}
\end{eqnarray}

This method makes several assumptions about emission from the isotopologues
\citep*[see, for example][for details]{myers1983,ladd1998}. In particular that
the beam efficiency, filling factor, and excitation temperature are the same
for all the isotopologues, and that the ratio of line widths to the line of
sight extent of the emitting gas is the same for all isotopologues.

\begin{figure*}
\vbox{
\hbox{
\includegraphics[width=8cm,angle=0]{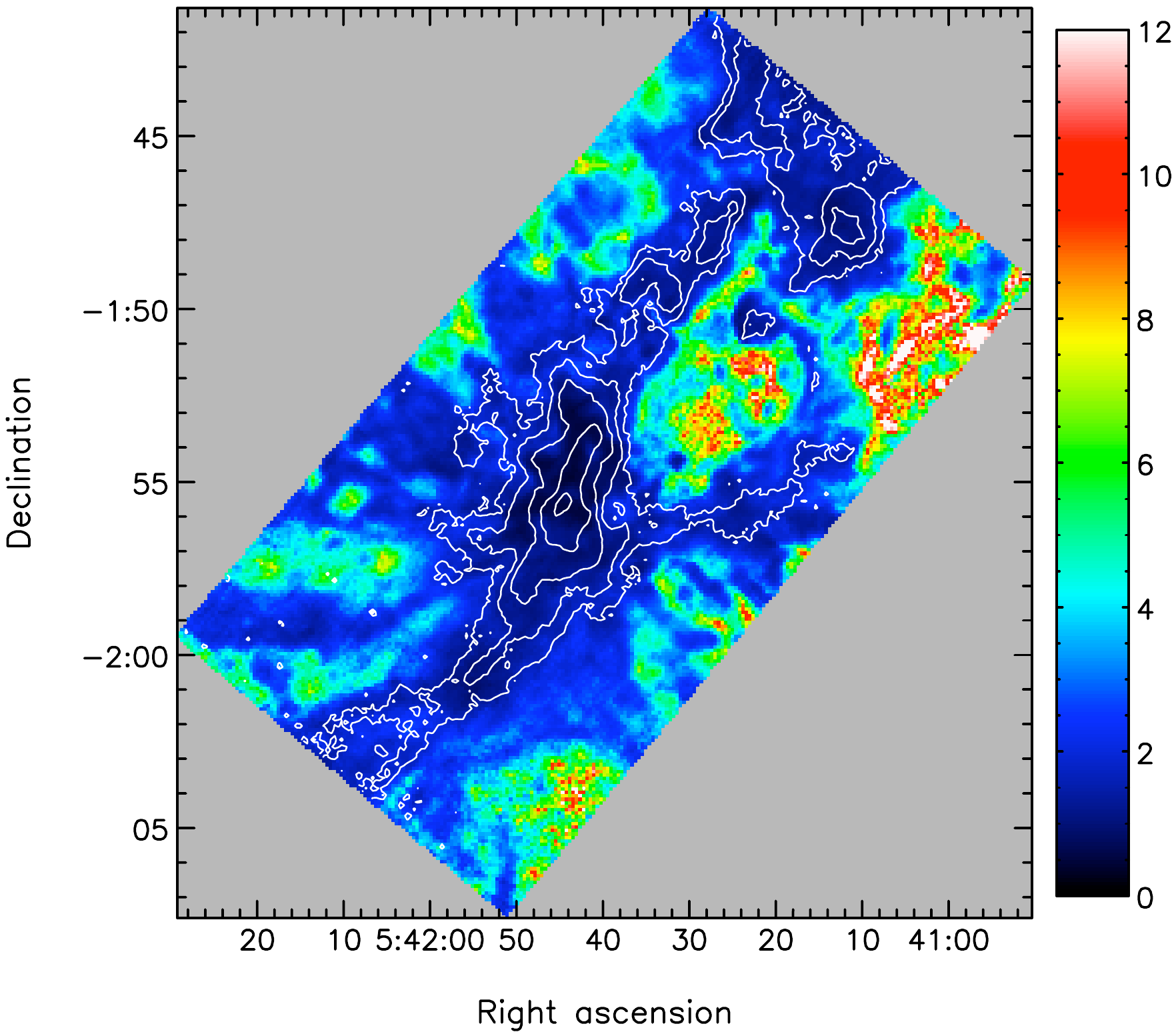}
\includegraphics[width=8cm,angle=0]{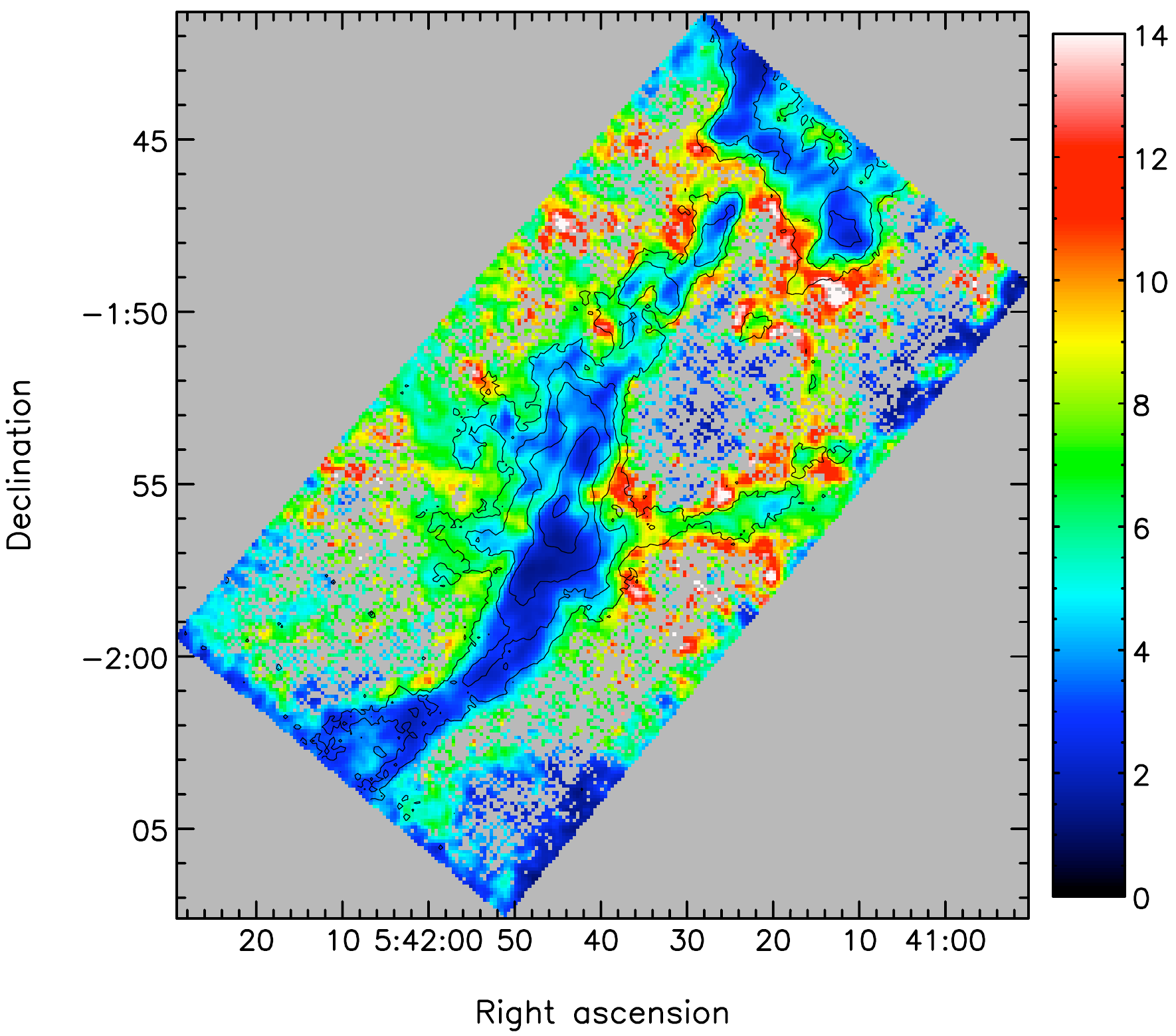}}
\hbox{
\includegraphics[width=8cm,angle=0]{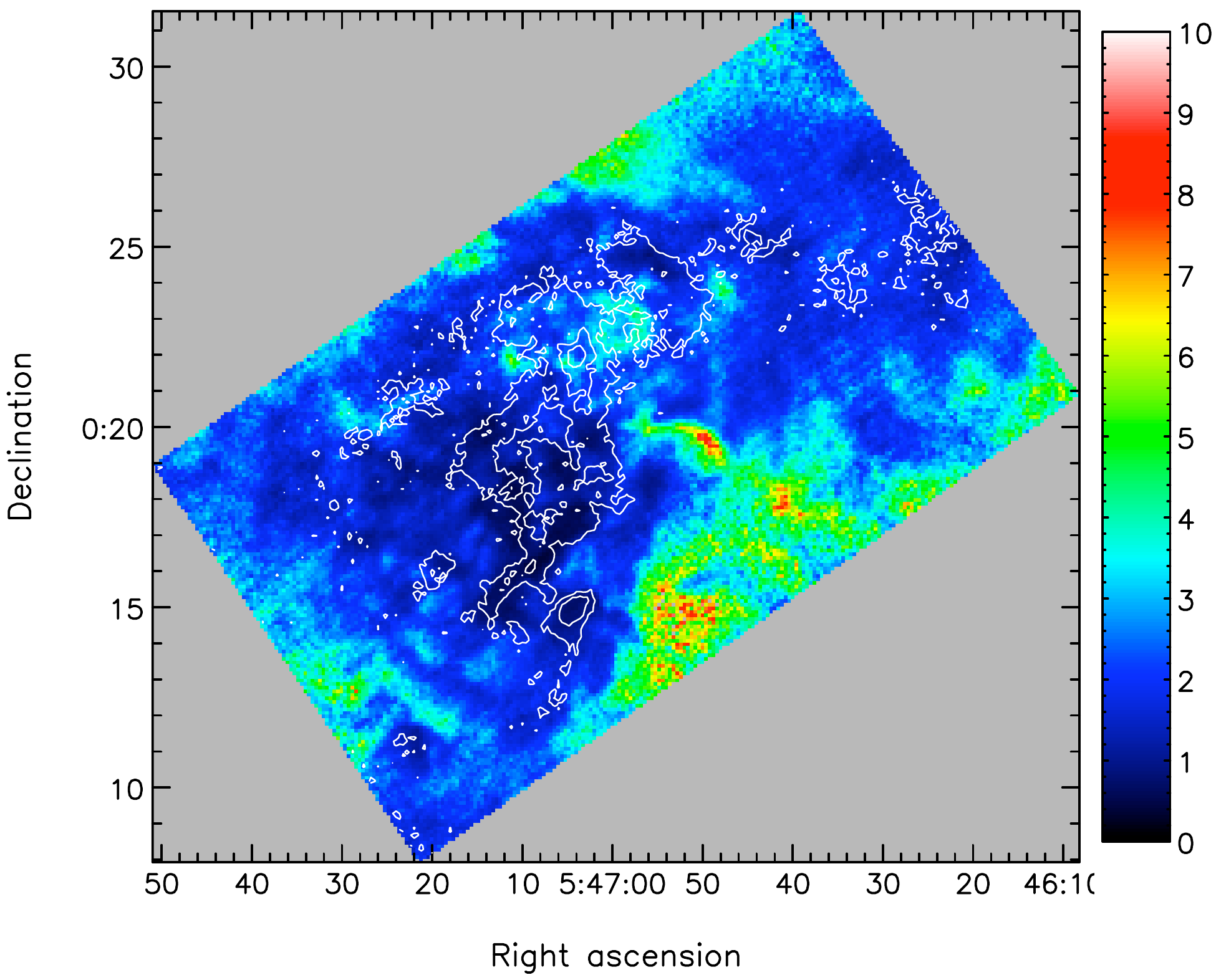}
\includegraphics[width=8cm,angle=0]{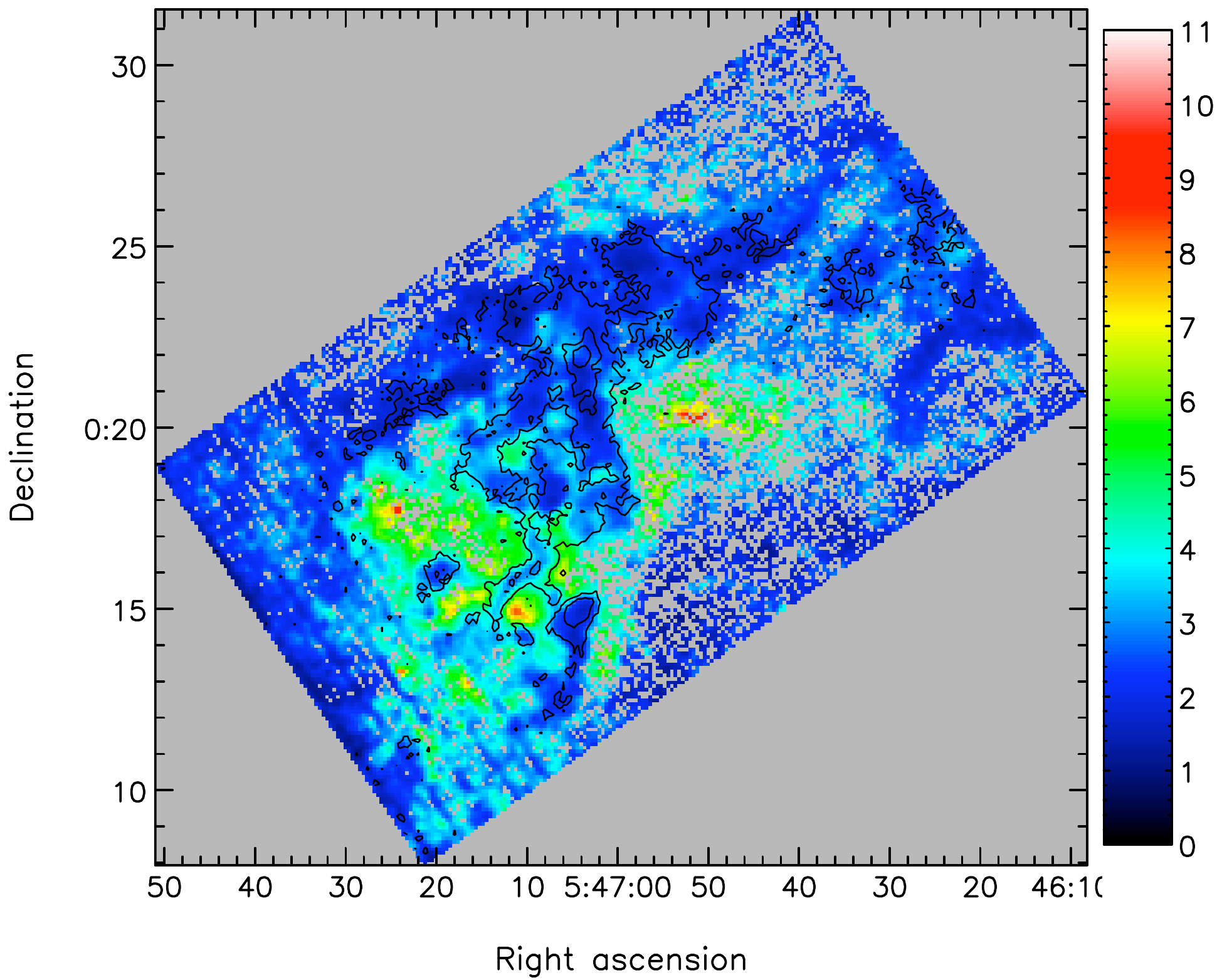}}}
\caption{\label{fig-ratio}The $^{12}$CO/$^{13}$CO (left) and $^{13}$CO/C$^{18}$O (right) peak intensity ratios for regions with detections $<$3$\sigma$ towards NGC 2024 (top) and NGC 2071 (bottom), overlaid with C$^{18}$O integrated intensity contours. The $^{13}$CO/C$^{18}$O ratio data has been smoothed by a Gaussian
with 12~arcsec FWHM to make the features more visible in these images.}
\end{figure*}

Where the lines are optically thin, the ratio $R$ should tend to the abundance
ratios, X[$^{13}$CO/C$^{18}$O]=8.4 and X[$^{12}$CO/$^{13}$CO]=70 for
$^{13}$CO/C$^{18}$O and $^{12}$CO/$^{13}$CO respectively
\citep*{frerking1982,wilson1999}. The $^{12}$CO/$^{13}$CO intensity ratio maps
for each source are shown in Fig.~\ref{fig-ratio}.

From the ratio maps, we can estimate the abundance ratios from the values the
ratios tend to towards the edge of detected emission regions, where we expect
the lines to be optically thin. At the edges of the C$^{18}$O emission regions,
$R_{\rm 1318}$ tends to  8--12 in NGC 2024, and we therefore adopt
X[$^{13}$CO/C$^{18}$O]=10 for NGC 2024. For NGC 2071, $R_{\rm 1318}$ tends to
6--9, so we similarly adopt X[$^{13}$CO/C$^{18}$O]=7.5. We detect $^{12}$CO and
$^{13}$CO across the full extent of the mapped region, and the integrated
intensity ratio does not approach the expected abundance ratio of $\sim$70. The
highest ratios that we detect are $\sim$20, suggesting that $^{12}$CO is
optically thick in the line core everywhere in both clouds, and so we use
X[$^{12}$CO/$^{13}$CO]=70.  Using the above equations and abundance ratios,
mean opacities in the clouds are $\bar\tau$($^{12}$CO)~$\sim$119, $\bar\tau$($^{13}$CO)~$\sim$1.7, and
$\bar\tau$(C$^{18}$O)~$\sim$0.17 in NGC 2024, while
$\bar\tau$($^{12}$CO)~$\sim$157, $\bar\tau$($^{13}$CO)~$\sim$2.24, and $\bar\tau$(C$^{18}$O)~$\sim$0.3 in
NGC 2071. These values suggest that C$^{18}$O is generally optically thin, while
$^{13}$CO is marginally optically thick, and $^{12}$CO is optically thick
throughout both clouds. Note that self-absorption in the line profiles, which
we see towards the densest regions, means that the opacity will be
over-estimated for $^{12}$CO, and to a lesser extent for $^{13}$CO.

\subsection{Temperature}
\label{sec-tex}

\begin{figure}
\vbox{
\includegraphics[width=8cm,angle=0]{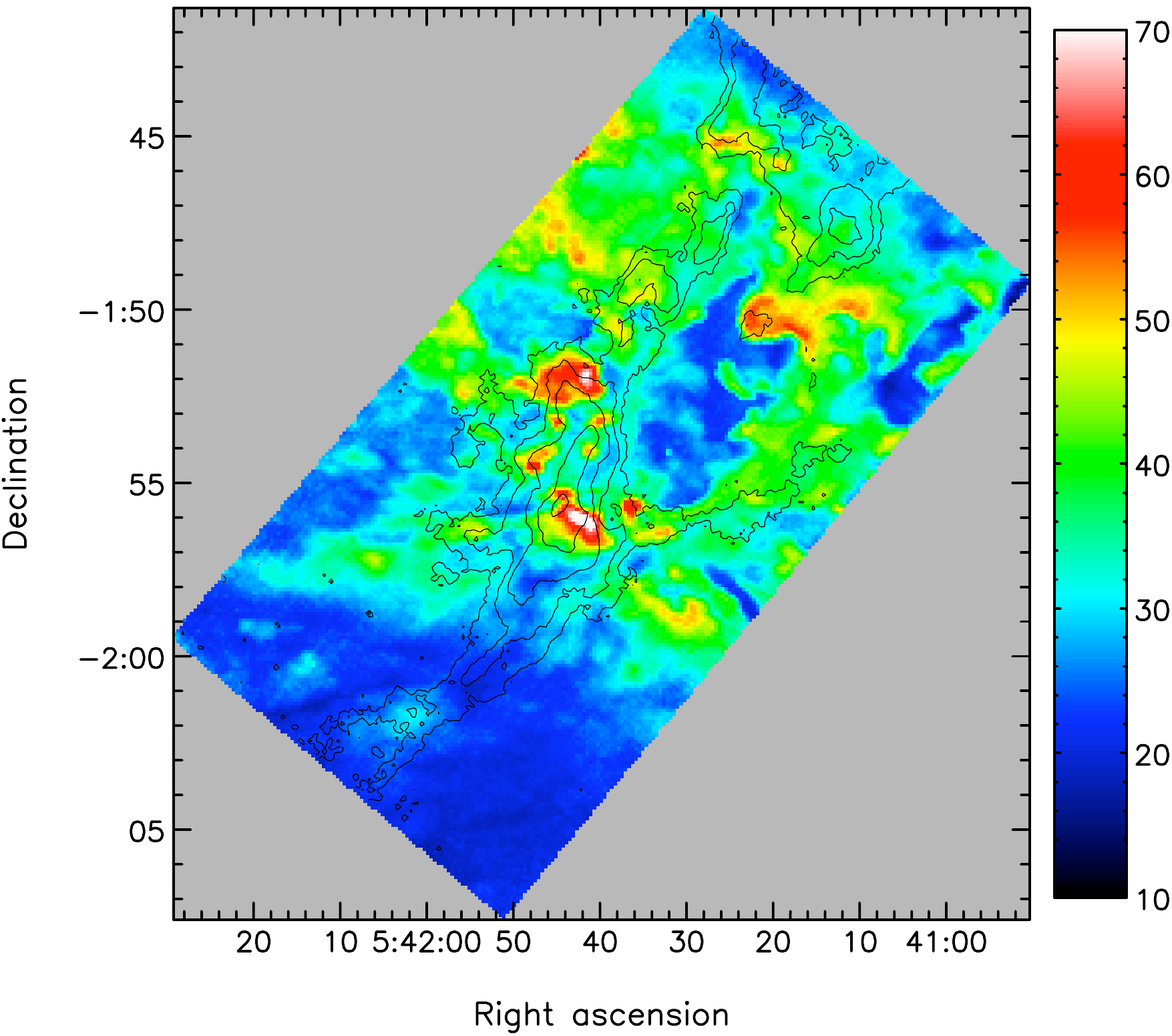}
\includegraphics[width=8cm,angle=0]{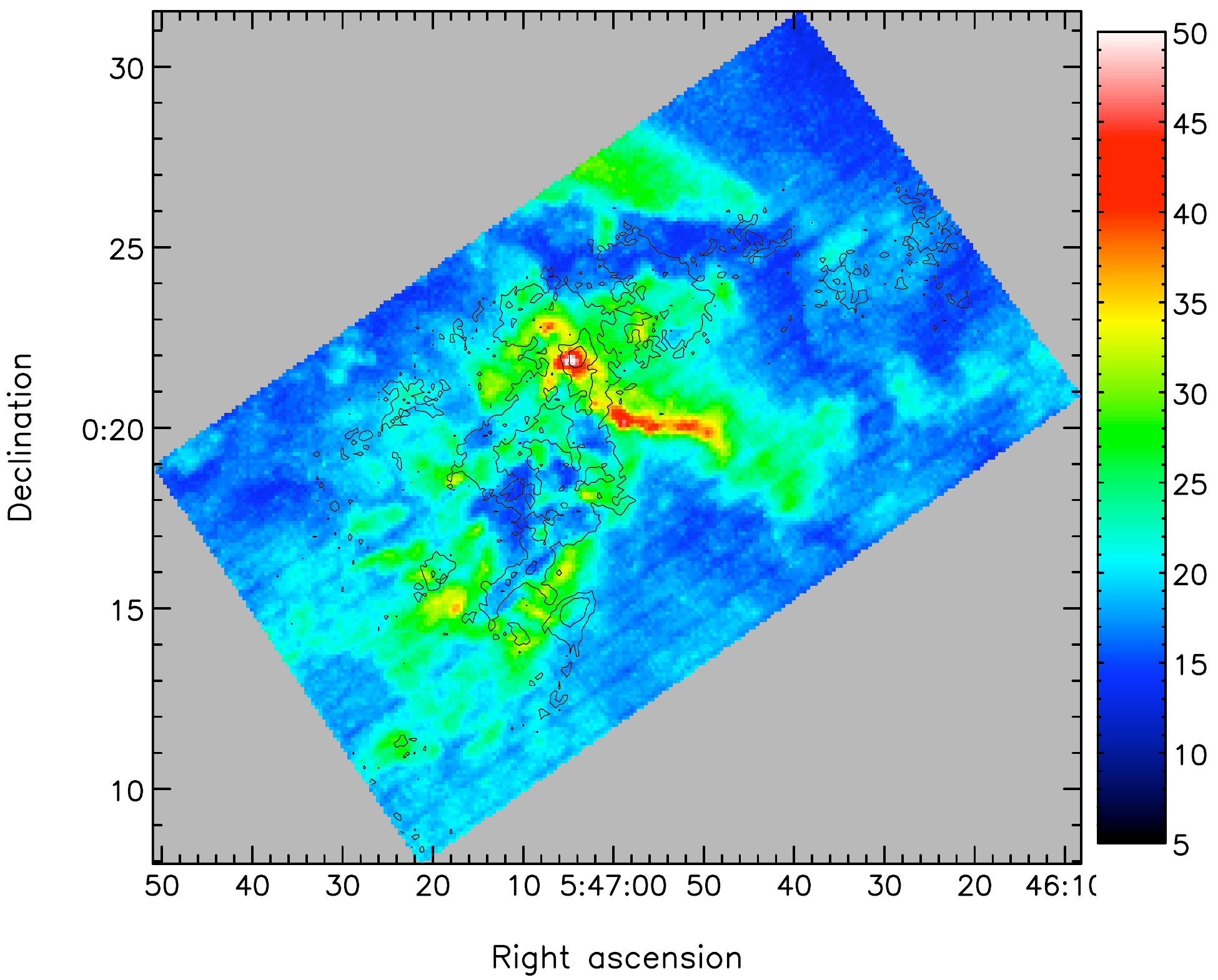}}
\caption{\label{fig-tex}Excitation temperature (in K) for NGC 2024 (top) and NGC 2071
(bottom). To indicate regions where the $^{12}$CO may be self-absorbed, and so
underestimating the excitation temperature, the contours show C$^{18}$O
integrated intensity.}
\end{figure}

Assuming local thermodynamic equilibrium (LTE), the excitation temperature
($T_{\rm ex}$) of the gas can be calculated from line peak temperature
($T_{\rm max}$) using the standard radiative transfer relation in an isothermal
slab. Also assuming that the $^{12}$CO emission is optically thick
($\tau\rightarrow\infty$) and fills the beam in the line core across the cloud,
then \citep*[e.g.][]{pineda2008}:

\begin{equation}
\label{eq-temp}
T_{\rm ex}(3\rightarrow2) = \frac{16.59 {\rm K}}{{\rm ln}\left[1+16.59 {\rm K}/\left( T_{\rm
max}({\rm ^{12}CO)}+0.036 {\rm K}\right)\right]},
\end{equation}

\noindent for regions where the $^{12}$CO line is not self-absorbed. We can
identify the regions where there is self-absorption by comparison of the
$^{12}$CO line profile with that of the $^{13}$CO line profile. Where the
$^{12}$CO is optically thick and self-absorbed, the $^{13}$CO line peak
temperature is higher than the $^{12}$CO line peak temperature. Towards these
regions, $^{13}$CO is optically thick, and we use the equivalent form of
Eq.~\ref{eq-temp} to calculate $T_{\rm ex}$ from $^{13}$CO. Fig. \ref{fig-tex}
shows $T_{\rm ex}$ across NGC 2024 (top) and NGC 2071 (bottom), calculated using
both $^{12}$CO and $^{13}$CO.

NGC 2024 generally reaches higher excitation temperatures than NGC 2071. Towards
NGC 2024, the mean excitation temperature is 31.8~K, which increases to
$\ga$50~K in the dense regions, and $\sim$70~K in regions which are known to
contain energetic outflows. Towards NGC 2071, the mean excitation temperature is
19.6~K, which increases to $\ga$30~K in the dense regions, and $\sim$50~K in
the regions associated with energetic outflows.

\begin{figure}
\vbox{
\includegraphics[width=8cm,angle=-90]{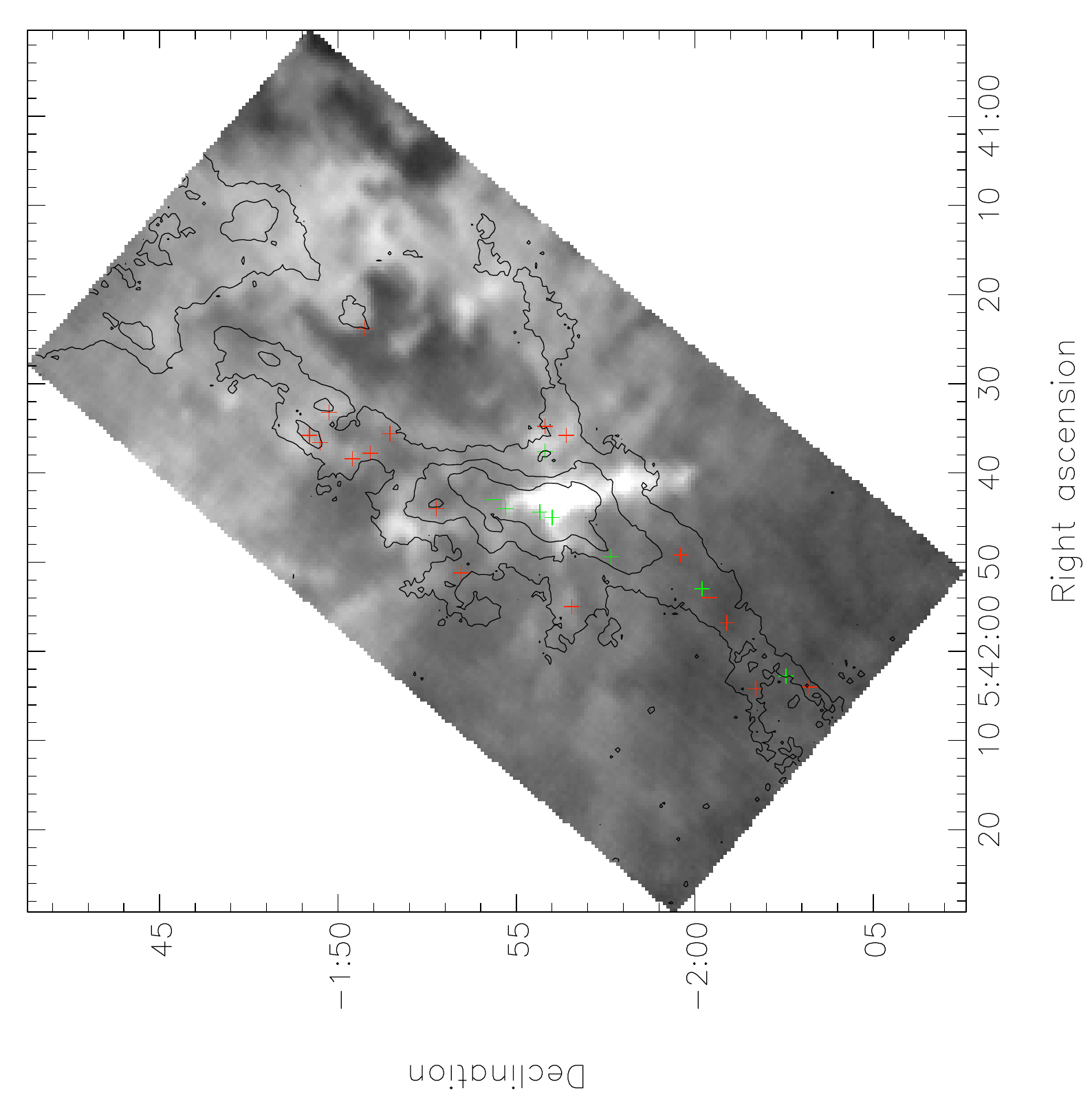}
\includegraphics[width=7.5cm,angle=-90]{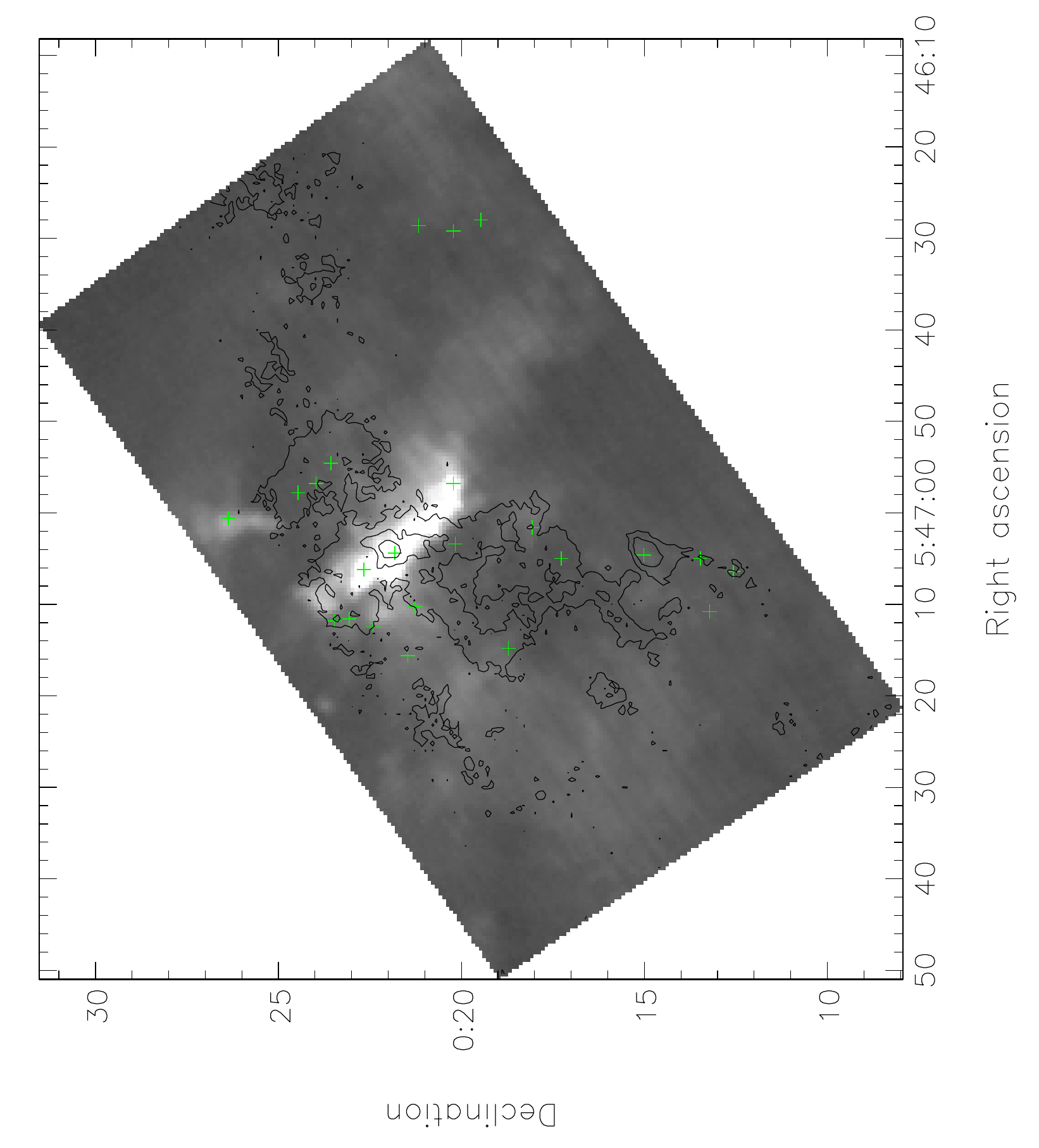}}
\caption{\label{fig-tex-comp}$^{12}$CO integrated intensity maps for NGC 2024
(top) and NGC 2071 (bottom), overlaid with C$^{18}$O contours. SCUBA cores from
J01, J06 are marked in green where the ratio $T_{\rm ex}$($^{12}$CO)/$T_{\rm
dust}$ $\leq$1.5 and in red otherwise.}
\end{figure}

Towards the dense cores of Orion B, kinetic temperatures have been published
using line ratios of H$_2$CO, a molecule known to trace gas kinetic
temperatures extremely well \citep{mangum,tothill,mangum1999}. Although only a
few of these sources overlap with the regions we have observed, the excitation
temperatures calculated from CO agree well with those calculated from H$_2$CO,
except towards NGC 2024 FIR5, the probable driving source of the energetic
outflow seen as a large red-shifted outflow lobe (Fig.~\ref{fig-out1}). Average
values for the small number of H$_2$CO dense cores, with the exception of
NGC 2024 FIR5, are 44~K from H$_2$CO, and 39~K from CO. Given that the H$_2$CO
transitions have a critical density of $\geq$3.9$\times$10$^6$~cm$^{-3}$, while
the critical density of the $^{12}$CO transition is
$\sim$3.5$\times$10$^4$~cm$^{-3}$, and that the velocity structure of these
regions is relatively complex, this is surprising, and suggests the combination
of $^{12}$CO and $^{13}$CO is a relatively good tracer of excitation conditions
in dense regions. We have also compared our CO excitation temperatures with
those derived from dust temperatures towards SCUBA cores in both clouds by J01,
J06.  Fig.~\ref{fig-tex-comp} shows the integrated intensity maps of $^{12}$CO,
overlaid with contours of C$^{18}$O, and marked with the positions of the SCUBA
dust cores from J01, J06. The cores are marked in green if $T_{\rm
ex}(^{12}$CO)/$T_{\rm dust}$ $\leq$1.5, and in red if the ratio is $>$1.5.
For the small number of cores that also have H$_2$CO measurements, the average
dust temperature is 41~K, similar to the H$_2$CO and CO values. Towards
NGC 2071, all of the SCUBA cores have CO and dust temperatures ratios $\leq$1.5,
while towards NGC 2024, the ratio is $\leq$1.5 only towards the high column density regions of
NGC 2024, along the molecular ridge. Towards NGC 2024, the cores where dust
temperatures are similar to temperatures derived from CO are spatially grouped
towards the highest column density regions. In NGC 2024, the differences in
$T_{\rm ex}(^{12}$CO) and $T_{\rm dust}$ towards the lower column density
regions may be evidence of sub-thermal excitation of CO. Given the complex
velocity structure of these regions, and the complications that opacity adds to
the line profile, observations utilizing all CO isotopologues so that opacity
corrections can be made appear to be relatively good tracers of the gas and
dust temperatures in energetic, densely clustered regions.

\subsection{Mass and energetics}
\label{sec:mass}

Since we have isotopic data, we can calculate the mass in the large-scale
cloud, using the $^{12}$CO and $^{13}$CO data, which we detect everywhere. In
the dense molecular ridge, where $^{12}$CO and $^{13}$CO line profiles are
affected by self-absorption, we can use the C$^{18}$O data, assuming LTE
conditions. We compute the masses assuming $T_{\rm ex}$ is the same for all
three isotopologues, and equal to the mean $^{12}$CO excitation temperature
(Sec.~\ref{sec-tex}). We correct for optical depth effects in the $^{12}$CO and
$^{13}$CO data using the mean opacities $\bar\tau(^{13}\rm CO)$, and
$\bar\tau(^{12}\rm CO)$ (Sec.~\ref{sec-opacity}). Following \citet{garden1991}
:

\begin{eqnarray}
\label{eq_ntot}
\rm N(C^{18}O)&=&8.26\times 10^{13}\exp\left(\frac{15.81}{T_{\rm ex}}\right)
\nonumber\\
&\times& \frac{T_{\rm ex} + 0.88}{1-\exp(\frac{-15.81}{T_{\rm ex}})}\int \tau
{\rm \,dv}~{\rm cm^{-2}}\\
\rm N(^{13}CO)&=& 8.28\times 10^{13}\exp\left(\frac{15.87}{T_{\rm ex}}\right)
\nonumber\\
&\times&\frac{T_{\rm ex} + 0.88}{1-\exp(\frac{-15.87}{T_{\rm ex}})}\int \tau
{\rm \,dv}~{\rm cm^{-2}}\label{eq-colden}\\
\rm N(^{12}CO)&=& 7.96\times 10^{13}\exp\left(\frac{16.6}{T_{\rm ex}}\right)
\nonumber\\
&\times&\frac{T_{\rm ex} + 0.92}{1-\exp(\frac{-16.6}{T_{\rm ex}})}\int \tau
{\rm \,dv}~{\rm cm^{-2}}
\end{eqnarray}

\noindent where we use the approximations:
\begin{eqnarray}
 \int \tau {\rm \,dv} &=&  \frac{1}{\left[J(T_{\rm ex}) - J(T_{\rm BG})\right]}\frac{\bar{\tau}}{1-e^{-\bar{\tau}}} \int T_{\rm mb} {\rm \,dv}~({\rm for}~\tau\geq 1)\\
 \int \tau {\rm \,dv} &=&  \frac{1}{\left[J(T_{\rm ex}) - J(T_{\rm BG})\right]}\int T_{\rm mb}{\rm \,dv}~({\rm for}~\tau\ll 1) \label{eq-thin} \\
J(T) &=& \frac{T_0}{\exp\left(\frac{T_0}{T}\right)-1}\\
T_o &=& \frac{h\nu}{k}
\end{eqnarray}

\noindent with $T_{\rm BG}$ = 2.7K. The mass is given by:
\begin{equation}
M_{\rm gas} = 1.13 \times 10^{-4} \umu_{H_2} m_H d^2 \Delta \alpha \Delta \beta {\rm X}
N({\rm CO})  ~~\msol \label{eq-mass}
\end{equation}

\noindent where $\Delta\alpha\Delta\beta$ is the pixel area in arcsec$^2$, $d$
is the distance in pc, v is in \kms and X is the isotopic abundance ratio of
CO and its isotopologues relative to H$_2$. We adopt a mean molecular weight
per H$_2$ molecule of $\umu_{H_2}$ = 2.72 to include helium, and
X[$^{12}$CO]=1.0$\times$10$^{-4}$, X[$^{13}$CO]=1.4$\times$10$^{-6}$,
X[C$^{18}$O]=1.7$\times$10$^{-7}$ \citep{frerking1982,wilson1999}. The results are
listed in Tab.~\ref{tab-mass}. The masses that we find for optically thick
$^{12}$CO are smaller than previous mapping observations of a 19~deg$^2$ region in Orion B, using a lower
excitation transition, where a total mass of 8.3$\times10^4$ \msol\ was measured
\citep{maddalena1986}. The total mass we derive for NGC 2024 and NGC 2071, covering a region $\sim$100 times smaller in area and using optically thick $^{12}$CO, is 8.5$\times10^3$~\msol. In $^{13}$CO, we find a total mass of
2.5$\times10^3$~\msol.  
The total mass of gas
that we detect in C$^{18}$O towards both clouds is 1000~\msol. 
Emission from the optically thin isotopologue
C$^{18}$O traces a smaller region within the cloud, and does not suffer from
the same uncertainties due to optical depth corrections. 
If the opacity corrections using mean values were accurate, we would perhaps expect the masses calculated from $^{12}$CO and $^{13}$CO emission to be similar to that calculated from C$^{18}$O emission. However, the differences may be due to physical conditions within the clouds affecting the transitions differently. Towards NGC 2024 in particular, the many velocity components seen in the spectra indicate that a single excitation temperature and density along the line of sight may not be a good approximation of the conditions. The clouds may not be in LTE, and emission from the isotopologues subthermally excited. As suggested by \citet{pineda2008}, the shielding of material within the cloud leads to physical and chemical properties which can affect $^{12}$CO and $^{13}$CO differently to the rarer isotopologue C$^{18}$O. We find a total mass of 1111~\msol\ using $^{13}$CO data, if we assume that
$^{13}$CO emission is also optically thin. The total mass calculated from
$^{12}$CO under the assumption of optically thin emission is 65~\msol,
indicating that emission from this isotopologue is, as calculated above, very
optically thick, and tracing only the surface regions of the clouds. NGC 2024
contains more mass than NGC 2071, measured in all of the isotopologues.

The gravitational potential energy, $W$, of the cloud can be calculated
assuming a density distribution $\rho~\propto r^a$
\citep*{maclaren1988,williams1994,bertoldi1992}. The kinetic energy, $E_{\rm
kin}$, can be calculated by estimating the three-dimensional velocity
dispersion, $\sigma_{v,3D}^2$, from the one-dimensional velocity dispersion of
the average spectrum, $\sigma_{^{12}{\rm CO}}$, $\sigma_{^{13}{\rm CO}}$ or
$\sigma_{\rm C^{18}{\rm O}}$:

\begin{eqnarray}
W &=& -\frac{3}{5}\gamma \frac{GM_{\rm gas}^2}{R} \label{eq-egrv}\\
\gamma &=& \frac{1 + a/3}{1 + 2a/5}\\
E_{\rm kin} &=& \frac{1}{2}M_{\rm gas} v^2\\
v^2 &=& \sigma_{v,3D}^2 = 3 \left[    \sigma_{{\rm CO}}^2   +  \frac{kT}{m_{H}} \left(
\frac{1}{\umu} - \frac{1}{m_{\rm CO}}             \right)  \right]
\label{eq-vdisp}
\end{eqnarray}

\noindent where $G$ is the gravitational constant, $k$ is Boltzmann's constant,
$T$ is the kinetic temperature of the gas, which we have taken to be equal to
the mean $^{12}$CO excitation temperature and $m_{{\rm CO}}$ is the atomic mass
of $^{12}$CO, $^{13}$CO, or C$^{18}$O. We assume the clouds follow $\rho
\propto r^{-2}$ density distributions, yielding $\gamma = 5/3$.  The LTE
masses, kinetic and gravitational potential energies of both clouds are listed
in Tab.~\ref{tab-mass}.

{\scriptsize
\begin{table*}
\caption{\label{tab-mass}Table of masses and energetics}
\begin{tabular}{lrrrrrrrrr}
\hline
Cloud &  \multicolumn{3}{c}{LTE mass ($\times10^3$
/\msol)}&\multicolumn{3}{c}{$E_{\rm kin}$ ($\times10^{40}$
/J)}&\multicolumn{3}{c}{$-W$ ($\times10^{40}$ /J)}\\
&$^{12}$CO&$^{13}$CO&C$^{18}$O&$^{12}$CO&$^{13}$CO&C$^{18}$O&$^{12}$CO&$^{13}$CO&C$^{18}$O\\
NGC 2024&4.8&1.6&0.6&37.5&5.9&1.6&9.5&1.1&0.2\\
NGC 2071&3.7&0.9&0.4&24.5&1.6&0.4&5.7&0.3&0.1\\
\hline
\end{tabular}
\end{table*}
}

NGC 2024, as well as containing more mass than NGC 2071, has greater kinetic and
gravitational potential energies. The gravitational potential energy calculated
from $^{12}$CO and $^{13}$CO is larger than the kinetic energy in both regions,
implying that these clouds are both gravitationally bound. The emission from
C$^{18}$O, tracing much smaller regions containing all of the currently
active outflows, has more kinetic energy.  The line profiles and PV diagrams of
$^{12}$CO and $^{13}$CO indicate that both of these isotopologues are tracing
some of the high-velocity, presumably outflowing, material. In
Sec.~\ref{sec:energy}, we compare these values to the energetics in the high-velocity material.

\subsection{Cloud kinematics}
\label{sec-ckin}

\begin{figure}
\vbox{
\includegraphics[width=8cm,angle=0]{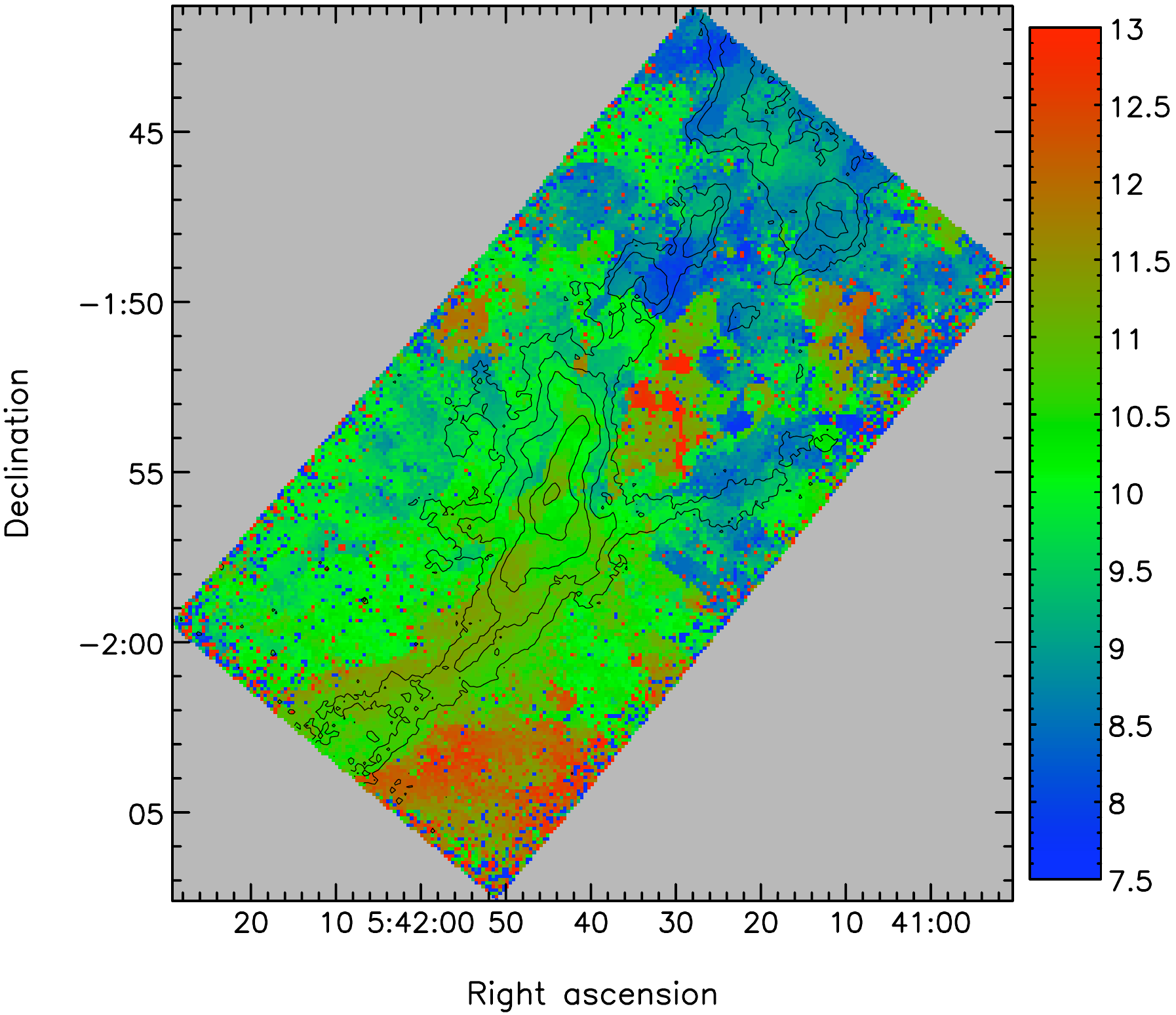}
\includegraphics[width=8cm,angle=0]{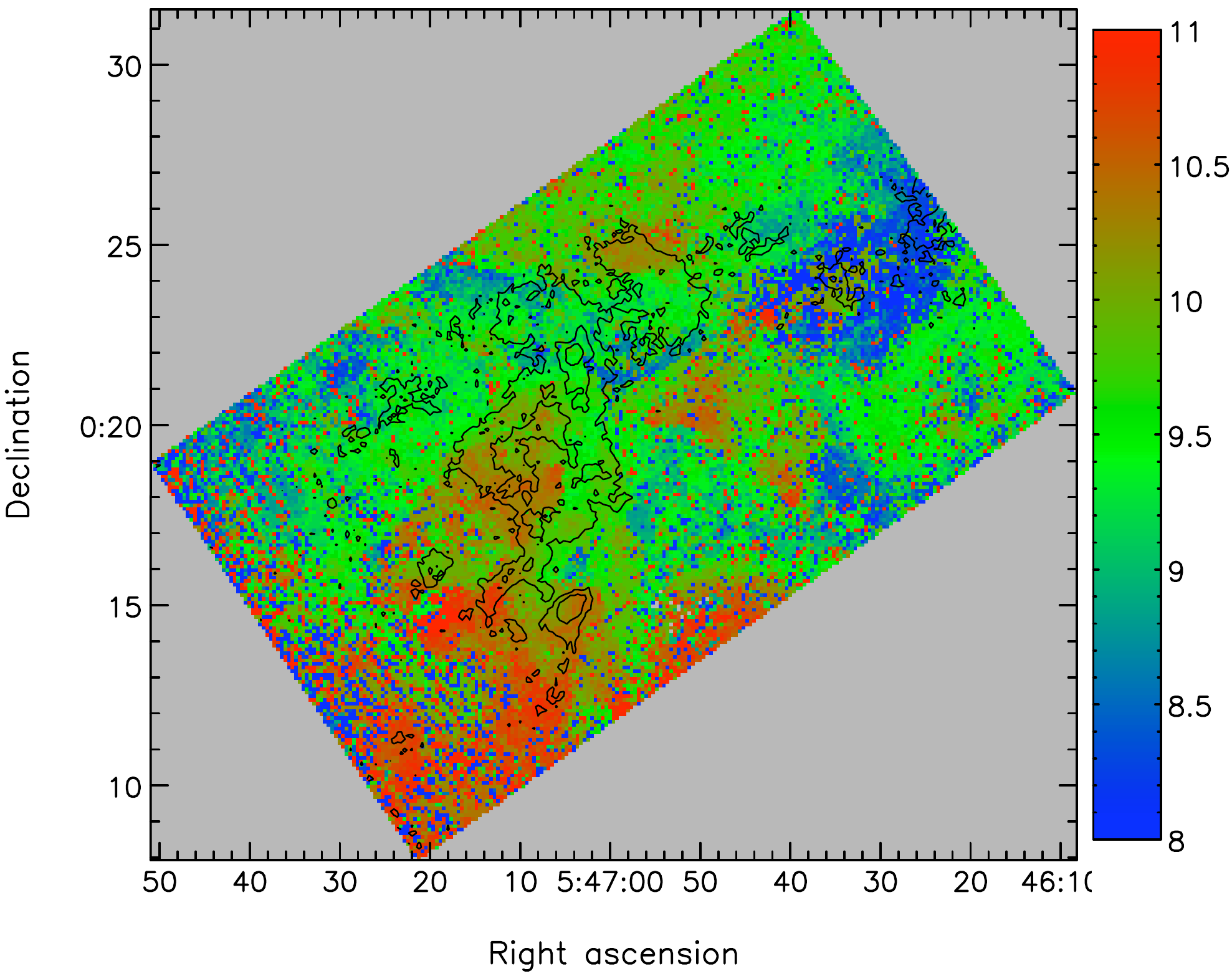}}
\caption{\label{fig-iwc1}NGC 2024 map of the velocity at the line peak (in \kms) for
C$^{18}$O where emission is above the 3$\sigma$ level, and $^{13}$CO at all
other regions (top). $^{13}$CO is detected above the 3$\sigma$ level at all
points in the map. The same Fig. for NGC 2071 (bottom).}
\end{figure}

Fig.~\ref{fig-iwc1} shows the velocity at the line peak of the optically-thin
gas. This has been created from the $^{13}$CO and C$^{18}$O data, using
$^{13}$CO in regions where the C$^{18}$O line peak falls below the 3$\sigma$
level. Towards NGC 2024, a velocity gradient runs across the image, from 8~\kms\
in the north, to 10~\kms\ in the south, with the molecular ridge in the south
seen in emission at 11.5~\kms. The dust cavity is seen in material at
$\ga$11.5~\kms, surrounded by material at 8.0--10.0~\kms, with a further higher
velocity band to the northwest. In the southernmost part of the map, a band of
high-velocity material appears, which is the tip of the NGC 2023 region. The
velocity gradient can also be clearly seen in the $^{12}$CO PV diagram
(Fig.~\ref{fig-ngc2024-pv}). Towards NGC 2071, there is no obvious underlying
velocity gradient. All of the velocity variations seen in the NGC 2071 map
appear to be associated with current star forming activity in the cloud. In the
south, there is a large region that is clearly red-shifted with respect to the
rest of the cloud, which is in the same direction as the very energetic
red-shifted outflow lobes seen towards NGC 2071 (Fig.~\ref{fig-out2})

\begin{figure}
\vbox{
\includegraphics[width=8cm,angle=0]{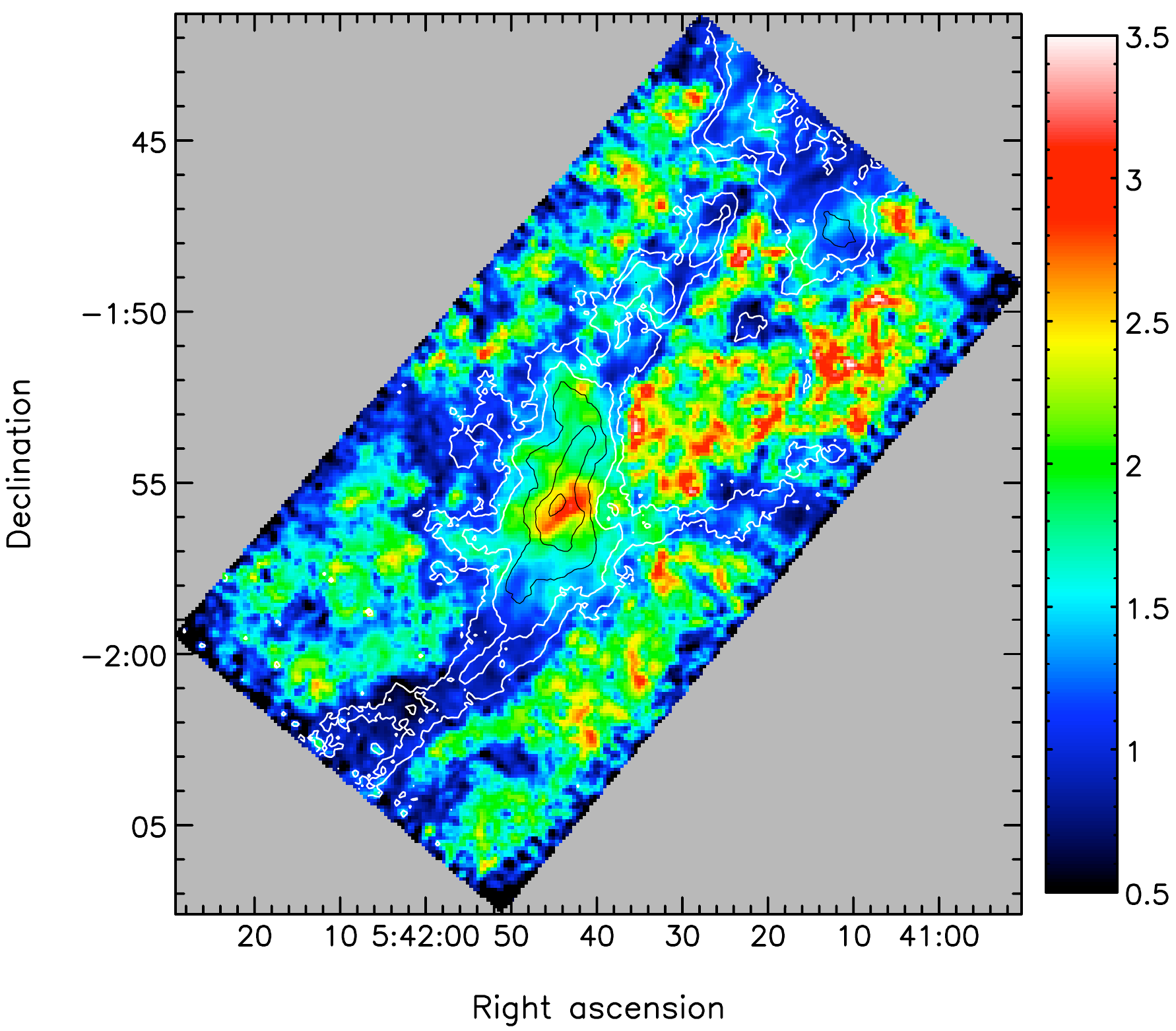}
\includegraphics[width=8cm,angle=0]{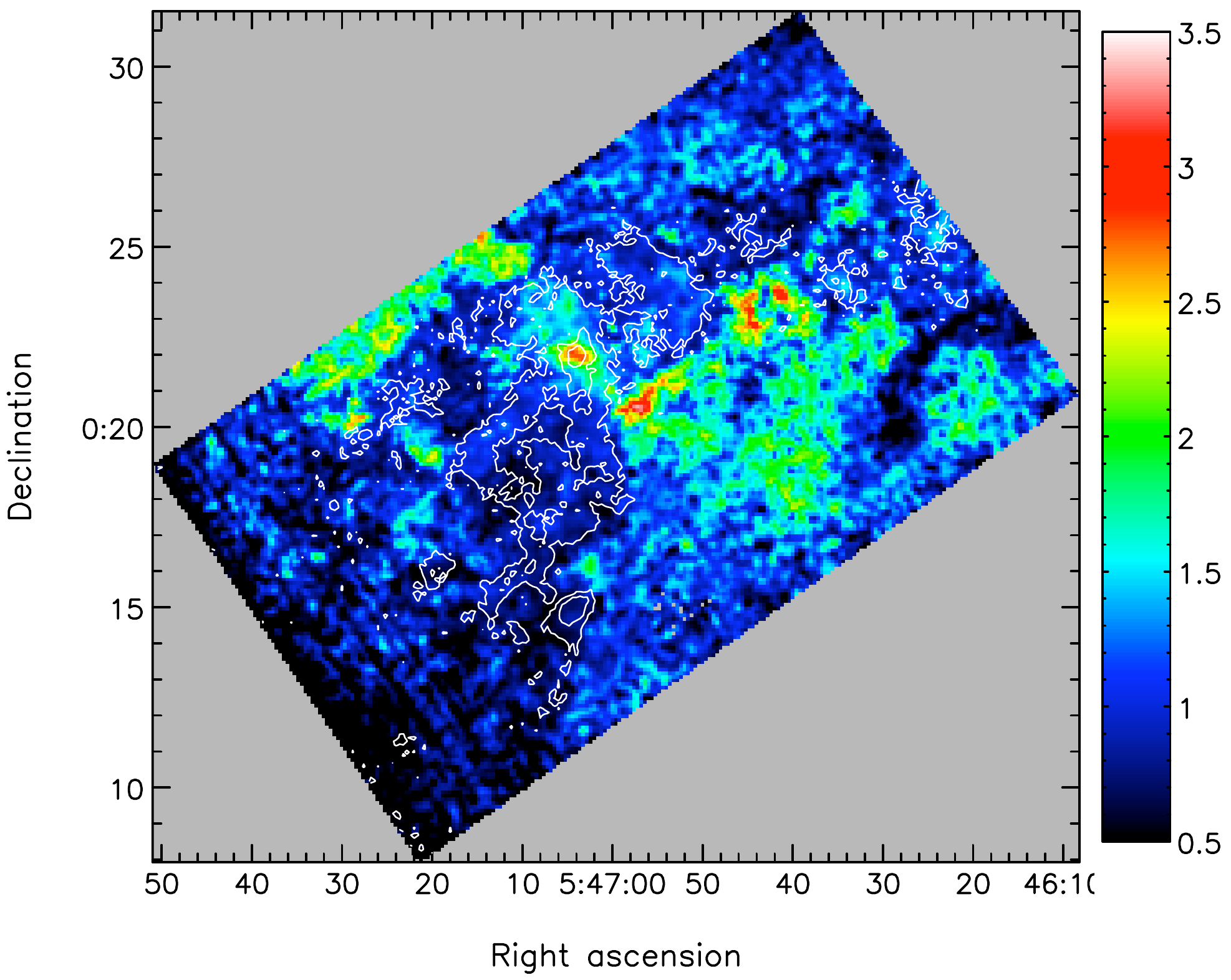}}
\caption{\label{fig-eqw1}NGC 2024 map of equivalent width (in \kms) for C$^{18}$O where
emission is above the 3$\sigma$ level, and $^{13}$CO at all other regions
(top). $^{13}$CO is detected above the 3$\sigma$ level at all points in the
map. The same Fig. for NGC 2071 (bottom). Data has been smoothed by a Gaussian
with 12~arcsec FWHM to make the features more visible in these images.}
\end{figure}

Fig.~\ref{fig-eqw1} shows the equivalent width ($ \int Tdv/T_{peak}$) for the
optically thin emission towards the two regions. The exterior edges of the
molecular ridge towards NGC 2024, traced by the outer contours of C$^{18}$O,
have lower velocity dispersions than the material within which it is embedded,
and the energetic region near the FIR sources. Towards NGC 2071, the material
with a lower velocity dispersion is perpendicular to the outflow direction. In
the regions extending along the outflow, a cone-shape of higher velocity
dispersion material can be seen.

\subsection{Mass and energetics of the high-velocity material}
\label{sec:energy}

Using the line centre of the average C$^{18}$O spectrum to identify the
velocity of the cloud, we calculate the total mass in the high-velocity
material generated from the outflows using $^{12}$CO emission, with red- and
blue-velocity extents across the line profile. We assume that the high-velocity
emission is optically thin, and at a similar excitation temperature to the bulk
of the gas, given by the mean $^{12}$CO excitation temperature, following the
method in Sec.~\ref{sec:mass}. We obtain masses of a few $\times$~1~\msol. This
is much lower than found using lower excitation transitions of $^{12}$CO and
$^{13}$CO \citep[e.g.][although these authors used a combination of
isotopologues across the velocity range for their
calculation]{snell1984,stojimirovic2008}. For $^{13}$CO, where the assumption
that the emission is optically thin in the line wings is more likely to be
correct, we obtain masses $\sim$ a few $\times$~10~\msol, in better agreement
with those previously found, which suggests the optically thin assumption for
$^{12}$CO line wings produced in outflows is not valid.

From the $^{13}$CO emission, we calculate the energy and momentum of this high
velocity material using :

\begin{eqnarray}
P&=& MV_{\rm char}\\
E_k&=& \frac{1}{2} MV_{\rm char}^2
\end{eqnarray}

\noindent where $V_{\rm char}$ is a characteristic velocity estimated as the
difference between the maximum velocity with detectable emission in $^{12}$CO
emission and the cloud velocity identified from the line centre velocity of a
Gaussian fit to the average C$^{18}$O line profile. For this reason, and since
we do not account for the inclination to the line of sight for any outflows,
the calculated values are upper limits. The mass, momentum and energy values
for the red- and blue-shifted emission for both clouds are listed in
Tab.~\ref{tab-hvmass}.

\begin{table*}
\caption{\label{tab-hvmass}Mass and energetics of high-velocity material}
\begin{tabular}{@{\,}l@{\,}crrrrr}
\hline
Cloud & Velocity& V$_{\rm char}$& ~$^{12}$CO Mass& $^{13}$CO Mass& Energy&
Momentum\\
&&\kms&\msol&\msol&$\times10^{40}$J&\msol~\kms\\
\hline
NGC 2024&red&32.3 &4.50&19.64&2.03&634\\
NGC 2024&blue&25.2 &3.38&14.59&0.92&367\\
NGC 2071&red&43.0 &6.75&37.34&6.86&1604\\
NGC 2071&blue&39.2 &5.66&28.01&4.29&1099\\
\hline
\end{tabular}
\end{table*}

The total energy of the high-velocity material within the cloud is 100 times
more than that calculated by investigating individual flows. Towards NGC 2071,
\citet{stojimirovic2008} found a total energy in the red and blue outflow lobes
of 9.67$\times10^{45}$~ergs, compared to our value of
11.15$\times10^{47}$~ergs. The values we calculate are likely to be
over-estimated, since we have used a maximum estimate for the characteristic
velocity. Comparing these values to the kinetic energy of the cloud as measured
from $^{13}$CO emission, the high-velocity material accounts for very little of
the total kinetic energy in NGC 2024, while the high-velocity material dominates
the kinetic energy in NGC 2071.

\section{Analysis of the gaseous condensations}
\label{sec:clumps}

The structure of molecular clouds has been characterized in numerous
ways in the literature, including identifying discrete objects in the
emission (often termed clumps). By studying the properties of such
clumps we hope to explore the properties of the cores, harboured by
some of the clumps, which will go on to form stars. Frequently,
automated clump-finding routines have been used, to look for clumps in
an unbiased manner, in both spectral-line and dust continuum
datasets. 
In this section, we undertake a statistical decomposition of the structure of the
$^{13}$CO $J=3\to2$ emission using the {\sc clumpfind} algorithm
\citep{williams1994}, implemented in Starlink's new CUPID package
\citep{berry2007}. The optically-thin C$^{18}$O $J=3\to2$ data will be a better
tracer of the mass in the star-forming cores, where the $^{13}$CO line is
saturated. Here we explore the $^{13}$CO $J=3\to2$ bulk cloud emission. Many different terminologies have been used to name
clumps in molecular clouds, here we call any individual object
identified using {\sc clumpfind} a `condensation', to emphasize that it may not collapse to form a star.  

The data were smoothed to 0.25~\kms\ spectrally to improve the
signal to noise ratio but still maintain sufficient resolution to
identify condensations with the narrowest anticipated line widths. At our
assumed cloud temperatures (50 and 30~K for NGC 2024 and NGC 2071
respectively) the $^{13}$CO line widths expected from thermal
broadening alone ($\sqrt{kT/m_H m_{CO}})$ are 0.4~\kms\ for NGC 2024
and 0.3~\kms\ for NGC 2071. At this resolution, the mean RMS noise
across the map is reduced to 0.14 and 0.20~K for NGC 2024 and NGC 2071 respectively.
We ran {\sc clumpfind} using a level spacing of $2
\sigma_{\rm RMS}$ (the value recommended by \citet{williams1994} to
reduce the contamination by noise) and a lowest level of $2
\sigma_{\rm RMS}$, which ensures the minimum peak intensity of a condensation is $4
\sigma_{\rm RMS}$. Additionally, condensations were rejected if they
touched any edge of the data array, contained fewer than 16 (3D)
pixels or were smaller than the beam size.  These parameters resulted
in a catalogue of 1561 and 1399 condensations in NGC 2024 and NGC 2071
respectively.

Previous studies of the clump distribution in the dust continuum emission of
Orion B have identified a factor of $\sim 10$ fewer condensations than
we do from the $^{13}$CO emission (J01, J06). This is probably because
the SCUBA observations are not sensitive to flux on large-scales
whereas the $^{13}$CO gas should trace this more ambient emission, which can account for the $^{13}$CO condensations identified which are not
spatially associated with a dust clump. Additionally, the spectral information in the $^{13}$CO data may break up multiple
condensations at distinct velocities along the line of sight, which may
be superposed in continuum data. For each dust clump
identified by J01 and J06 in our observing fields, we found the mean number of $^{13}$CO condensations whose peak
lies within the radii of the dust clumps to be $\sim 7.4$ for
NGC 2024 and $\sim 3.7$ for NGC 2071.

 \subsection{Condensation properties}

In the catalogue generated by CUPID, the `size' of each
condensation along each axis (right ascension, declination, velocity) is given
by the standard deviation of the pixel co-ordinate values about the centroid
position, weighted by the pixel values, then corrected to remove the effect of
instrumental smoothing.  We define the radius of a condensation as the
geometric mean of the size along axes 1 and 2 ($R_1$ and $R_2$):
\begin{equation}
R = \sqrt{R_1 R_2}\rmn{,}
\end{equation}
where $R_1$ and $R_2$ have been already been deconvolved with the beam
size.

For NGC 2024 the condensation radii range from $1.3\times10^{-3}$~pc to 0.053~pc
with a mean value of $0.018\pm 0.009$~pc.  For NGC 2071 the values are
similar; the radii range from $1.3\times10^{-3}$~pc to 0.075~pc with a mean
value of $0.017\pm 0.008$~pc.

We estimate the three-dimensional velocity dispersion of each condensation from
the one-dimensional $^{13}$CO velocity dispersion, $\sigma_{^{13}{\rm
    CO}}$ (the size along axis 3 calculated by CUPID), as described in
Sec.~\ref{sec:mass}. For NGC 2024, the velocity dispersions range from 0.66 to 2.0~\kms, with a
mean value of  $0.99\pm$0.23~\kms.  Again the values are similar for NGC 2071,
ranging from 0.50 to 2.5~\kms\ with a mean of $0.88\pm$0.19\kms.

Assuming the clouds are in LTE and the $^{13}$CO $J=3\to 2$ transition is optically
thin, the masses of the $^{13}$CO clumps can be derived following
Sec.~\ref{sec:mass}. At moderately high densities (which we expect the
$^{13}$CO $J=3\to 2$ transition to probe, given its critical density)
the cloud gas and dust should be coupled and at the same
temperature \citep{burke1983}. The mean temperatures of the dust
clumps associated with our $^{13}$CO condensations, computed by J06 and
J01, are $45 \pm 11$~K and $26\pm 8$~K for NGC 2024 and NGC 2071
respectively. We therefore adopt excitations temperatures of $T_{\rm
  ex} = 50$ and 30~K for NGC 2024 and NGC 2071 respectively in our mass
calculations, which are
also consistent with the temperatures we derived in Sec.~\ref{sec-tex} (see
Fig.~\ref{fig-tex}). 

In NGC 2024, we find the masses of the condensations range from $2.6 \times
10^{-3}~\msol$ to $13~\msol$, with a mean value of  $0.52\pm1.2~\msol$.
For NGC 2071 the condensations are less massive, ranging from $3.4 \times 10^{-3}~\msol$ to $5.3~\msol$, with a mean
value of  $0.22\pm0.34~\msol$.  These masses are much less that those derived
from the dust analysis of J06 and J01, who found clumps as massive as
$90~\msol$ in NGC 2024 and $30~\msol$ in NGC 2071.  As mentioned above, this
could be because there are several $^{13}$CO condensations associated with each
dust clump.  Furthermore, opacity effects and freeze-out of CO in dense regions
could reduce the observed intensity and hence the mass estimate.

The virial mass of a spherical condensation with a density profile of $\rho
\propto r^a$ is calculated using \citep{maclaren1988,williams1994}:

\begin{equation}
M_{\rm vir} = \frac{5R\sigma_{v,3D}^2}{3\gamma G}
\end{equation}

\noindent where $R$ is the radius of the condensation, $\sigma_{v,3D}$ is the
three dimensional velocity dispersion (Eq.\ref{eq-vdisp}) and $G$ is the
gravitational constant. Assuming the condensations have density distributions
of $\rho \propto r^{-2}$, $\gamma = 5/3$. The virial masses for NGC 2024 range
from 0.21 to 30~$\msol$ with a mean of $5.1\pm4.6~\msol$,
while for NGC 2071 $M_{\rm vir}$ ranges from 0.13 to 32~$\msol$
with a mean, $3.4\pm2.7~\msol$.  The virial masses are much higher
than the LTE masses for \emph{every} condensation, but this is partly
because the LTE masses are underestimates.

 \subsection{Correlation between the condensation properties}
 \label{params}

\begin{figure}
\includegraphics[width=85mm]{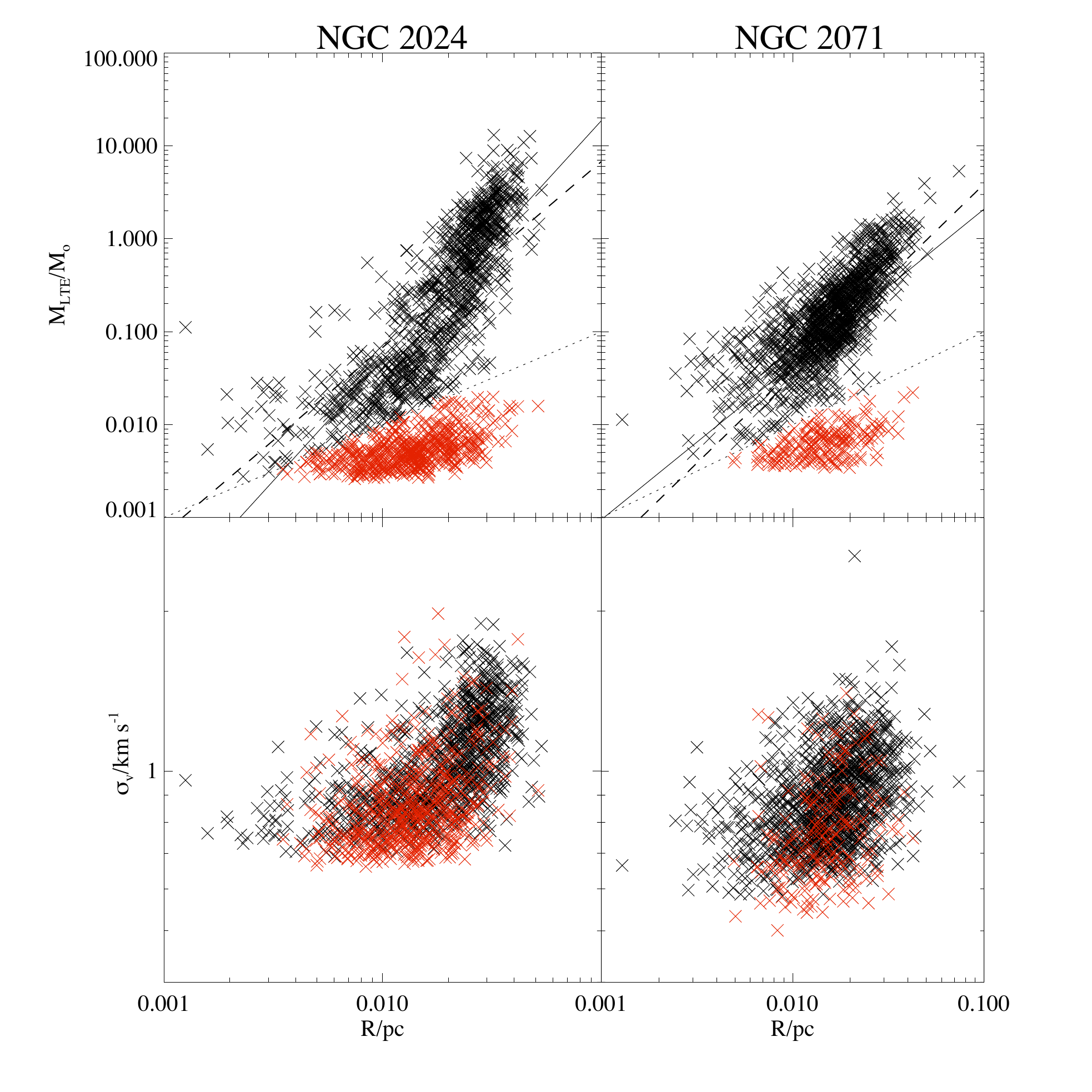}
\caption{$M_{\rm LTE}-R$ (top panels) and $\sigma_{\rm v}-R$ (bottom panels)
relations for NGC 2024 (left) and NGC 2071 (right).  Condensations with $M_{\rm
LTE}/\msol < R/{\rm pc}$ are plotted in red because they appear to form a
separate population.  In the top panels the dotted lines show $M_{\rm
LTE}/\msol = R/{\rm pc}$, the dashed lines show $M_{\rm LTE} \propto R^2$
(Larson's Law), and the solid lines show the lines of best fit: $M_{\rm LTE}
\propto R^{2.6}$ for NGC 2024 and $\propto R^{1.7}$ for NGC 2071.
}
\label{corr}
\end{figure}

In Fig.~\ref{corr} we plot the $M_{\rm LTE}-R$ and $\sigma_{\rm v}-R$
relations for both clouds.  In the plots of $M_{\rm LTE}$ versus $R$, there
appears to be a separate population of low-mass condensations.   These low-mass condensations are roughly separated by the line
$M_{\rm LTE}/\msol=R/{\rm pc}$ and are plotted in red.  The remaining
condensations (plotted in black) appear to have a strong correlation between
mass and radius, with correlation coefficients of 0.8 for both clouds.  The
linear regression coefficient ($s_{xy}/s_{xx}$) of $\log (M_{\rm LTE}/\msol)$
versus $\log (R/{\rm pc})$ is 2.6 for NGC 2024 and 1.7 for NGC 2071, implying
that the mass-radius relationship is of the form $M_{\rm LTE}/\msol \propto
(R/{\rm pc})^{\sim2.6}$ and $\propto (R/{\rm pc})^{\sim1.7}$ for NGC 2024 and
NGC 2071 respectively.
 Given the scatter in the plots and the uncertainties in the values of $M_{\rm
LTE}$ and $R$, the relation for both clouds is consistent with the one found by
\citet*{kramer1996} for their $^{13}$CO condensations in the southern part of Orion
B, which had a power law index of 2.2.
 The relations are also consistent with Larson's Law relating mass and radius,
which is of the form $M_{\rm LTE} \propto R^2$ \citep{larson1981}.
In Fig.~\ref{corr} the linear regression fits are shown with the solid black
line, and the Larson relation is shown with the broken line.

The distinct population of low-mass condensations, on the other hand, only have a
weak correlation between mass and radius (with correlation coefficients of 0.5
in both clouds). Since these populations have such low masses and behave
differently to the rest of the condensations it is possible they have
some contamination from noise or they may be a population of transient
objects in the ambient cloud, not related to the active star formation.

The plots of the velocity dispersion against radius in Fig.~\ref{corr} do not show an obvious
correlation. This is reflected in the low values of the correlation
coefficients (0.2 for each cloud).
In these plots the populations of low-mass condensations mentioned above are
also plotted in red, but here they are not distinct from the other
condensations.
Several other surveys have also found very weak or no correlation between
$\sigma_{\rm v}$ and $R$ for condensations in molecular clouds
\citep[e.g.][]{kramer1996,onishi2002}.  For a larger range of molecular cloud
sizes, \citet{larson1981} found that $\sigma_{\rm v} \propto R^{\sim 0.4}$.
The lack of correlation found here could be because of the small range in $R$
and $\sigma_{\rm v}$ and the large scatter in the values.

\begin{figure}
\includegraphics[width=85mm]{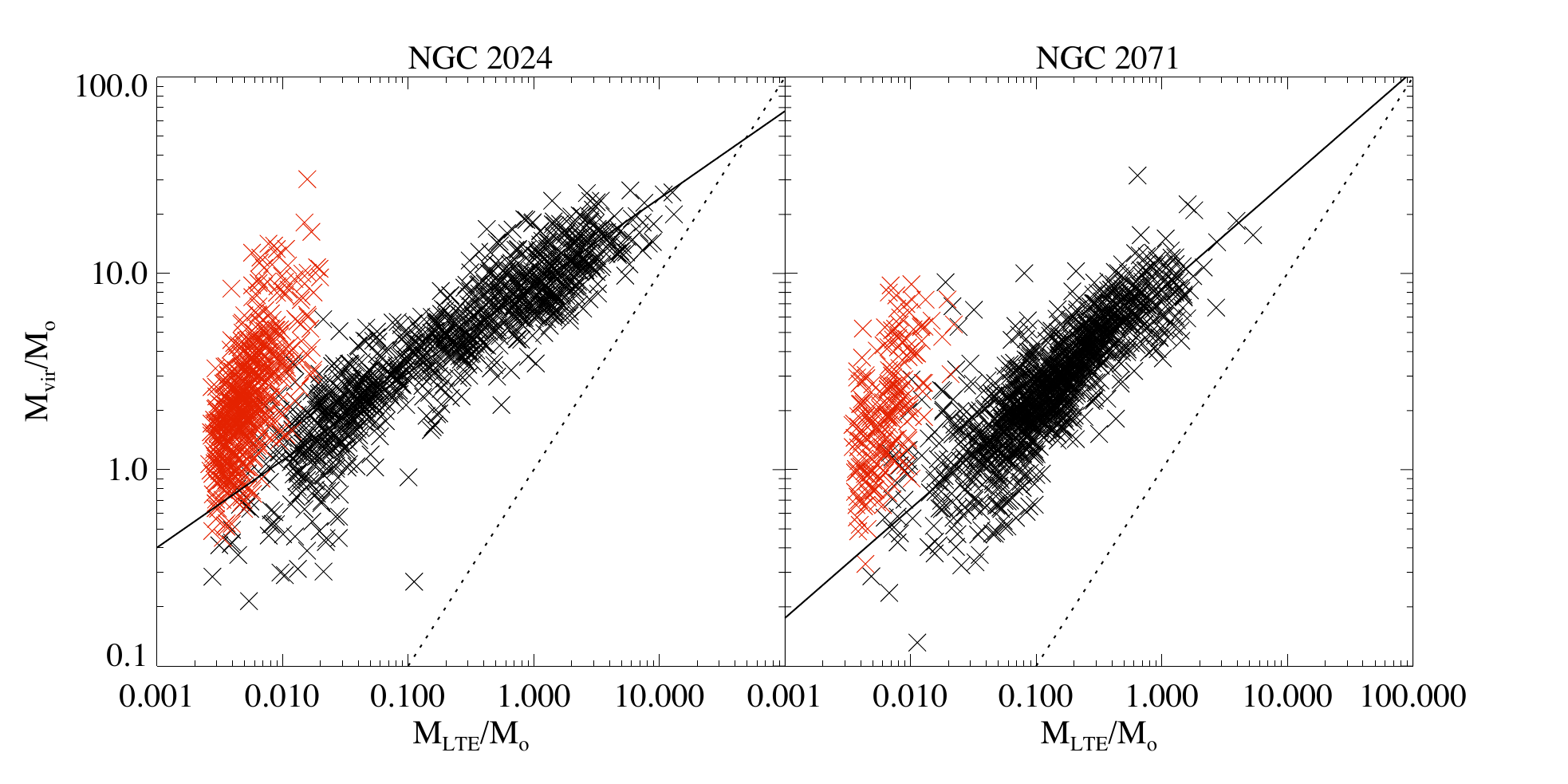}
\caption{Plots of virial mass versus LTE mass for NGC 2024 (left) and NGC 2071
(right).  Condensations with $M_{\rm LTE}/\msol < R/{\rm pc}$ are plotted in
red.  The solid lines on each plot show the lines of best fit ($M_{\rm vir}
\propto M_{\rm LTE}^{0.4}$ for NGC 2024 and $\propto M_{\rm LTE}^{0.6}$ for
NGC 2071) and the dotted lines show $M_{\rm vir} = M_{\rm LTE}$. }
\label{mvir}
\end{figure}

Fig.~\ref{mvir} shows the relationship between $M_{\rm vir}$ and $M_{\rm LTE}$.
Again, the low-mass condensations identified above are plotted in red,
and they appear as a distinct population in this plot.
 The rest of the condensations are relatively well-correlated, with correlation
coefficients of 0.9 for NGC 2024 and 0.8 for NGC 2071.  The plots appear to be
well-fitted with a power law (solid line), $M_{\rm vir} \propto M_{\rm
LTE}^\beta$, with $\beta = 0.4$ for NGC 2024 and 0.6 for NGC 2071, calculated
from the regression coefficient in log-log space.
 \citet*{ikeda2009} found $M_{\rm vir}/M_{\rm LTE} \propto M_{\rm LTE}^{-0.33}$
for H$^{13}$CO$^+$ condensations in Orion B (i.e. $\beta = 0.67$), which is
very close to our value for NGC 2071.
 The dotted line on each plot shows where the condensations are in
approximate equipartition ($M_{\rm vir}=M_{\rm LTE}$).  The condensation virial
masses are in fact \emph{all} much higher than their LTE masses, implying that the condensations
are unbound.  However, there are large uncertainties in both the virial
and LTE masses (because of, for example,  uncertainties in the distances to the
cloud, the fractional abundance of $^{13}$CO, the excitation temperature, and
the assumed density  profile), therefore we cannot make any definite
conclusions about the fate of any particular condensation from this
plot.  However, the relative positions of the condensations to each
other should be more robust and suggests that condensations from the low-mass population (in red)
are in general less bound than the rest (i.e.\ they have higher
$M_{\rm vir}/M_{\rm LTE}$ ratios). 

 \subsection{Condensation mass function}
 The differential condensation (or core) mass function (CMF) is usually fitted
by the power law
\begin{equation}
\frac{dN}{dM} \propto M^{-\alpha}.
\label{CMF}
\end{equation}
If these condensations are the direct precursors of stars, the form of
the CMF may provide insights into the origin of the stellar initial mass
function (IMF), which
is described by Equation~\ref{CMF} with $\alpha = 2.35$ \citep{salpeter1955}
over a large range of environments. For instance the similarity of the
CMF to the IMF for the earliest stages of protostellar evolution,
may rule out models of star formation where the shape
of the CMF is set at later stages \citep[e.g.][]{bate}.

\begin{figure}
\includegraphics[width=85mm]{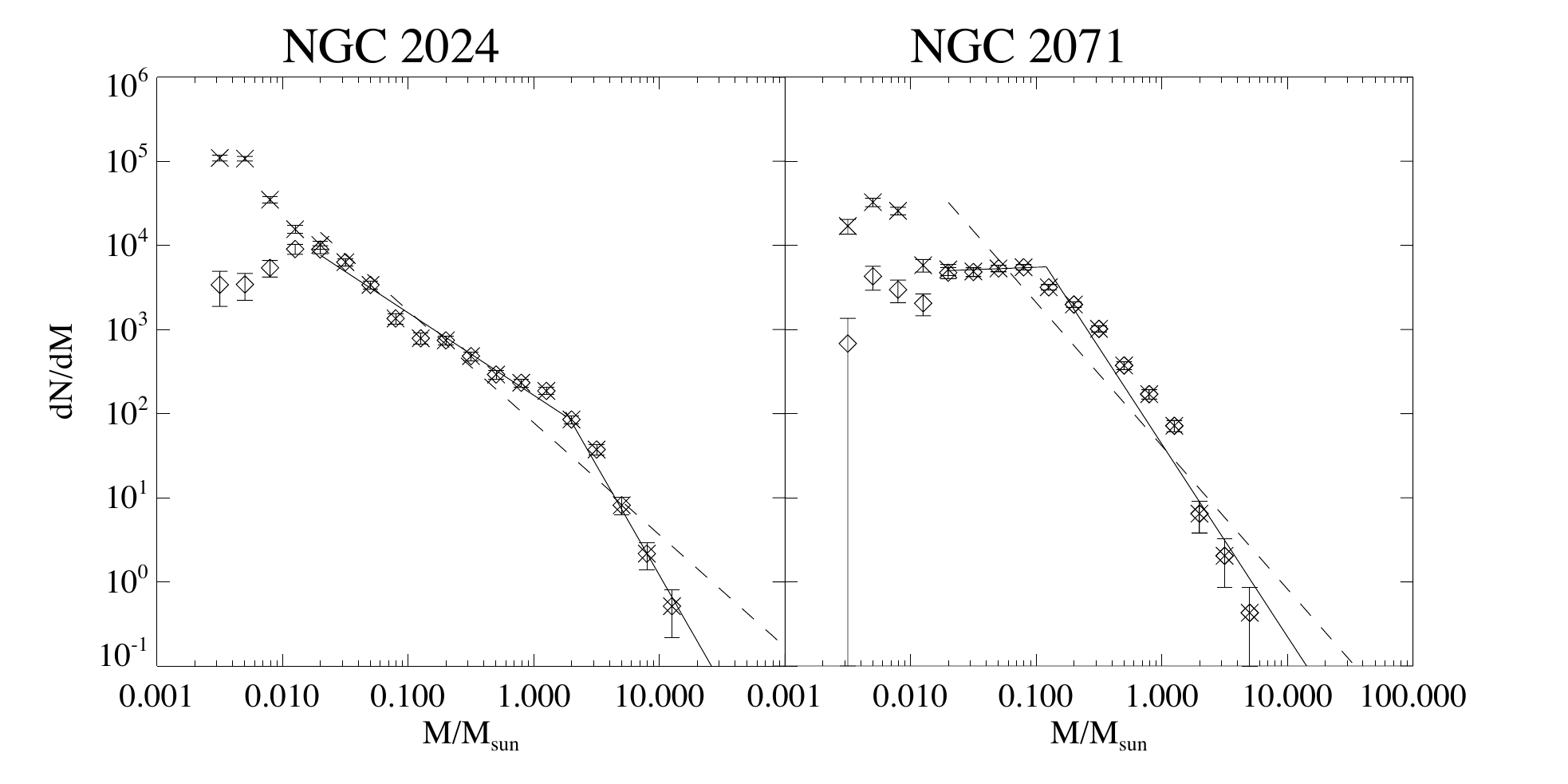}
\caption{The differential mass functions of $^{13}$CO condensations in NGC 2024
(left) and NGC 2071 (right).  The crosses mark the mass function including all
of the condensations and the diamonds mark the mass function with condensations
with $M_{\rm LTE}/\msol < R/{\rm pc}$ removed.
The errors plotted show the statistical uncertainties of $\sqrt{N}$, where $N$
is the sample number in each mass bin.
The dashed lines show the single power law fits ($\alpha = 1.3$ for NGC 2024 and
$\alpha =1.7$ for NGC 2071), and the solid lines show the double power law fits
($\alpha_{\rm low} = 1.0$ for $0.02 < M_{\rm LTE}/\msol < 2.0$ and
$\alpha_{\rm high} = 2.6$ for $M_{\rm LTE} > 2 \msol$ for NGC 2024 and
$\alpha_{\rm low} = 0.06$ for $0.02< M_{\rm LTE}/\msol < 0.12$ and
$\alpha_{\rm high} = 2.3$ for $M_{\rm LTE} > 2 \msol$ for NGC 2071).}
\label{MF}
\end{figure}

Fig.~\ref{MF} compares the plots of $dN/dM$ versus $M_{\rm LTE}$ for both
clouds.  The crosses mark the mass function including all of the condensations
and the diamonds mark the mass function with the population of low-mass
condensations, discussed in Sec.~\ref{params}, removed.
The errors plotted show the statistical uncertainties of $\sqrt{N}$, where $N$
is the sample number in each mass bin.
The two mass functions for each cloud are the same for $M_{\rm LTE}\ga
0.02~\msol$.  Below this mass, the total mass function continues to increase
with decreasing mass, whereas the mass function with the low-mass population
removed has a turning point and begins to decrease with decreasing
mass.  Due to the increasing incompleteness in the mass function at these low masses
arising from our detection thresholds, we have only attempted to
fit the mass function for $M_{\rm LTE}\ga 0.02~\msol$.

Reduced-squared fitting of a single power law mass function to NGC 2024 gives
$\alpha \sim 1.3$ for $M_{\rm LTE}\ga 0.02~\msol$ (plotted with the dashed
line on Figure~\ref{MF}).
The data suggest that there is a break in the power law at $M_{\rm LTE} \sim
2.0~\msol$.
Fitting a double power law gives $\alpha_{\rm low} \sim 1.0$ for $0.02 < M_{\rm
LTE}/\msol < 2.0$ and
$\alpha_{\rm high} \sim 2.6$ for $M_{\rm LTE} > 2~\msol$.

For NGC 2071, a single power law gives $\alpha \sim 1.7$ for $M_{\rm LTE}\ga
0.02~\msol$, but it is clear that the mass function flattens off for $0.02
<M_{\rm LTE}/\msol < 0.12$.  Fitting a double power law gives $\alpha_{\rm
low} \sim 0.06$ for $0.02< M_{\rm LTE}/\msol < 0.12$ and $\alpha_{\rm high}
\sim 2.3$ for $M_{\rm LTE} > 2 \msol$.  The data suggest there could be
another break in the power law at $M_{\rm LTE} \sim 1.0~\msol$.

The single power law values for $\alpha$ are similar to the values of $1.6 -
1.8$ derived using isotopologues of CO by \citet{kramer1998} for seven
molecular clouds, including Orion B South.
For the double power law fits, the values of $\alpha_{\rm high}$ are similar to
those derived by J01 for the dust condensation of Orion B North  and by
\citet{ikeda2009} for the H$^{13}$CO$^+$ condensations of Orion B, which had
values of ranging from 2.3 to 3.0.
These values are also very similar to the slope of the initial mass function
($\alpha=2.35$), which has been interpreted as evidence for a physical link
between the CMF and IMF.

\citet{ikeda2009} also found $\alpha_{\rm low} \sim 0.06$, just as we have
found for NGC 2071.  They suggested the flattening could be  a confusion effect
where low-mass cores are misidentified as parts of high-mass cores.
The turnover point found by \citet{ikeda2009} was much higher however, at
$6~\msol$, compared to our value of $0.12~\msol$.  Again this could be
because the $^{13}$CO LTE masses are underestimated due to opacity and
freeze-out.

From our analysis on the $^{13}$CO condensations in NGC 2024 and NGC 2071, we
have found that NGC 2024 has a wider mass range than NGC 2071 (with masses up to
$\sim 13~\msol$ for NGC 2024, compared to $\sim 5~\msol$ for NGC 2071).
Compared to the virial masses and condensation masses derived from other
tracers, the $^{13}$CO mass estimates appear to be underestimated, possibly due
to optical depth and freeze-out effects.  The relationships between the
condensation properties are similar to those seen in previous surveys of
condensations of CO isotopologues and H$^{13}$CO$^+$. The slopes of the CMF at
the high-mass ends ($\sim 2.6$ for NGC 2024 and $\sim 2.3$ for NGC 2071)  are
very similar to the slope of the initial mass function.

\section{Summary}
\label{sec:summ}

We have offered a first look at the data on Orion B being collected  by
HARP/ACSIS on the JCMT for the Gould Belt Legacy Survey. Our observations of
CO, $^{13}$CO and C$^{18}$O $J=3\rightarrow2$ provide a comprehensive
determination of the characteristic physical properties and dynamics of star
forming regions.  We have observed two large regions in Orion B,
(10.8$\times$22.5)~arcmin$^2$ in NGC 2024, and (13.5$\times$21.6)~arcmin$^2$ in
NGC 2071. Towards NGC 2024, $^{12}$CO and $^{13}$CO are detected throughout the
region, while C$^{18}$O is detected in an extended ridge, which follows the
dust emission, and also in a fragmented ring surrounding a bright optical
nebula. The $^{12}$CO line profiles indicate multiple line of sight components.
Comparisons of the intensity ratios of the three isotopologues indicate that
the $^{12}$CO is optically thick throughout the cloud, and also in the
lower-velocity outflow material, while the C$^{18}$O is generally optically
thin. These opacity results also apply to NGC 2071.

Towards NGC 2071, the molecular gas is concentrated in a region almost
completely surrounding a bright optical nebula. The $^{12}$CO emission is
heavily self-absorbed, and $^{13}$CO also shows self-absorption in the densest
regions. The main outflow is very energetic, and is seen to produce extended
wings even in the C$^{18}$O line profiles.

Both clouds show a complex clumpy and filamentary structure. In the regions
traced by C$^{18}$O, the equivalent widths suggest the bulk of the quiescent
material has a lower velocity dispersion than the material in which it is
embedded, increasing only in the regions of current star formation. The
energetic outflow activity is contributing a few percent of the kinetic energy
of NGC 2024, while the outflows dominate the kinetic energy towards NGC 2071.
Towards NGC 2024, the low-velocity red-shifted material is extended and
filamentary, while the low-velocity blue-shifted material is clumpy and
compact. 

A {\sc CLUMPFIND} analysis on the $^{13}$CO data finds multiple $^{13}$CO
condensations spatially associated with SCUBA dust cores, $\sim$7.4
towards NGC 2024, and $\sim$3.7 towards NGC 2071. The condensations in
NGC 2024 have higher LTE and virial masses than in NGC 2071, and
towards both clouds, $M_{\rm vir} \ge M_{\rm LTE}$. The slopes of the
condensations' mass function at the high{\bf -}mass ends are in
agreement with the slope of the initial mass function. Towards NGC 2024, we also detect
a group of condensations that appear to be less bound, which may be
tracing transient clumps that do not go on to form stars.

These data are still being collected, and the Gould Belt Survey has a
final 1$\sigma$ noise requirement of of 0.3~K in 1.0~\kms\ for
$^{12}$CO, and 0.3~K in 0.1~\kms\ for C$^{18}$O per 7.5~arcsec
pixel. $^{13}$CO is obtained simultaneously with C$^{18}$O, and the
C$^{18}$O requirement sets a 1$\sigma$ noise level of 0.25~K in
0.1~\kms\ for $^{13}$CO. When the full HARP data set has been
obtained, we will carry out more detailed analyses of the
individual cores and outflow properties in the Orion B clouds. The
completeness limit for the $^{13}$CO results will be improved, although
overall condensation masses, and the slope of the condensations' mass
function will not be significantly affected.  With datasets of the
same sensitivity, the properties of the Orion B clouds will be
compared to the other star-forming clouds being observed by the GBS.
This data set will be highly complementary to the SCUBA-2 and POL-2
observations planned as part of this survey, and to planned {\it Herschel}
key programmes.

\section*{Acknowledgements}
The James Clerk Maxwell Telescope is operated by The Joint Astronomy Centre on
behalf of the Science and Technology Facilities Council of the United Kingdom,
the Netherlands Organisation for Scientific Research, and the National Research
Council of Canada. J.F.R. would like to thank E. Rosolowsky for help and
advice on analyzing the dense cores. J.F.R. acknowledges the support of the MICINN under grant number
ESP2007-65812-C02-C01.

\label{lastpage}

\end{document}